\documentclass[english,aps,amsmath, showpacs, 11pt]{revtex4}
\usepackage[T1]{fontenc}
\usepackage[utf8x]{inputenc}
\usepackage{float}
\usepackage{bm}
\usepackage{amsmath}
\usepackage{graphicx}
\usepackage{amssymb}
\usepackage{esint}

\makeatletter
\@ifundefined{textcolor}{}
{%
 \definecolor{BLACK}{gray}{0}
 \definecolor{WHITE}{gray}{1}
 \definecolor{RED}{rgb}{1,0,0}
 \definecolor{GREEN}{rgb}{0,1,0}
 \definecolor{BLUE}{rgb}{0,0,1}
 \definecolor{CYAN}{cmyk}{1,0,0,0}
 \definecolor{MAGENTA}{cmyk}{0,1,0,0}
 \definecolor{YELLOW}{cmyk}{0,0,1,0}
 }

\usepackage{subfigure}\usepackage{bm}\usepackage{epstopdf}%opening

\makeatother

\usepackage{babel}

\makeatother

\usepackage{babel}

\makeatother

\usepackage{babel}

\begin{document}

\title{Spin dynamics of hard-soft magnetic multi-layer systems: effect of
Exchange, Dipolar and Dzyaloshinski-Moriya interactions}

\author{A. F. Franco and H. Kachkachi}

\affiliation{Laboratoire PROMES CNRS UPR8521, Université de Perpignan Via Domitia,
Rambla de la Thermodynamique - Tecnosud, 66100 Perpignan, France}

\email{andres.franco@univ-perp.fr, hamid.kachkachi@univ-perp.fr}
\begin{abstract}
We investigate the effect of coupling (intensity and nature), applied
field, and anisotropy on the spin dynamics of a multi-layer system
composed of a hard magnetic slab coupled to a soft magnetic slab through
a nonmagnetic spacer. The soft slab is modeled as a stack of several
atomic layers while the hard layer, of a different material, is either
considered as a pinned macroscopic magnetic moment or as an atomic
multi-layer system. We compute the magnetization profile and hysteresis
loop of the multi-layer system by solving the Landau-Lifshitz equations
for the net magnetic moment of each (atomic) layer. We study the competition
between the intra-layer anisotropy and exchange interaction, applied
magnetic field, and the inter-slab exchange, dipolar or Dzyaloshinski-Moriya
interaction. Comparing the effects on the magnetization profile of
the three couplings shows that despite the strong effect of the exchange
coupling, the dipolar and Dzyaloshinski-Moriya interactions induce
a slight (but non negligible) deviation in either the polar or azimuthal
direction thus providing more degrees of freedom for adjusting the
spin configuration in the multi-layer system. 
\end{abstract}
\pacs{75.10.-b General theory and models of magnetic ordering - 75.78.-n Magnetization
dynamics  - 75.10.Hk Classical spin models}
\maketitle

\section{Introduction}

Although the {}``nano-rush'' tends to dominate the realm of technological
applications, especially magnetic recording, multi-layer magnetic
systems benefit from a growing interest in this area mainly due to
their high performance \citep{knehaw91i3e,vicshe05i3e,suessetal05j3m}.
On the other hand, magnetic multi-layer systems and thin films benefit
from the acquired long-standing experience and know-how both
in growth and characterization leading to a good control of the relevant
intrinsic parameters (dimensionality and anisotropy). Moreover, there
are many well-established techniques for precise measurements, such
as FMR \citep{urquhartetal88jap,heicoc93ap,heinrich94springer,farle98rpp,wolbac07prl},
BLS \citep{cochran94springer,mathieuetal98prl}, and the ever developing
optical techniques \citep{beaurepaireetal96prl,koopmansetal00prl,vankampenetal02prl,stosie06springer},
to cite a few.

The magnetization dynamics of laterally confined elements of alternating
magnetic and nonmagnetic layers exhibits a large variety of interesting
phenomena for both applications and fundamental research. In this
context, theory has to play its usual role of providing reasonable
models for interpreting the observed phenomena and suggesting new
experiments. An issue of particular interest in this context concerns
the coupling in multi-layer systems \citep{bruno95prb}. The inter-layer
coupling determines the mechanisms of transport and propagation of
a stimulus applied at one end of the structure and the mechanism of
the adjustment of magnetic configurations in metallic multi-layer
systems by an external magnetic field. 
In the context of perpendicular magnetic recording with exchange-spring media,
improved write-ability is achieved by appropriately tuning the coupling between
the soft and hard magnetic layers \cite{vicshe05i3e,
wangetal05apl, suessetal05apl,suessetal05i3e, schuermannetal06jap, bergeretal08apl}.
Therefore, it is of paramount importance to fathom the nature of interaction acting at the
interface and the role it plays in conveying any perturbation through the multi-layer
system. There is a great amount of published work investigating the
inter-layer coupling and its effects on the magnetic properties
\cite{hilletal93jap,varalt00prb,xiaetal97prb,henrichetal03prl, kakazeietal05jap,
skubicetal06prl,wolbac07prl,madamietal08jap,sunetal11jap}. 
For instance, in Ref. \onlinecite{wolbac07prl} the authors investigated
the magnetization dynamics due to spin currents in magnetic double
layers and argued that transport in this structure is governed by
a kind of long-range interaction called the \emph{dynamic exchange
coupling}. The microscopic mechanism underlying this effective coupling
and the way it affects the collective behavior of magnetic hybrid
structures require further investigation. While the RKKY coupling
provides an interpretation of experiments on very thin conducting
spacers assuming a high degree of perfection \citep{lacroixetal90tsf},
rough surfaces may induce strong stray fields and thereby dipolar
coupling in multi-layer systems. For a magnetic bi-layer several configurations
are considered in Refs.
\onlinecite{varalt00prb,wangetal90prl,albiretal95jmmm,edwardsetal95mseb}. In
particular, in Ref. \onlinecite{varalt00prb} the calculations confirm the
fact that in the absence of roughness the dipolar coupling between
two perfectly flat infinite planes vanishes and that for planes of
finite dimensions the dipolar coupling may give rise to a ferromagnetic
or an anti-ferromagnetic coupling. On the other hand, one cannot exclude
the Dzyaloshinski-Moriya interaction, which is an anti-symmetrical
exchange interaction and mainly stems from a combination of low symmetry
and spin-orbit coupling \citep{dzyaloshinsky58jpcs,moriya60prl}.
In the presence of disorder, especially at the interface of thin films
or multi-layer systems, the Dzyaloshinski-Moriya interaction has been shown to
play an important
role since local symmetry is broken by surface effects. Indeed, it
leads to large anisotropy and may even change the magnetic order,
see Ref. \citep{dzyaloshinsky58jpcs,moriya60prl,crelac98jmmm,moriya60pr}. In particular,
it has been shown that the Dzyaloshinski-Moriya interaction is induced by
spin-orbit coupling between
two ferromagnetic layers separated by a paramagnetic layer \citep{xiaetal97prb}.

In the present work, we bring a new contribution that attempts to
further clarify the role of three inter-slab couplings, namely exchange,
dipolar and Dzyaloshinski-Moriya, in the spin dynamics and magnetization
profile (MP). It is also essential to compare these interactions with
respect to their efficiency in the realignment of the spin configurations
and eventually the magnetization reversal. More precisely, we investigate
the effect of coupling (intensity and nature), applied field, and
anisotropy on the spin dynamics of a multi-layer system composed of
a hard magnetic slab coupled to a soft magnetic slab through a nonmagnetic
spacer. The soft slab is modeled as an array of several atomic layers
while the hard layer, of a different material, is either considered
as a pinned macroscopic magnetic moment or as an array of atomic planes.
We compute the magnetization profile and hysteresis loop of the whole
system by solving the (coupled) Landau-Lifshitz equations for the
net magnetic moment of each (atomic) layer.

This paper is organized as follows. After the introduction we define
the system we study in section II. In section III we deal analytically
and numerically with the rigid interface. Section IV deals with the
soft interface. It presents the results for the effects of applied
field, inter-slab couplings, and their comparison. It ends with the
results of hysteresis loops. We finally summarize the main results
in the conclusion and give a few perspectives for future investigations.

\section{Statement of the problem}

\begin{figure}[!htbp]
\begin{centering}
\subfigure[Rigid interface]{\includegraphics[width=4cm]{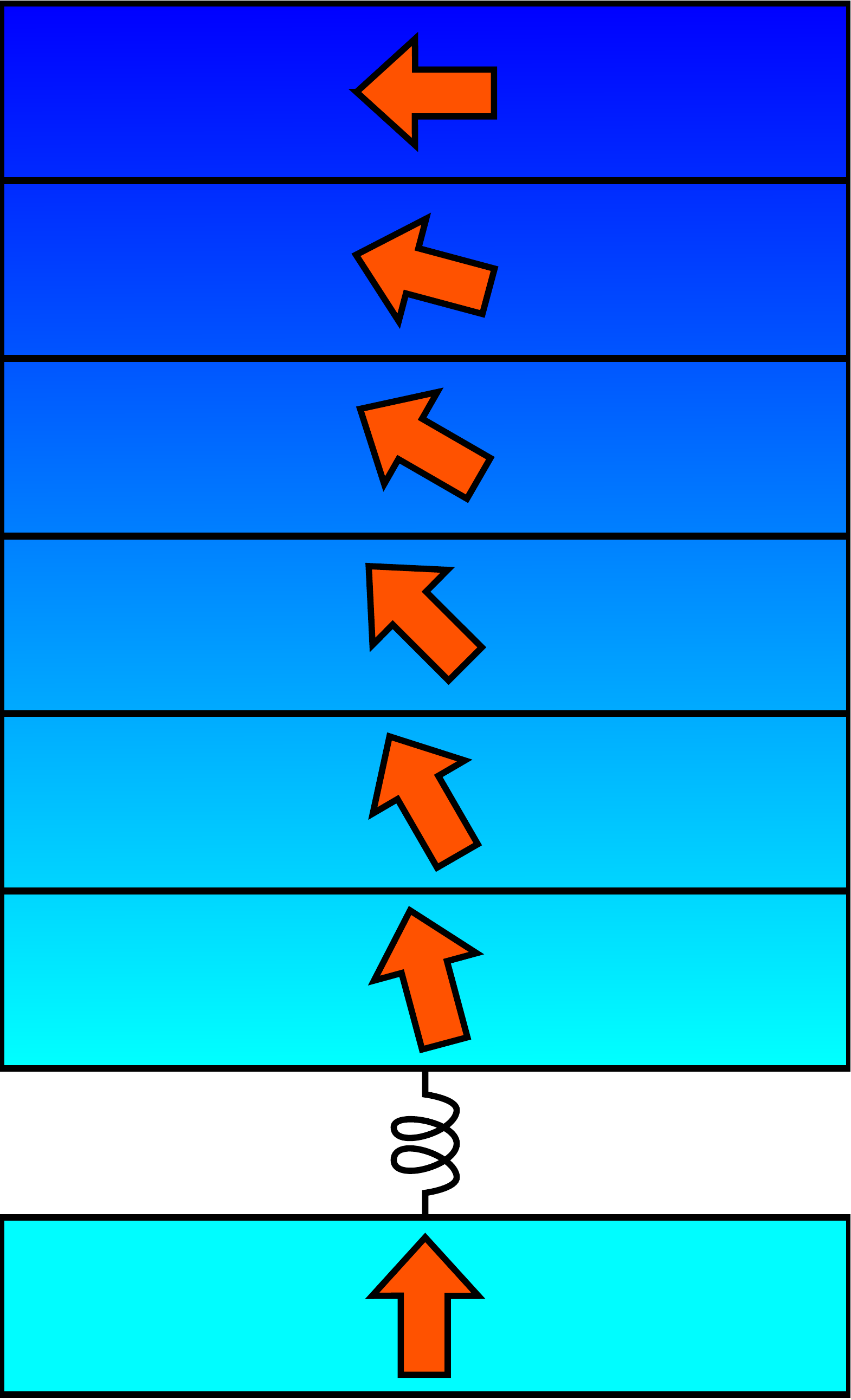}\label{fig:HSBLRigid}}
\subfigure[Soft interface]{\includegraphics[width=4cm]{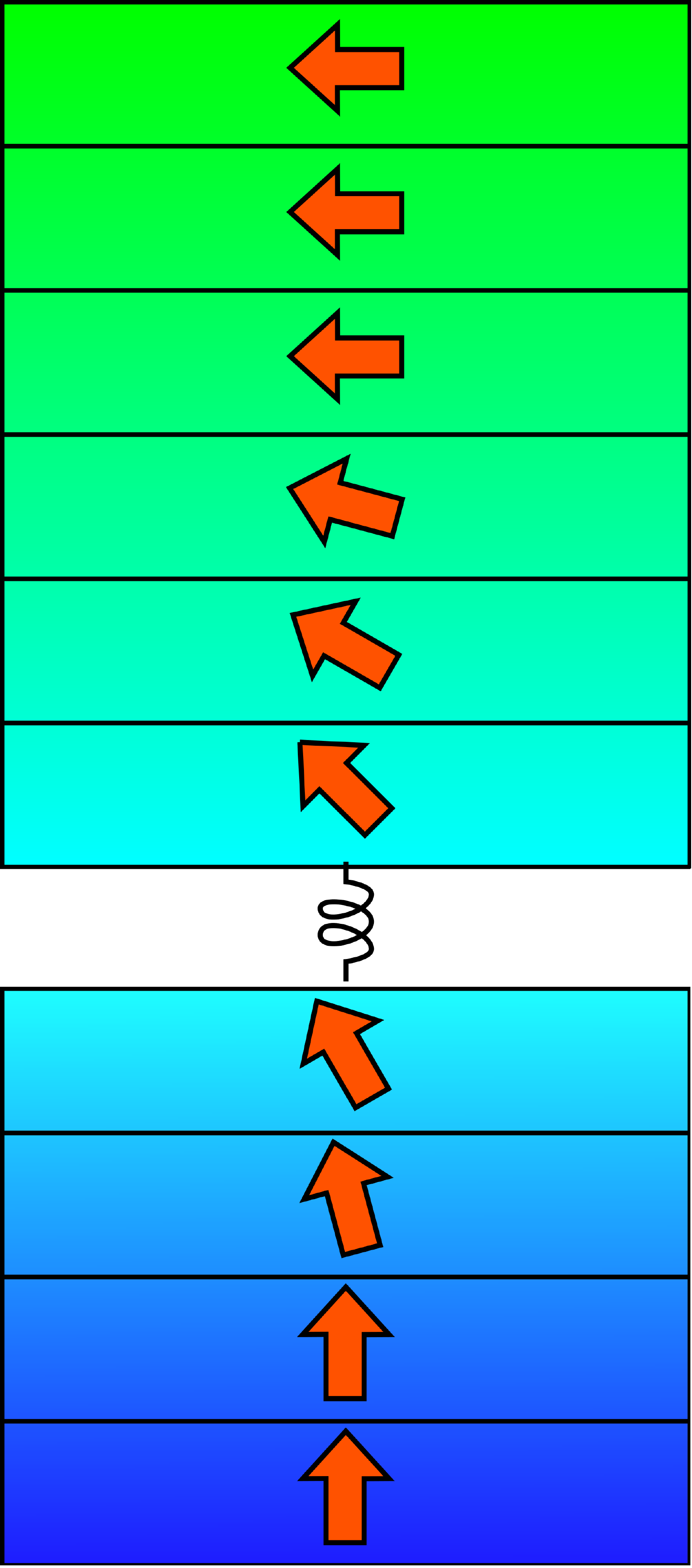}\label{fig:HSBLRelaxed}}
\caption{Scheme of a hard/soft coupled bi-layer system}

\par\end{centering}

\centering{}\label{fig:HSBL} %
\end{figure}

We consider the coupling between two slabs: one is a sufficiently
thin hard magnetic system (HMS) with strong out-of-plane anisotropy
and, on top of it, a thick soft magnetic system (SMS) with in-plane
anisotropy. The SMS slab will be modeled as a multi-layer system of
atomic planes of e.g. Fe, whereas the HMS slab will be modeled either
as i) a single macroscopic magnetic moment in which case we have a
rigid interface (RI), see Fig. \ref{fig:HSBLRigid} or as ii) a multi-layer
system of atomic planes of e.g. FePt, and in this case we get a relaxed
or soft interface (SI), see Fig. \ref{fig:HSBLRelaxed}. In both cases,
i.e. SMS /HMS-RI or SMS/HMS-SI we have an exchange-spring system with
a characteristic MP that depends on various physical parameters such
as the SMS-HMS coupling, the anisotropy and exchange coupling within
each slab, and the applied field. Accordingly, in this work, we will
investigate the effects of these parameters on the MP in the whole
system and on the hysteresis loop for the two configurations. In particular, we
will consider
and compare the effects of three kinds of inter-slab couplings, which
are exchange, dipole-dipole and Dzyaloshinski-Moriya. For later reference,
we denote by EI-ES, DDI-ES, and DMI-ES the exchange-spring system
where SMS and HMS are respectively coupled by exchange interaction
(EI), dipole-dipole interaction (DDI), or Dzyaloshinski-Moriya interaction
(DMI).

In Ref. \citep{francoetal11prb} the relatively simpler
case was considered where both the SMS and HMS were represented by their
respective net magnetic moments. The relaxation processes of the corresponding
magnetic dimer were then studied. In particular, the relaxation rates related
with the reversal of its magnetization were computed. Moreover, the effects
of the three couplings mentioned above were investigated for different
configurations of the anisotropy easy axes, and at finite temperature. 
In the present work, we are interested in the exchange-spring effect and in the
MP at equilibrium and at zero temperature. Therefore, the main objective
here is to investigate how the magnetic structure/state of the SMS
adapts to a change in the physical parameters, especially at the interface.

\section{Rigid interface : numerical versus analytical calculations}

\subsection{Magnetization profile of the multi-layer system}

\begin{figure}[!htbp]
\begin{centering}
\includegraphics[scale=0.25]{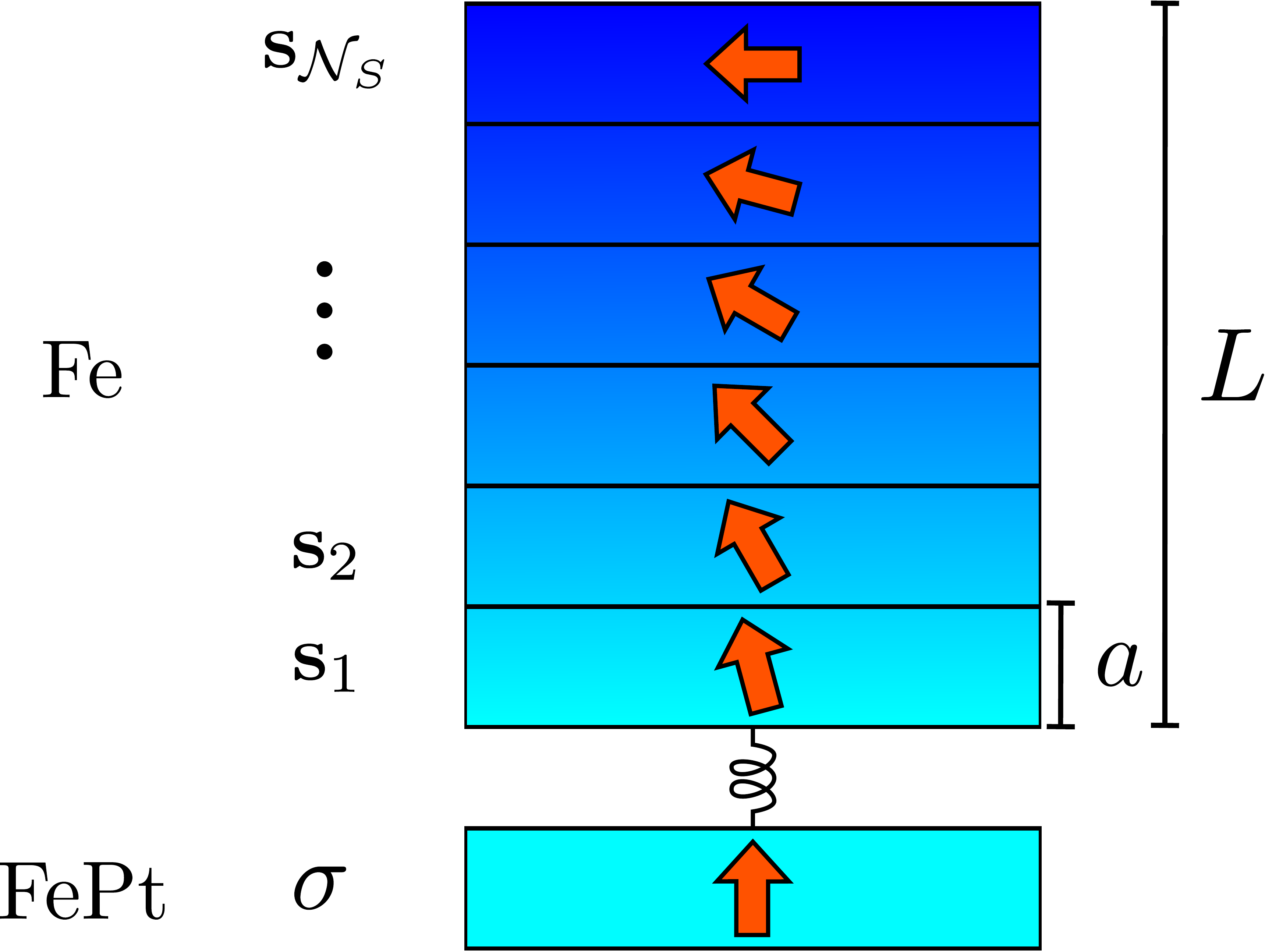} 
\par\end{centering}

\caption{\label{fig:RigidInterface}Setup of the SMS/HMS system with rigid
interface.}
\end{figure}

The exact solution for the MP has been obtained
for the system with RI, in the absence of the external magnetic field
\citep{sousaetal10prb}. For definiteness, the SMS-HMS lies on the
$xy$ plane and the out-of-plane anisotropy in the HMS slab points
in the $z$ direction. We denote the (net) magnetic moments of HMS
by $\bm{\mathbf{\sigma}}$, and the $\mathcal{N}_{s}$ magnetic moments
of the SMS by $\mathbf{s}_{i},i=1...,\mathcal{N}_{s}$. Therefore,
there is a total of $\mathcal{N}_{s}+1$ magnetic moments in the system
pictured here as an open chain of coupled magnetic moments. Indeed,
it is assumed that all atomic magnetic moments of each sub-layer $i$
in the SMS are parallel to each other so that their magnetic state
can be represented by the net (normalized) magnetic moment $\mathbf{s}_{i}$.
Hence, we do not include boundary effects in this study. For the HMS,
$\bm{\mathbf{\sigma}}$ represents the net magnetic moment of the
whole slab. The system setup is shown in Fig. \ref{fig:RigidInterface}.

If we denote by $\mathbf{e}_{h}$ the anisotropy easy axis (along
the $z$ axis) of HMS and $\mathbf{e}_{s}$ the anisotropy easy axis,
taken here along the $x$ axis, of the SMS, the total energy of the
system can be written as 
\[
E=E_{\mathrm{SMS}}+E_{\mathrm{HMS}}+E_{\mathrm{Int}},\]

with \begin{align*}
E_{\mathrm{SMS}} &
=-D_{s}\sum_{i=1}^{\mathcal{N}_{s}}\left(\mathbf{s}_{i}^{x}\right)^{2}-J_{s}
\sum_{i=1}^{\mathcal{N}_{s}-1}\mathbf{s_{i}}\cdot\mathbf{s}_{i+1}
\end{align*}
being the energy of the SMS slab. The first term is the in-plane
anisotropy with constant $D_{\mathrm{s}}$ (s stands for soft), the
second term is intra-SMS nearest-neighbor exchange coupling, of intensity
$J_s$. Similarly, the energy of the HMS slab contains the out-of-plane
anisotropy contribution and the exchange coupling. However, in the
present case of a RI, the energy coupling is dropped since we assume
that all atomic magnetic moments within the slab are already tightly
bound together by the exchange coupling. On the other hand, the anisotropy
contribution is also dropped because it is assumed to be strong enough to pin
the macroscopic magnetic moment $\bm{\mathbf{\sigma}}$ in the $z$ direction.

Finally, the interaction between the two slabs will be denoted in
the sequel by $\lambda$ and will be of three different origins~:
exchange, dipolar or Dzyaloshinski-Moriya. In the present case, $\lambda$
is the exchange coupling and will be set to $J_{0}$, so that

\[
E_{\mathrm{Int}}=-\lambda\bm{\mathbf{\sigma}}\cdot\mathbf{s}_{1}=-J_{0}\bm{
\mathbf{\sigma}}\cdot\mathbf{s}_{1}.
\]

The analytical solution of the problem requires an additional simplification
that consists in setting $J_{0}=J_s\equiv J$. Therefore, we can write
the energy (in units of exchange coupling) as 
\begin{eqnarray}
\mathcal{E}\equiv\frac{E}{J} & = &
-d_{s}\sum_{i=1}^{\mathcal{N}_{s}}\left(\mathbf{s}_{i}^{x}\right)^{2}-\sum_{i=1}
^{\mathcal{N}_{s}-1}\bm{s}_{i}\cdot\mathbf{s}_{i+1}-\bm{\mathbf{\sigma}}
\cdot\mathbf{s}_{1}\label{eq:energyRigidInterface}
\end{eqnarray} 
 where we have introduced the dimensionless parameter $d_{s}\equiv D_{s}/J_s$.

Then, we introduce the coordinate $z$ of each atomic sub-layer along
the $z$ direction, such that $z=0$ and $z=L$ represent the sub-layers
at the interface and the top of the SMS slab, with $L$ being the
thickness of the SMS slab and $a$ that of its atomic sub-layers,
\emph{i.e.} $L=a(\mathcal{N}_{s}-1)$. Note that $i=0$ represents
the HMS slab and a slight shift of $0.5a$ in the reference frame
between the numerical and the analytical calculations should be taken
into account. Indeed, the analytical calculations were performed in
the continuum limit. Moreover, the angle deviation with respect to
the $z$ axis $\theta_{i}-\theta_{i+1}$ of two adjacent magnetic moments
$\bm{s}_{i},\bm{s}_{i+1}$ is supposed to be small. For convenience, we denote
$\xi_i\equiv \theta_i-\theta_e$, 
where $\theta_e$ is the equilibrium polar angle of the multilayer system as
defined in \cite{sousaetal10prb}. For typical materials $d_s\ll1$,
thus $\theta_e\simeq0$.
In such a case one can write down the (finite-difference) equation
for $\xi$ from the energy minimization. The solution of this equation
can be written as \cite{sousaetal10prb} 
\begin{equation}
\begin{array}{lll}
z & = & \int_{0}^{\xi}\!\frac{d\eta}{\sqrt{C_{L}+d_{s}\cos2\eta}},\\
\\C_{L} & = &
\left(d_{s}\sin2\xi_{L}\right)^{2}-d_{s}\cos2\xi_{L},\end{array}
\label{eq:magnetProfileContinuum}
\end{equation} 
 $\xi_{L}$ being the angle deviation of the SMS top sub-layer, \emph{i.e.}
for $z=L$ and is given through the relation 
\begin{equation}
L=\intop_{0}^{\xi_{L}}\frac{d\xi}{\sqrt{C_{L}+d_{s}\cos2\xi}}.\label{eq:SMSTopLayer}
\end{equation}

There is a minimal number of sub-layers of the SMS slab, $\mathcal{N}_{s}^{\mathrm{min}}$,
or length of the chain, $L^{\mathrm{min}}$, necessary for an onset
of noncolinearities of the magnetic moments $s_{i}$. In Ref. \onlinecite{sousaetal10prb},
$\mathcal{N}_{s}^{\mathrm{min}}$ was found to be given by 
\begin{equation}
\mathcal{N}_{s}^{\mathrm{min}}\simeq\frac{\pi}{2\sqrt{2d_{s}}}.
\label{eq:MinimalNumber}
\end{equation}
This yields, for instance, $\mathcal{N}_{s}^{\mathrm{min}}\simeq11$ for
$d_{S}=0.01$.

\subsection{Magnetization profile of the exchange spring: numerical solution}

In order to obtain the MP for the system described above, we solve the set of
coupled Landau-Lifshitz equations (LLE) 
\begin{equation}
\frac{1}{\gamma}\frac{d\mathbf{s}_{i}}{dt}=\mathbf{s}_{i}\times\mathbf{h}_{i}^{
\mathrm{eff}}-\alpha\mathbf{s}_{i}\times\left(\mathbf{s}_{i}\times\mathbf{h}_{i}
^{\mathrm{eff}}\right),\label{eq:LLE}
\end{equation} 
where $\mathbf{h}_{i}^{\mathrm{eff}}=-\delta\varepsilon/\delta\mathbf{s}_{i}$
is the (dimensionless) effective field comprising the anisotropy and
exchange contributions, together with the interaction at the interface.
$\alpha$ is the phenomenological damping parameter, set here to $0.01$.
In these zero-temperature calculations, damping is used to drive the
system into the equilibrium state, \emph{i.e.} the state of minimal
energy. We start from a state of homogeneous magnetization along the
HMS anisotropy axis and allow the system to relax towards a minimum
that yields the MP of the system.

\begin{figure*}[!htbp]
 \includegraphics[width=5.5cm]{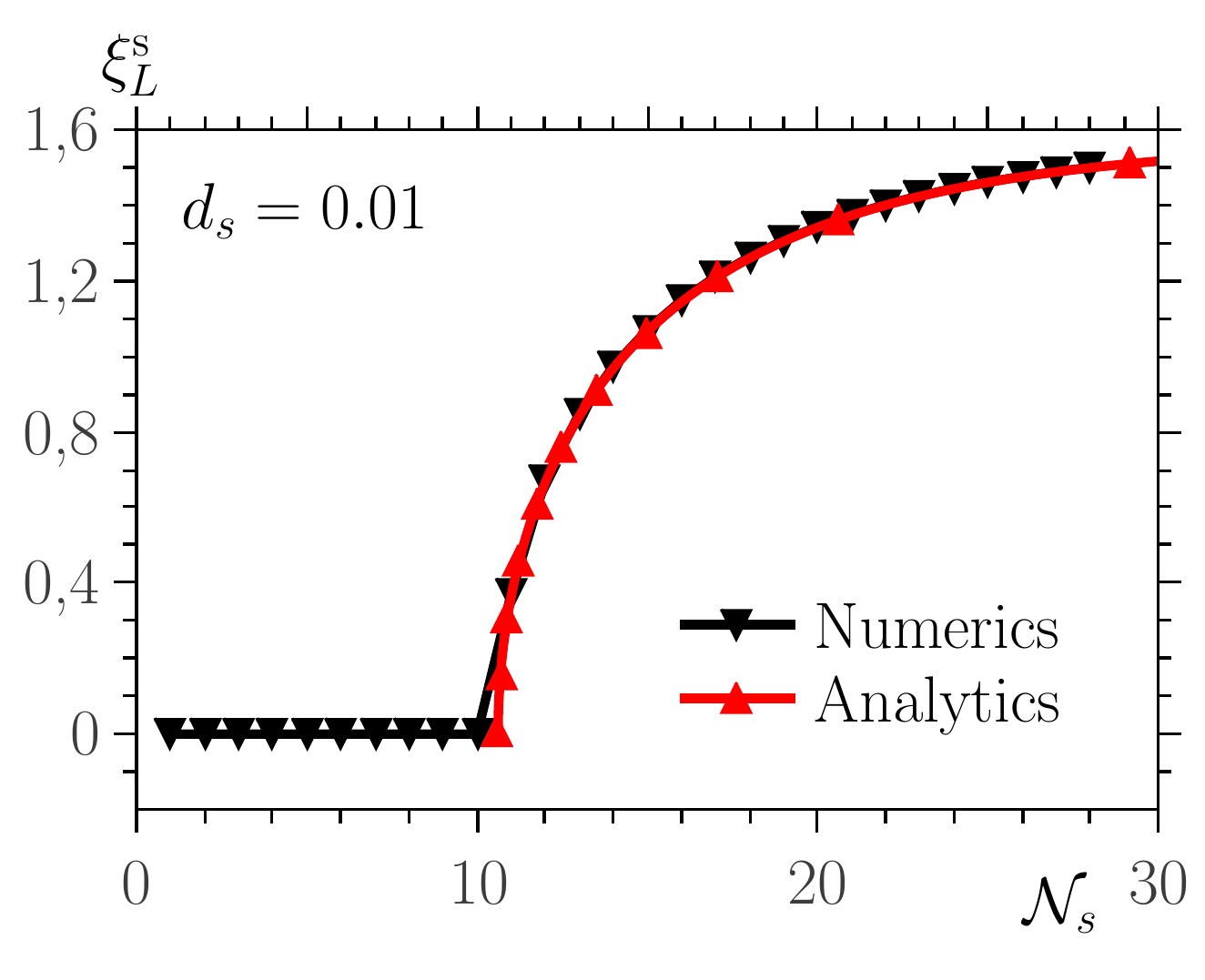} \includegraphics[width=5.5cm]{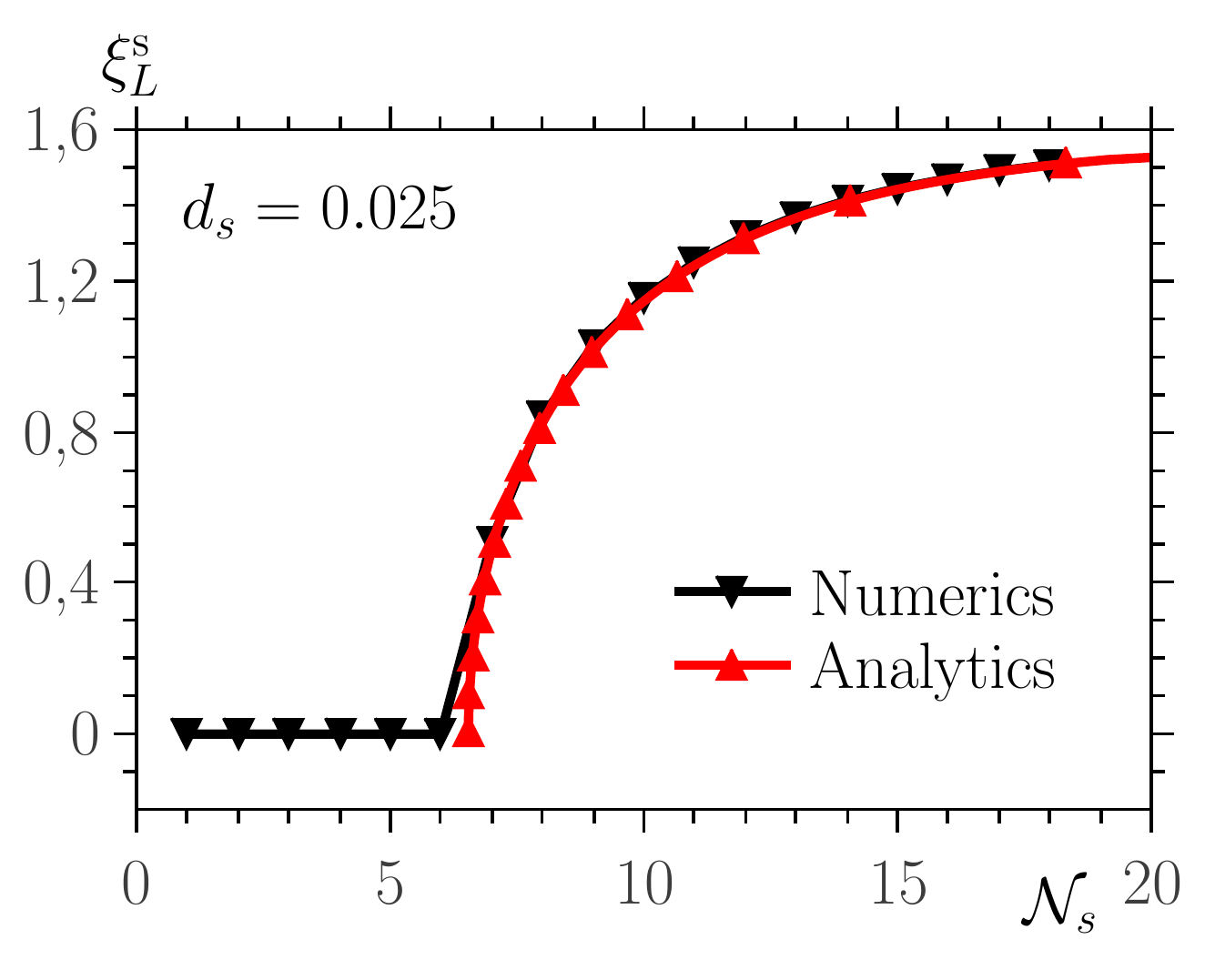}
\includegraphics[width=5.5cm]{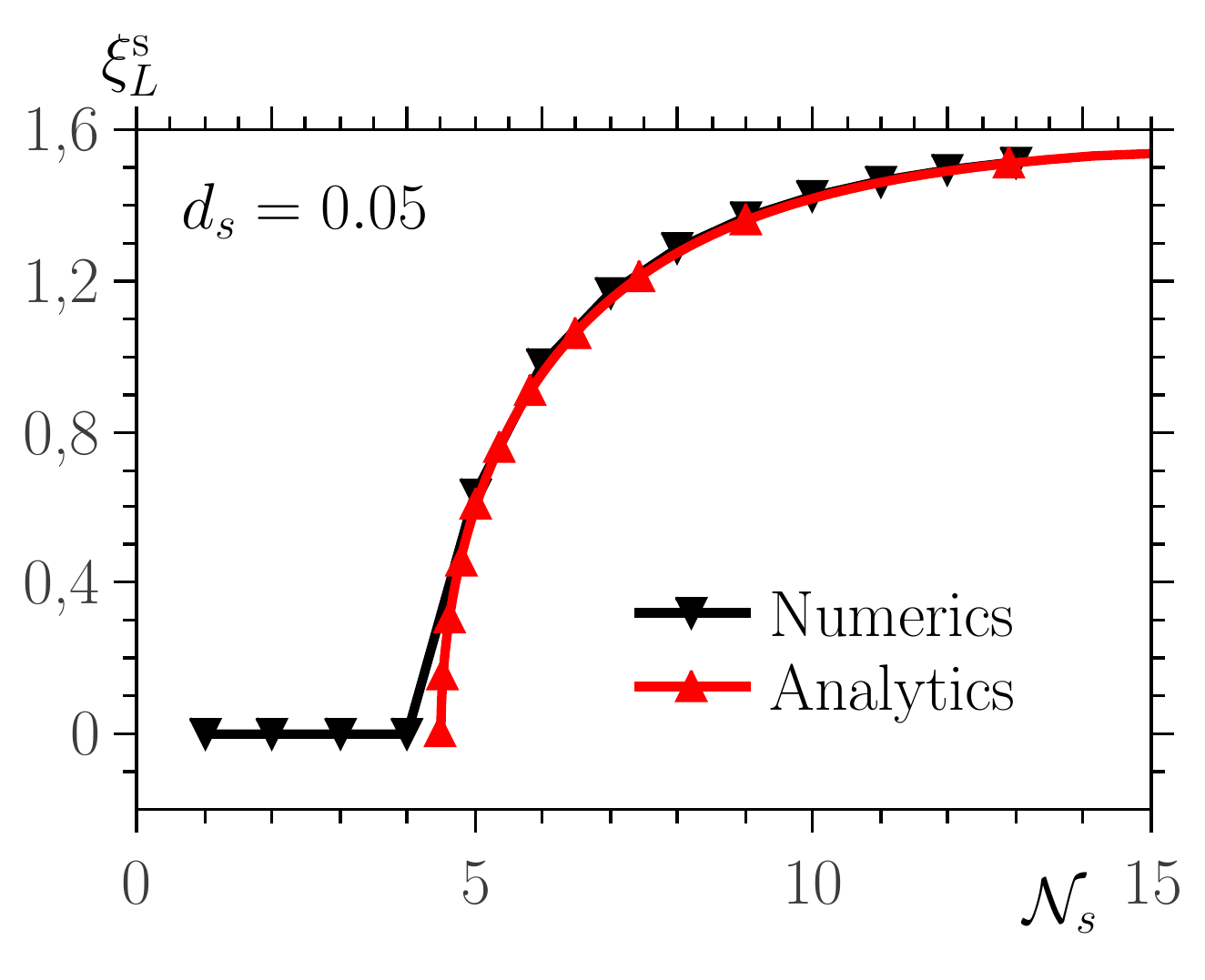} \caption{Comparison between analytical and numerical calculations of the magnetization
of the top sub-layer $\xi_{L}^{\mathrm{s}}$ for various values of
the SMS in-plane anisotropy, $d_{s}=0.01$ (left), $d_{s}=0.025$
(middle), $d_{s}=0.05$ (right).}

\label{fig:AvsNxiL} %
\end{figure*}

In Fig. \ref{fig:AvsNxiL} we plot the angle deviation $\xi^{\mathrm{s}}_L$ of the magnetic moment of the top SMS sub-layer against the 
number of sub-layers in the SMS slab. The results for three values of the in-plane
anisotropy $d_{s}$ show a good agreement between our numerical results and Eq. (\ref{eq:SMSTopLayer}). 

One can also see how $\mathcal{N}_{\mathrm{min}}$ depends on $d_{s}$
through Eq. (\ref{eq:MinimalNumber}). For a weak $d_{s}$ the system
behavior is dominated by the fixed magnetic moment of the HMS slab
through the interface exchange coupling. Adding more sub-layers to
the SMS slab attenuates this effect since the interface interaction
is counter-balanced by the anisotropy and intra-layer interaction
of each SMS sub-layer, leading eventually to a deviation of
the magnetic moment of the top SMS sub-layer for $\mathcal{N}\geq\mathcal{N}_{\mathrm{min}}$.
On the other hand, when $d_{s}$ increases, the contribution of each
SMS sub-layer to the energy of the whole system increases, thus leading
to a smaller $\mathcal{N}_{\mathrm{min}}$.

\section{Soft interface}

In the previous section, we defined a model for a system with rigid
interface and energy given by Eq. (\ref{eq:energyRigidInterface}).
The agreement with the previously published analytical work validates the
numerical approach used here, which can now be employed to deal with more
complex systems that are difficult to tackle analytically.

\begin{figure}[!htbp]
\begin{centering}
\includegraphics[scale=0.25]{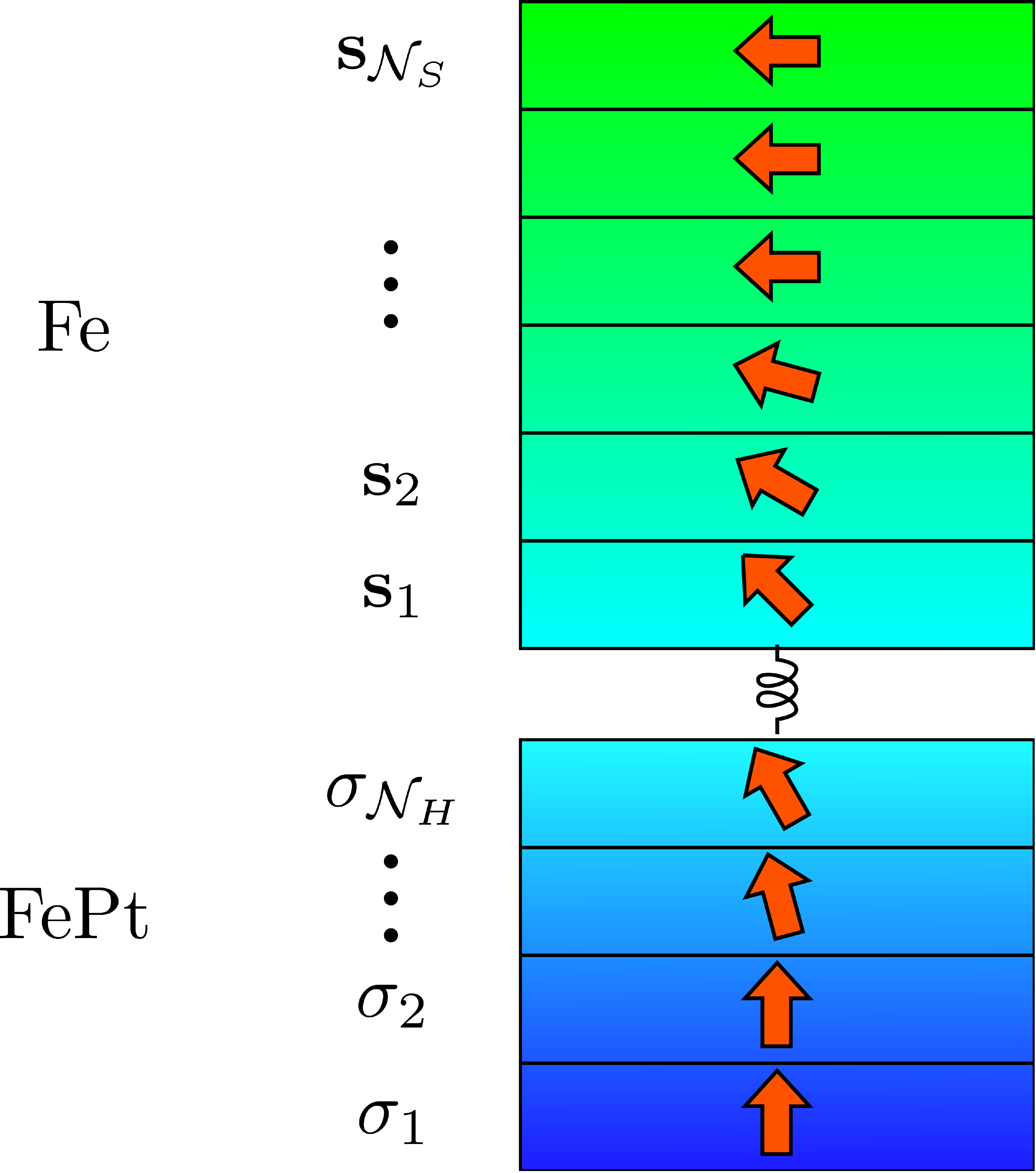} 
\par\end{centering}

\caption{\label{fig:SoftInterface}Setup of the SMS/HMS system with soft interface.}
\end{figure}

More precisely, we extend the previous model by replacing the macroscopic
magnetic moment $\bm{\mathbf{\sigma}}$ of the HMS, that was pinned
along the anisotropy direction, by $\mathcal{N}_{h}$ sub-layers
each with an out-of-plane (in the $z$ direction) uniaxial anisotropy
of intensity $D_{h}$. Moreover, in the present case we apply an external
magnetic field $H$ at an angle $\theta_{H}$ with respect
to the $z$ axis. Now, within the HMS slab, and similarly to the SMS
slab, each sub-layer of index $k$ carries a magnetic moment $\bm{\sigma}_{k}$
with $k=1,\ldots,\mathcal{N}_{h}$, see Fig. \ref{fig:SoftInterface}.

The energy of this model then reads \[
E=E_{\mathrm{SMS}}+E_{\mathrm{HMS}}+E_{\mathrm{Int}},\]
 with \begin{align*}
E_{\mathrm{SMS}} & =-\mu_{s}H\sum_{i=0}^{\mathcal{N}_{s}}\mathbf{s}_{i}\cdot\mathbf{e}_{H}-D_{s}\sum_{i=1}^{\mathcal{N}_{s}}\left(\mathbf{s}_{i}^{x}\right)^{2}-J_{s}\sum_{i=1}^{\mathcal{N}_{s}-1}\mathbf{s}_{i}\cdot\mathbf{s}_{i+1}\\
\end{align*}
 which is the energy (\ref{eq:energyRigidInterface}) to which we
have added the Zeeman contribution. \begin{align*}
E_{\mathrm{HMS}} &
=-\mu_{h}H\sum_{k=1}^{\mathcal{N}_{h}}\bm{\mathbf{\sigma}}_{k}\cdot\mathbf{e}_{H
}-D_{h}\sum_{k=1}^{\mathcal{N}_{h}}\left(\sigma_{k}^{z}\right)^{2}-J_{h}\sum_{
k=1}^{\mathcal{N}_{h}-1}\bm{\mathbf{\sigma}}_{k}\cdot\bm{\mathbf{\sigma}}_{k+1}.
\end{align*}
$\mu_{s}$ and $\mu_{h}$ are the atomic magnetic moments of the soft and hard
materials, respectively.

The interaction contribution is a function of $\mathbf{s}_{1}$ and
$\bm{\sigma}_{\mathcal{N}_{h}}$ and its form depends on the nature
of the coupling between the two slabs, \emph{i.e.} the top layer of
the HMS slab and the bottom layer of the SMS slab.
Here again, we use the same normalization as in Eq.~(\ref{eq:energyRigidInterface})
and divide by the exchange coupling $J_{s}$ within the SMS slab,
and introduce the following dimensionless parameters (with obvious
notation).

Hence, for the interaction energy we write \[
\mathcal{E}_{\mathrm{Int}}=\lambda\mathcal{F}\left(\mathbf{s}_{1},\bm{\sigma}_{
\mathcal{N}_{h}}\right)\]
where for exchange we have 
\begin{equation}
\mathcal{E}_{\mathrm{Int}}=-\lambda_{\mathrm{EI}}\bm{\sigma}_{\mathcal{N}_{h}}
\cdot\mathbf{s}_{1},\quad\lambda_{\mathrm{EI}}\equiv\frac{J_{hs}}{J_{s}},
\label{eq:DM-J}
\end{equation} 
for DMI \begin{equation}
\mathcal{E}_{\mathrm{Int}}=-\lambda_{\mathrm{DMI}}\mathbf{e}_{{\rm DMI}}\cdot\left(\bm{\sigma}_{\mathcal{N}_{h}}\times\mathbf{s}_{1}\right),\quad\lambda_{\mathrm{DMI}}\equiv\frac{D}{J_{s}},\label{eq:DM-DM}\end{equation}
and for DDI \begin{equation}
\mathcal{E}_{\mathrm{Int}}=\lambda_{\mathrm{DDI}}\bm{\sigma}_{\mathcal{N}_{h}}
\cdot\mathbf{\mathcal{D}}_{12}\mathbf{s}_{1},\quad\lambda_{\mathrm{DDI}}
\equiv\frac{\mu_{0}}{4\pi}\frac{\mu_{s}^{2}/d^{3}}{J_{s}}\label{eq:DM-DDI}
\end{equation} 
with \begin{equation}
\mathbf{\mathcal{D}}_{12}\equiv3\left(\leftarrow\cdot\mathbf{e}_{z}\mathbf{e}_{z
}\cdot\rightarrow\right)-1.\label{eq:DDI-tensor}\end{equation}
being the DDI tensor.

$h\equiv\mu_{s}H/J_{s}$ is the reduced magnetic field; $J_{hs}\equiv J_h/J_s$
is the inter-layer exchange interaction between the HMS and SMS, $D$
is the magnitude of the DM vector, and $d$ is the distance between
the SMS and HMS. Furthermore, we assume that the magnetic moment
of each sub-layer, within each slab, has the same magnitude.

Now, we are ready to discuss the results of several numerical studies
investigating the effects of the applied magnetic field $h$, the
in-plane anisotropy $d_{s}$ of the SMS slab, and the nature of the
inter-slab coupling and its intensity. In each case we compute the
magnetization profile, namely the angle deviation with respect to
the vertical ($z$) axis of each layer as we move upwards from the
bottom layer of the HMS slab to the top layer of the SMS slab. We
also compute in each situation the angle deviation of the bottom layer
of the HMS $\xi_{h}^{1}$ and the top layer of the SMS slab $\xi_{s}^{L}$,
as a function of the applied field.

The parameters used for the calculations are $d_{h}=0.02$, $j_{h}=1.44$
and $\mathcal{N}_{h}=10$. Note that the change from HMS to SMS occurs
when the sub-layer index $n$ passes from the top HMS layer to
the bottom SMS layer, with $n=0,...,\mathcal{N}_{h}+\mathcal{N}_{s}-1$
being the sub-layer index that labels the sub-layers of the whole system
(SMS+HMS). $\mathcal{N}_{s}=41$ for the calculations presented in
section \ref{sec:HmagHDir} and $\mathcal{N}_{s}=11$ for those presented
in sections \ref{sec:EI} to \ref{sec:DMI}. In order to better understand
the effects of each variable we set to zero all those that we are
not studying, with only two exceptions, namely $d_{s}$, which takes
the value of $0.01$ for all the profile plots and various values
for $\xi_{L}^{\mathrm{s}}$ and $\xi_{1}^{\mathrm{h}}$ variation.
$\lambda_{\mathrm{EI}}=1.44$ for calculations with varying $h$ and
$\theta_{H}$, but zero otherwise, except when we study the effect
of a varying exchange inter-slab coupling. The values of the parameters considered
here are typical of the Fe/FePt multi-layer systems {[}see Ref. \onlinecite{sousaetal10prb}
and references therein{]}.

\subsection{\label{sec:HmagHDir}Effects of the applied field and in-plane anisotropy}

\begin{figure*}[!htbp]
 \subfigure[ ]{\includegraphics[width=6cm]{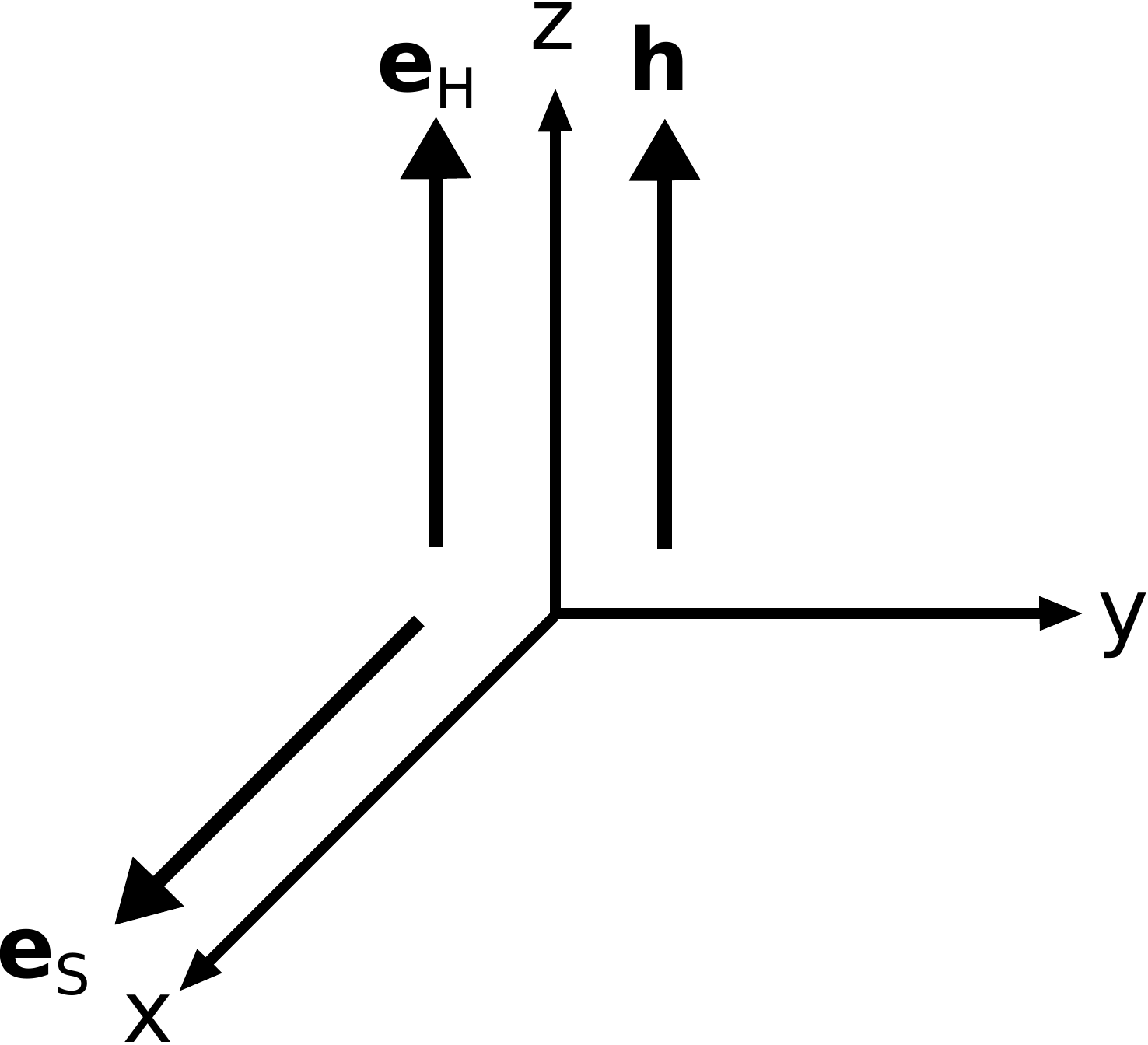}\label{fig:systemConfigESHMag}}
\subfigure[ ]{\includegraphics[width=8cm]{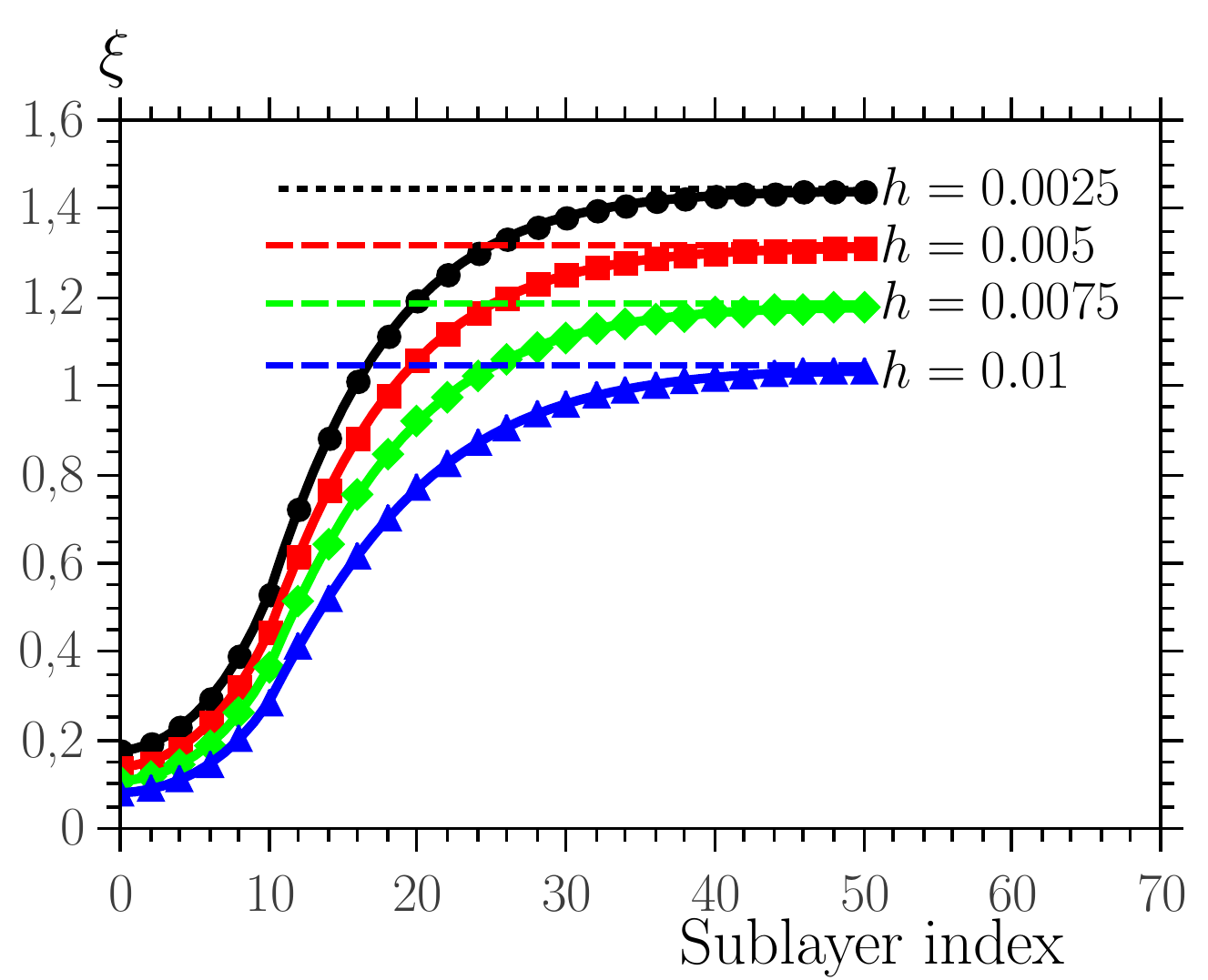}\label{fig:HMP}}
\subfigure[ ]{\includegraphics[width=8cm]{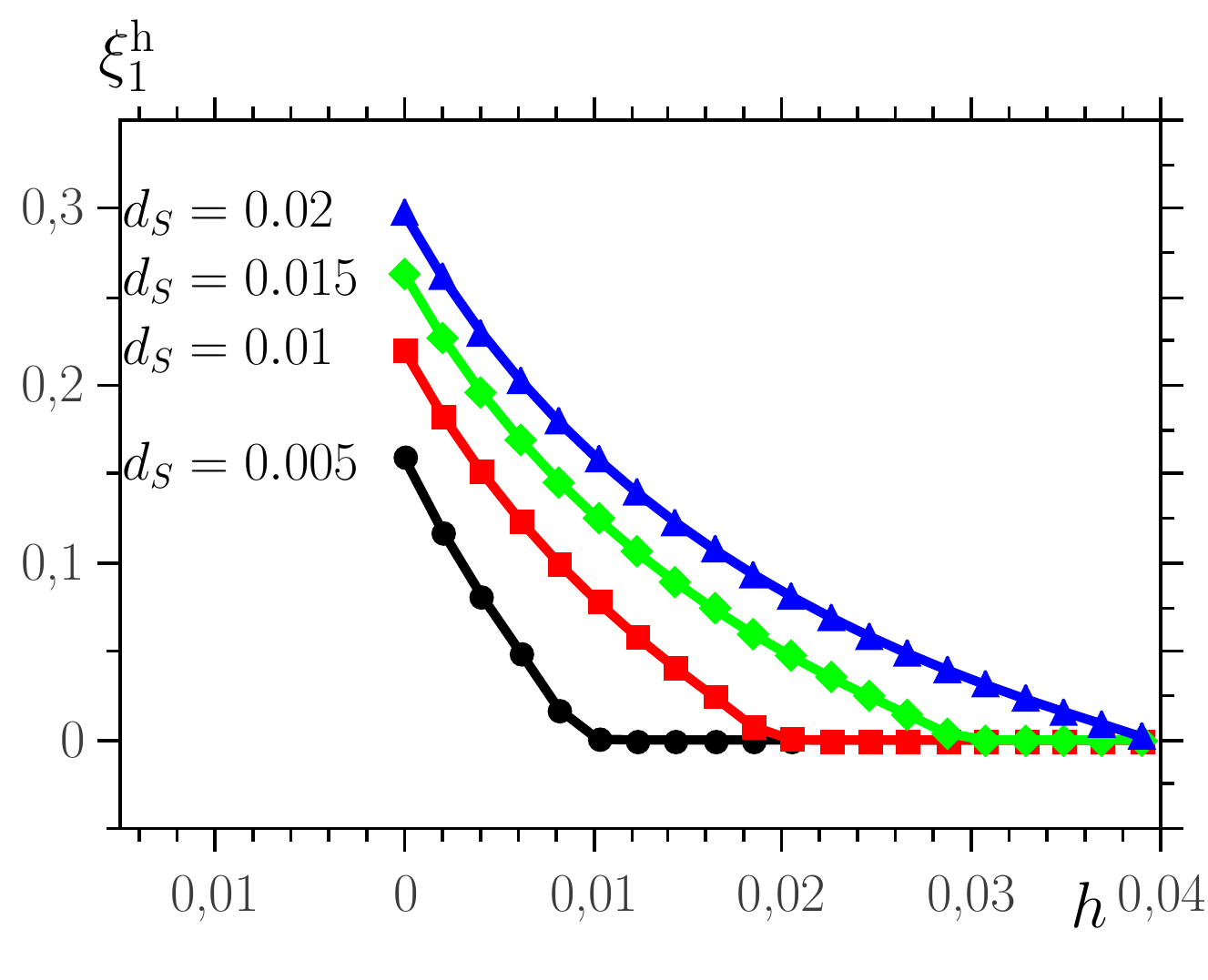}\label{fig:Hxi0}}
\subfigure[ ]{\includegraphics[width=8cm]{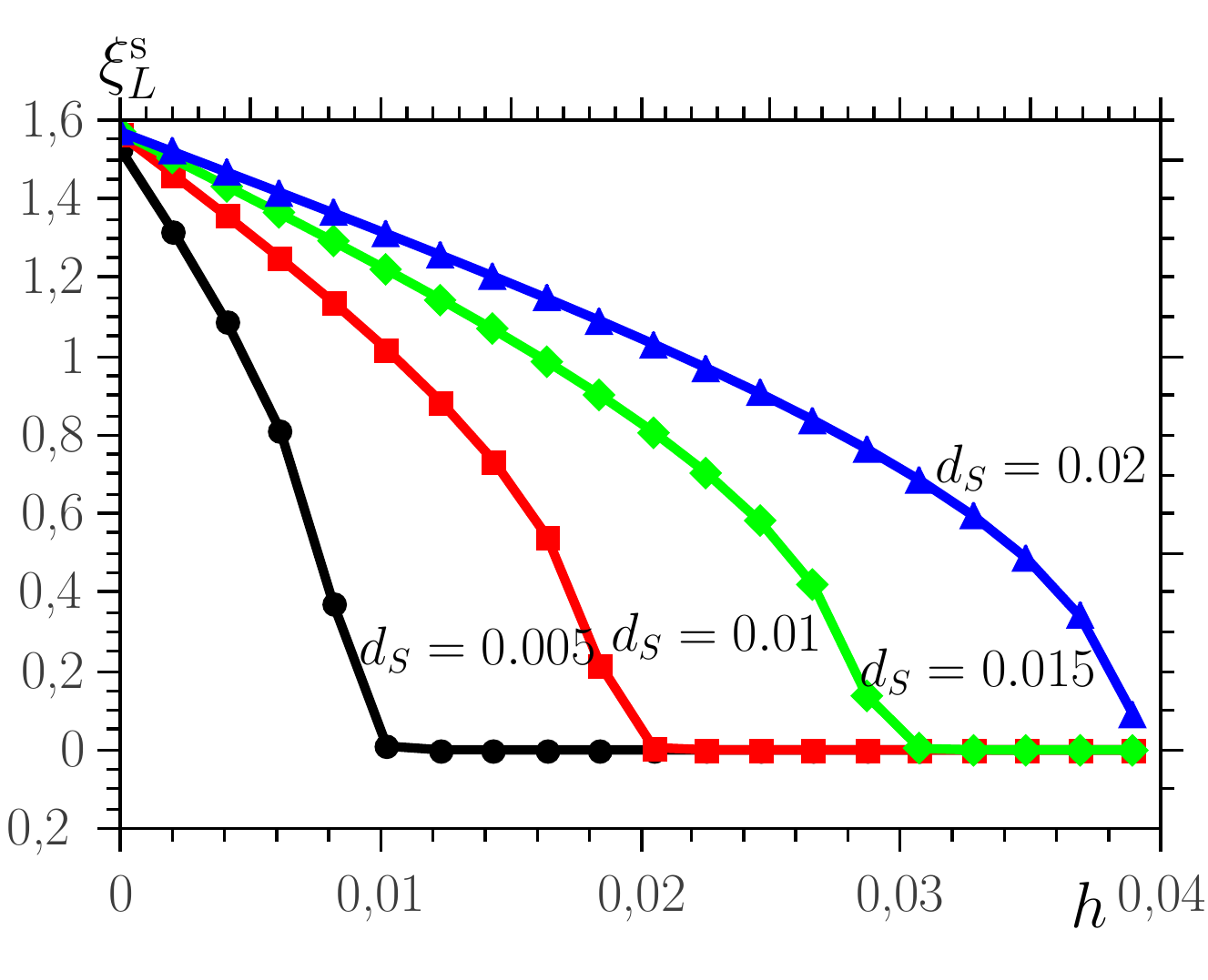}\label{fig:HxiL}}
\caption{(a) System setup with inter-slab exchange interaction and applied
field along the $z$ axis; (b) variation of the MP; (c) the deviation
of the magnetic moment of the bottom HMS sub-layer $\xi_{1}^{\mathrm{h}}$ and
(d) of the top SMS sub-layer $\xi_{L}^{\mathrm{s}}$ with the magnitude
of the applied magnetic field with $\theta_{H}=0$. $\mathcal{N}_{h}=10$
and $\mathcal{N}_{s}=41$.}

\label{fig:HMPandxiL} %
\end{figure*}

The angle deviations, $\xi_{1}^{{\rm h}}$ for the HMS bottom layer
and $\xi_{L}^{{\rm s}}$ for the top layer of the SMS slab, are computed
for various values of the applied field $h$ with a direction $\theta_{H}=0$
and $\lambda_{\mathrm{EI}}=1.44$; the results are shown in Fig. \ref{fig:HMPandxiL}.

It is clear that the external field competes with the anisotropy and
tends to align all the magnetic moments of the system parallel to
its direction. Indeed, with enough layers in the SMS slab, $\xi_{L}^{{\rm s}}$
can be obtained from the Stoner–Wohlfarth equation \begin{equation}
h\sin\left(\theta-\theta_{h}\right)-d_{s}\sin2\theta=0.\label{eq:SWAsymptote}\end{equation}
 In the specific case of $\theta_{h}=0$, we get $\theta=\arccos\left(h/2d_{s}\right)$,
in agreement with the asymptotes (dashed lines) in Fig. \ref{fig:HMP}.
Indeed, Figs. \ref{fig:Hxi0} and \ref{fig:HxiL} show that there
is a critical value of the field, i.e. $h_{c}=2d_{s}$, at which all
magnetic moments are aligned along the direction of the field, i.e.
$\theta_{H}=0$. In this case, as the applied field is along the HMS
anisotropy, $h_{c}$ is not affected by the latter. Hence, for a system
with a sufficient number of layers in the SMS slab, $h_{c}$ will
depend only on $d_{s}$. In general, however, $h_{c}$ should depend
on $d_{h}$, $d_{s}$, and the applied field orientation. In cases
where the system has less layers than the necessary number to attain
the asymptotic value of $\xi_{1}^{{\rm h}}$ and $\xi_{L}^{\mathrm{s}}$ given
by the SW equation (\ref{eq:SWAsymptote}), the magnitude of $h_{c}$
should decrease.

\begin{figure*}[!htbp]
 \subfigure[ ]{\includegraphics[width=6cm]{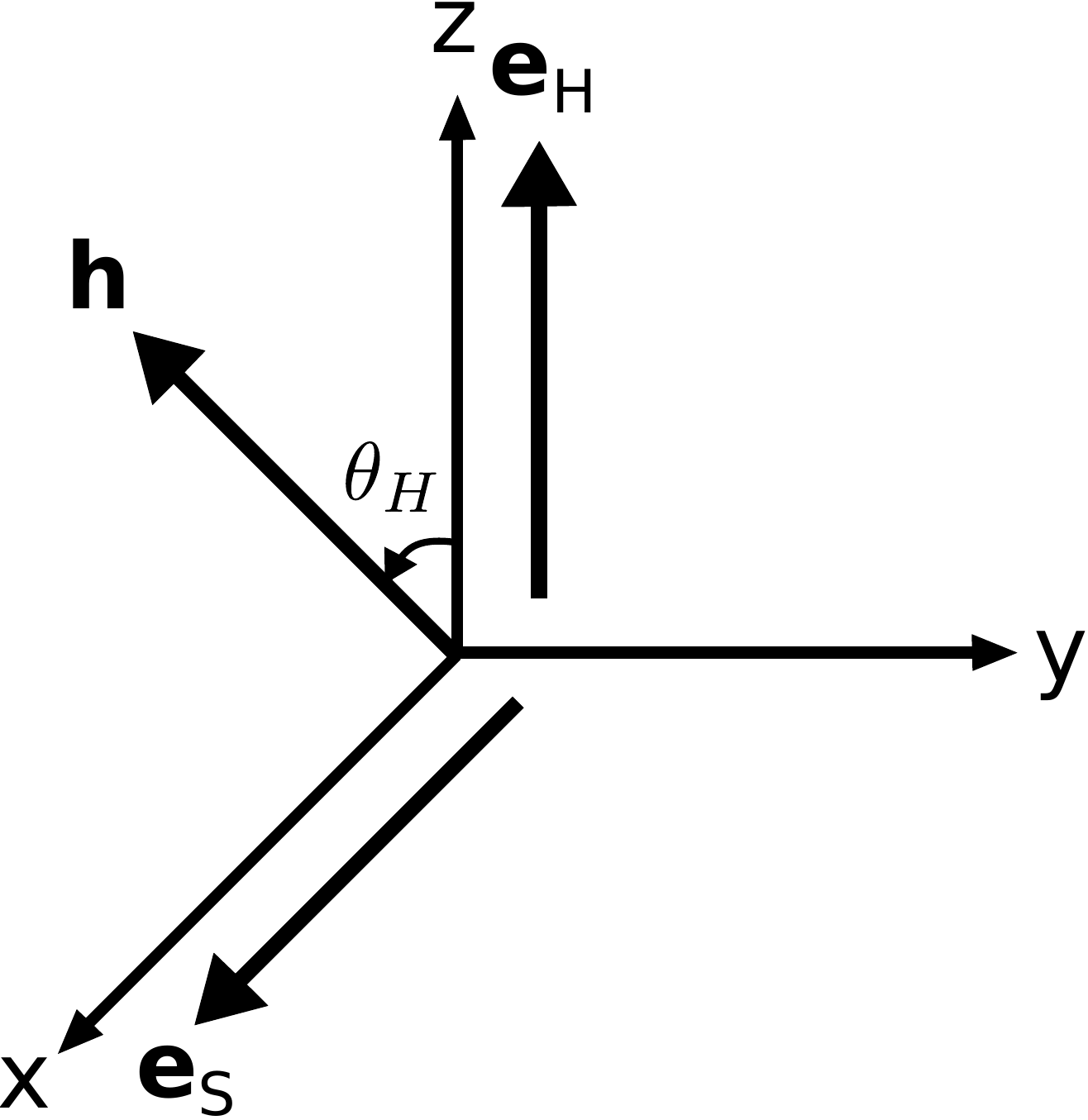}\label{fig:systemConfigESHDir}}
\subfigure[ ]{\includegraphics[width=8cm]{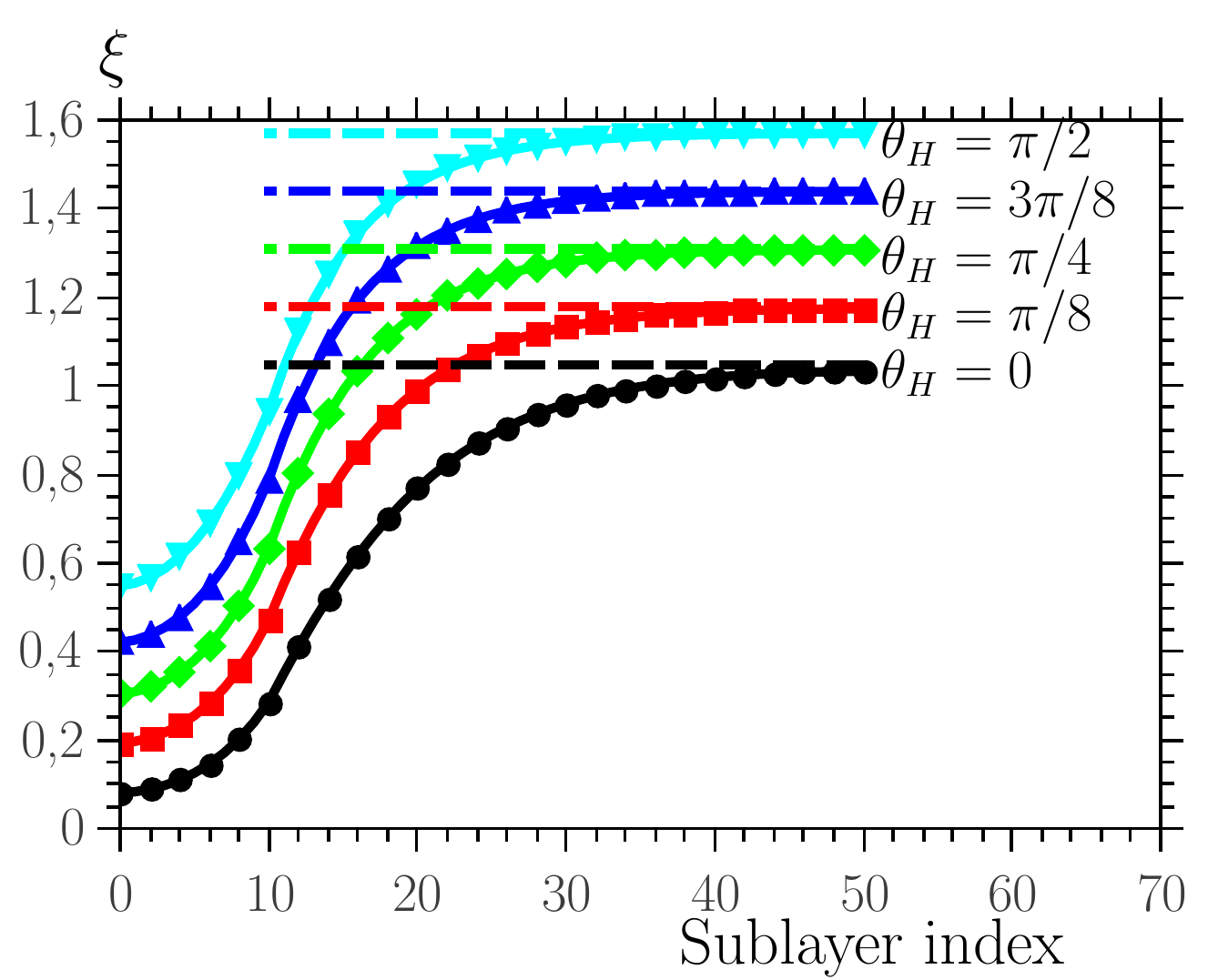}\label{fig:HDirMP}}
\subfigure[ ]{\includegraphics[width=8cm]{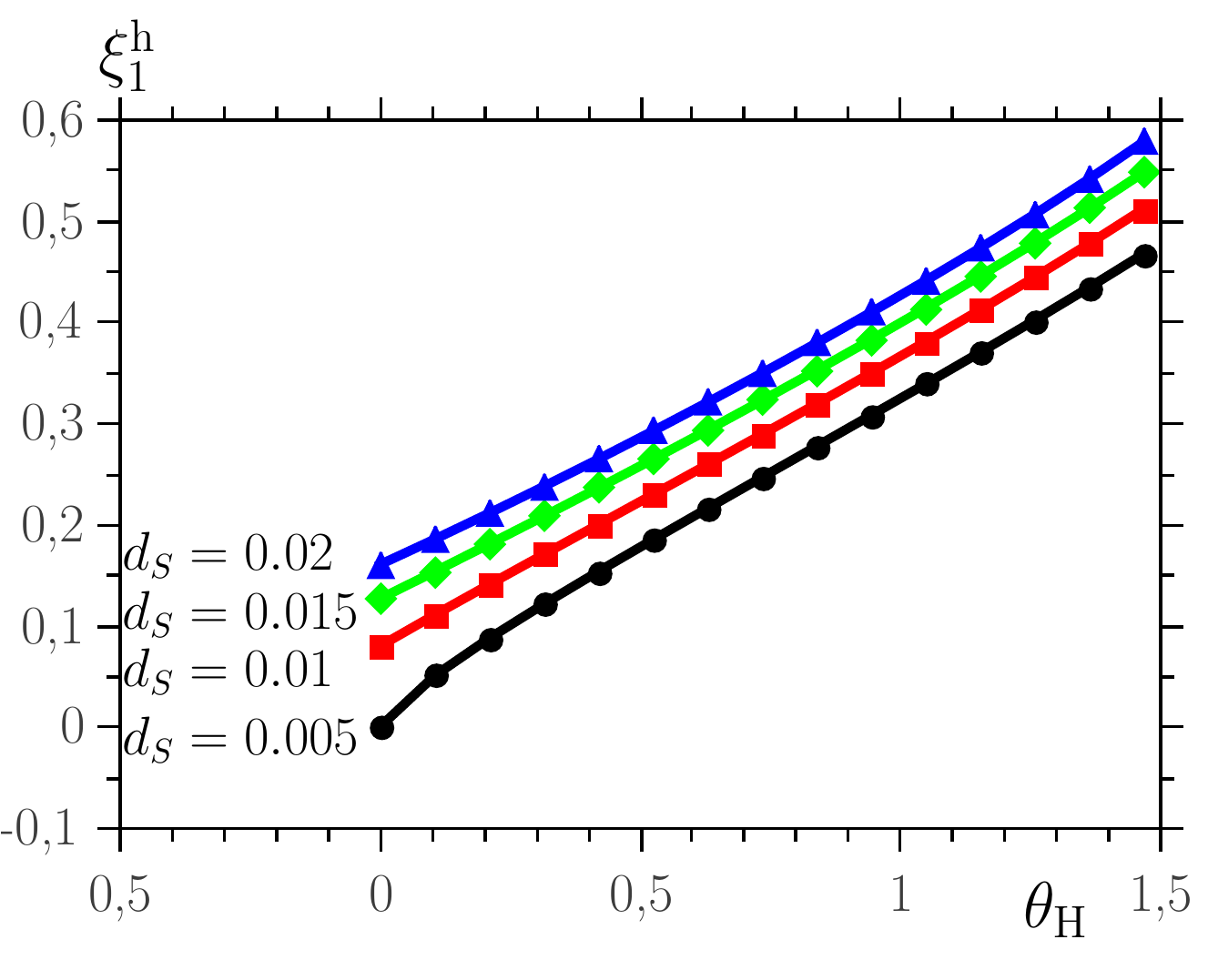}\label{fig:HDirxi0}}
\subfigure[ ]{\includegraphics[width=8cm]{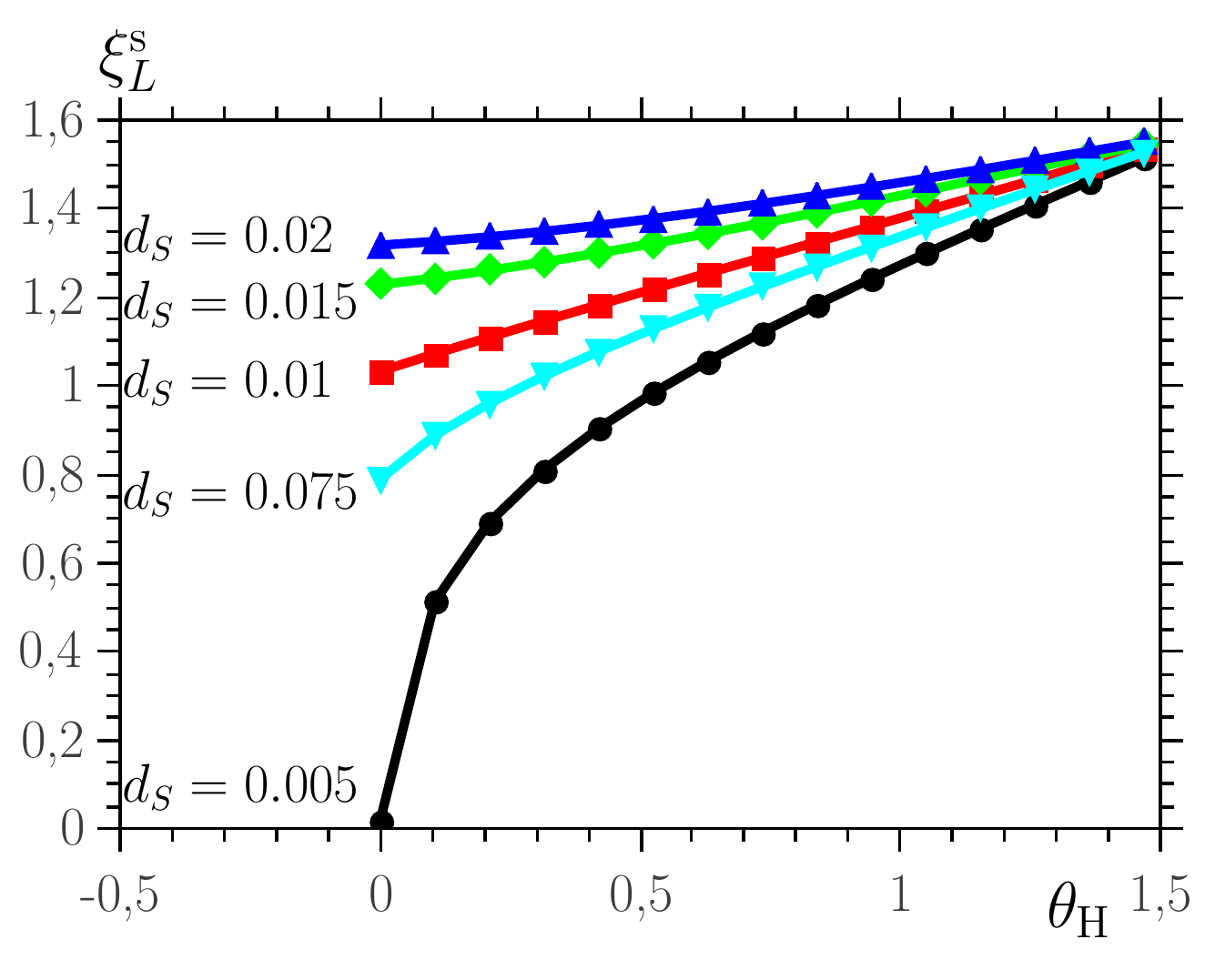}\label{fig:HDirxiL}}
\caption{(a) System setup with inter-slab exchange interaction and applied
field of varying direction; (b) variation of the MP; (c) deviation
of the magnetic moment of the bottom HMS sub-layer $\xi_{1}^{\mathrm{h}}$
and (d) of the top SMS sub-layer $\xi_{L}^{\mathrm{s}}$ with the field
direction $\theta_{H}$. $h=0.01$, $\mathcal{N}_{h}=10$ and $\mathcal{N}_{s}=41$.}

\label{fig:HDirMPandxiL} %
\end{figure*}

Fig. \ref{fig:HDirMPandxiL} shows the same plots for a variable field
orientation and the SMS intra-plane anisotropy $d_{{\rm s}}$, for
$h=0.01$ and $\lambda_{\mathrm{EI}}=1.44$. We see that as the field
is turned towards the SMS easy axis, the MP obviously shifts upwards
with no noticeable change in shape, reaching a deviation at the top
sub-layer that is again given by the numerical solution of the SW equation
(Fig. \ref{fig:HDirMP} - dashed lines). Figs. \ref{fig:HDirxi0}
and \ref{fig:HDirxiL} suggest that beyond a given $\theta_{H}$ the
increase in the deviation is almost linear with a slope that depends
on the anisotropy $d_{s}$. This implies a constant rate of change
of $\xi_{1}^{{\rm h}}$ as it is mainly affected by $d_{h}$ (constant
here), whereas the rate of change of $\xi_{L}^{{\rm s}}$ varies with
$d_{s}$.

\subsection{Effect of inter-slab coupling}

Now, we investigate the effect of the inter-slab coupling, considering
successively EI, DDI, and DMI. In the end we compare their effects
on the magnetization profile.

\subsubsection{\label{sec:EI}Exchange interaction}

\begin{figure*}[!htbp]
 \subfigure[ ]{\includegraphics[width=8cm]{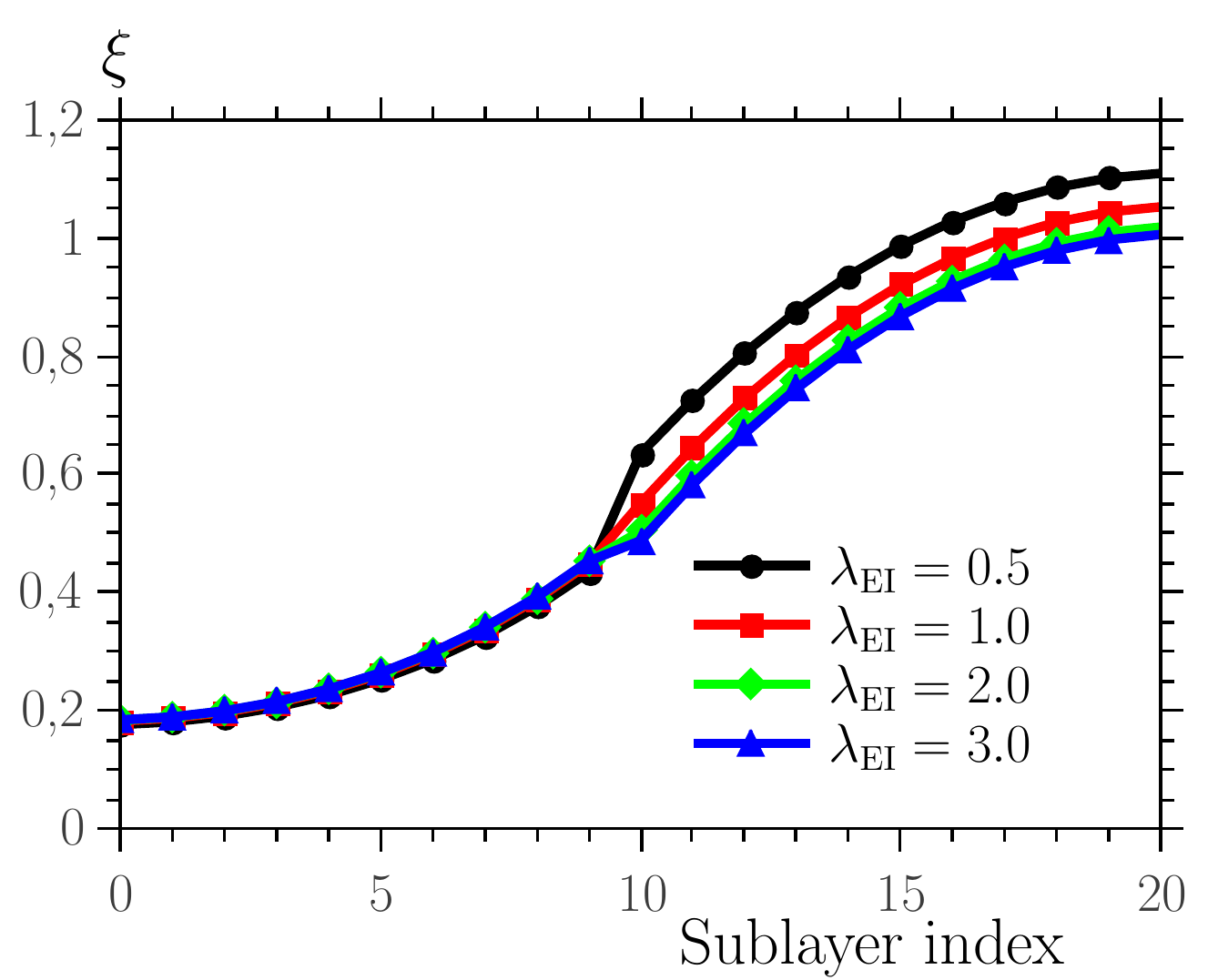}\label{fig:JMP}}

\subfigure[ ]{\includegraphics[width=8cm]{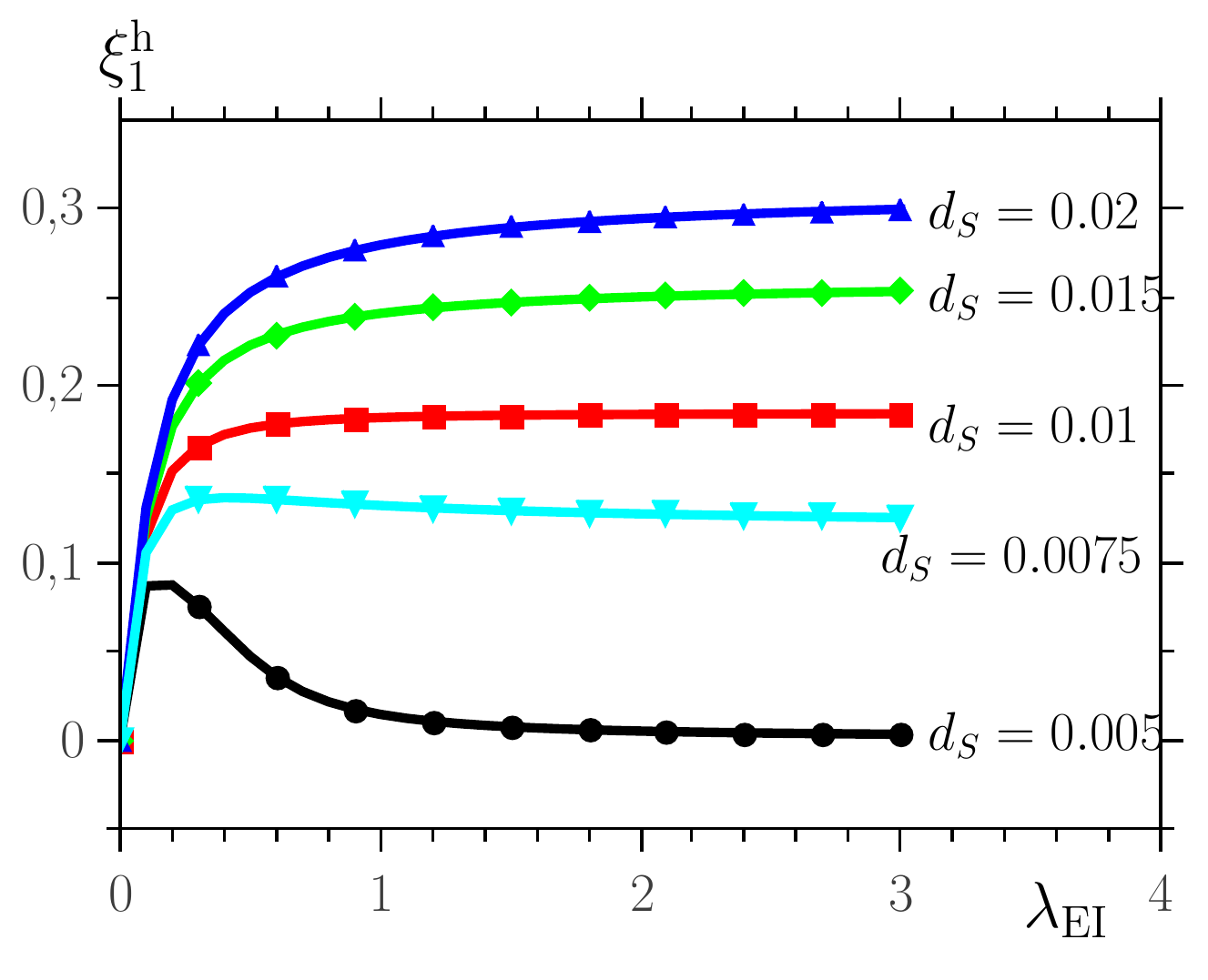}\label{fig:Jxi0}}
\subfigure[ ]{\includegraphics[width=8cm]{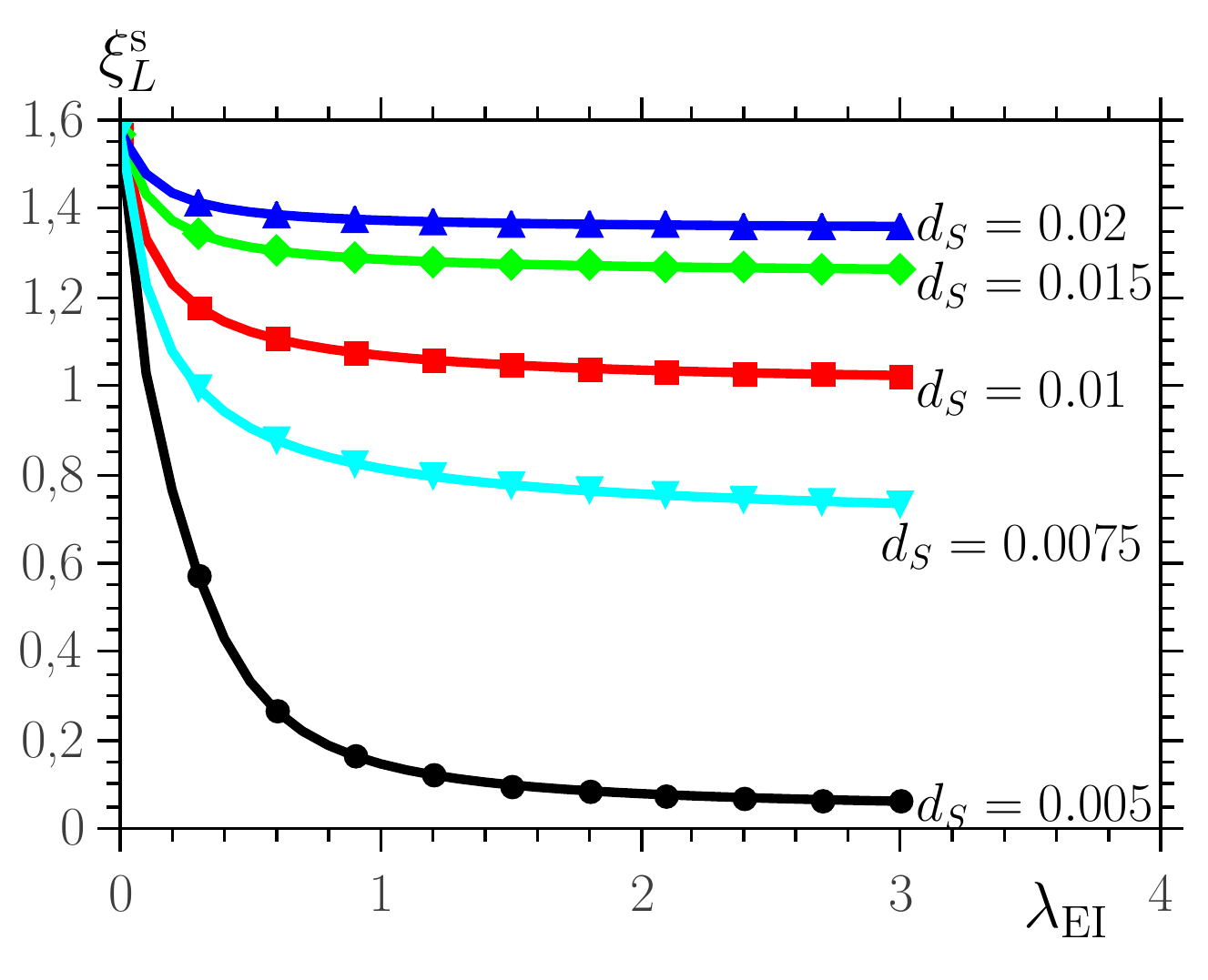}\label{fig:JxiL}}
\caption{(a) Variation of MP and (b) deviation of the magnetic moment of the
bottom HMS sub-layer $\xi_{1}^{\mathrm{h}}$ and (c) of the top SMS sub-layer
$\xi_{L}^{\mathrm{s}}$ with the inter-slab EI $\lambda_{\mathrm{EI}}$.
$\mathcal{N}_{h}=10$ and $\mathcal{N}_{s}=11$.}

\label{fig:JMPandxiL} %
\end{figure*}

The results in Fig. \ref{fig:JMP} show that due to the strong anisotropy
of the HMS slab, varying the inter-slab exchange coupling only affects
the sub-layers in the SMS slab that starts here at $n=11$. On the
other hand, the inter-slab exchange coupling competes with the intra-slab
exchange coupling $J_{s}$ and anisotropy $D_{s}$ of the SMS slab.
Thus the weaker $\lambda_{\mathrm{EI}}$ the stronger is the deviation.

Figs. \ref{fig:Jxi0} and \ref{fig:JxiL} show the deviation $\xi_{1}^{{\rm h}}$
and $\xi_{L}^{{\rm s}}$ as functions of $\lambda_{\mathrm{EI}}$,
respectively. As it could be expected, with increasing EI we achieve
higher $\xi_{1}^{{\rm h}}$ and lower $\xi_{L}^{{\rm s}}$ deviations.
As we effectively increase the rigidity of the interface, the HMS
deviation at the interface thereby increases, whereas that of the
SMS decreases. This change in deviation is then conveyed through all
the sub-layers by means of the intra-layer EI, leading to the observed
changes in $\xi_{1}^{{\rm h}}$ and $\xi_{L}^{{\rm s}}$. For low
values of $d_{s}$, the $\xi_{1}^{{\rm h}}$ increase is much slower
than the $\xi_{L}^{{\rm s}}$ decrease. However, as $d_{s}$ approaches
$d_{h}=0.02$, the two become comparable as the two systems
are close to being identical.

\subsubsection{\label{sec:DDI}Dipolar Interaction}

In the present work we consider the system setup where the dipolar
coupling is assumed to induce a ferromagnetic coupling between the
two magnetic slabs.

\begin{figure*}[!htbp]
 \subfigure[ ]{\includegraphics[width=6cm]{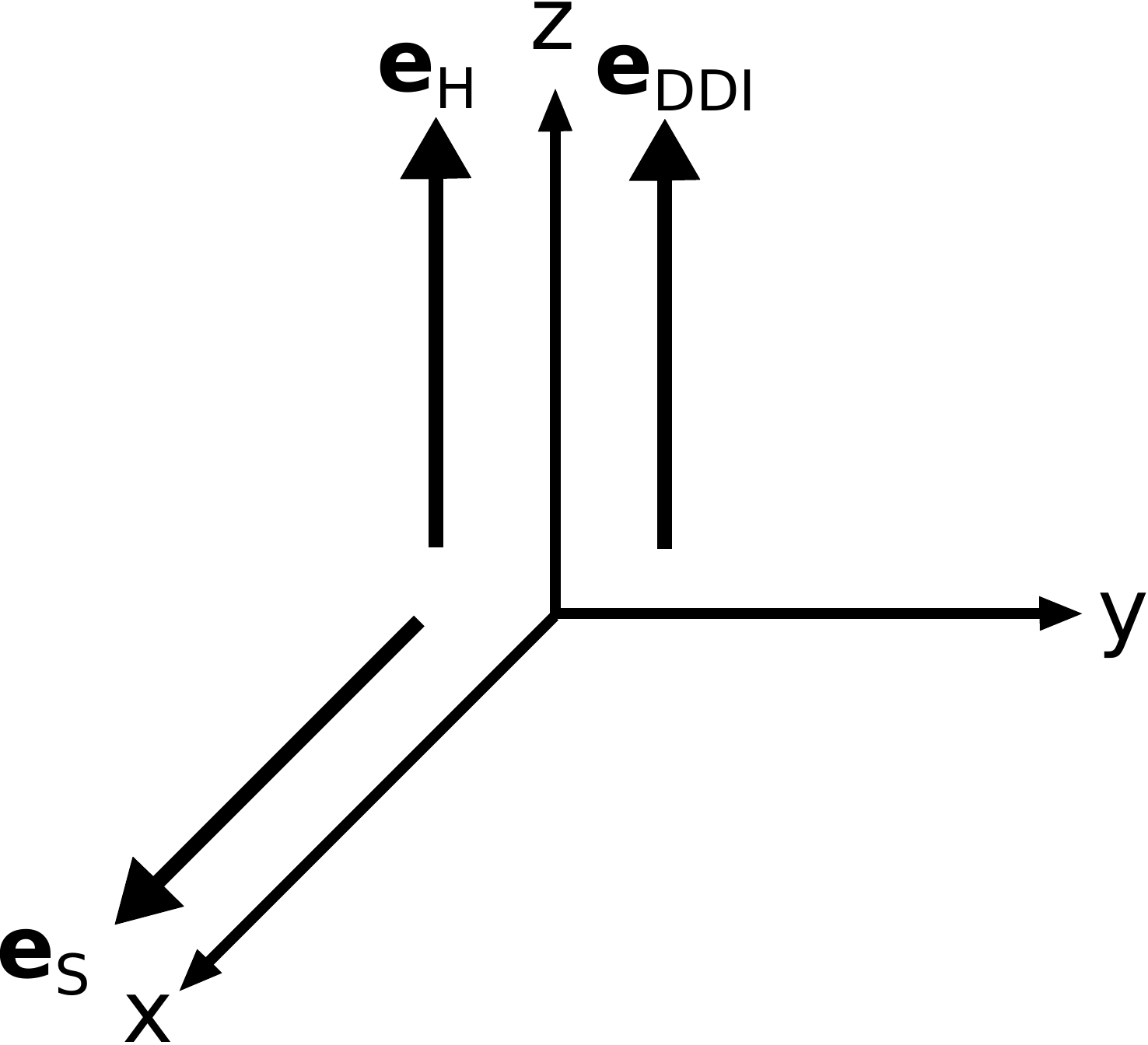}\label{fig:systemConfigESDDI}}
\subfigure[ ]{\includegraphics[width=8cm]{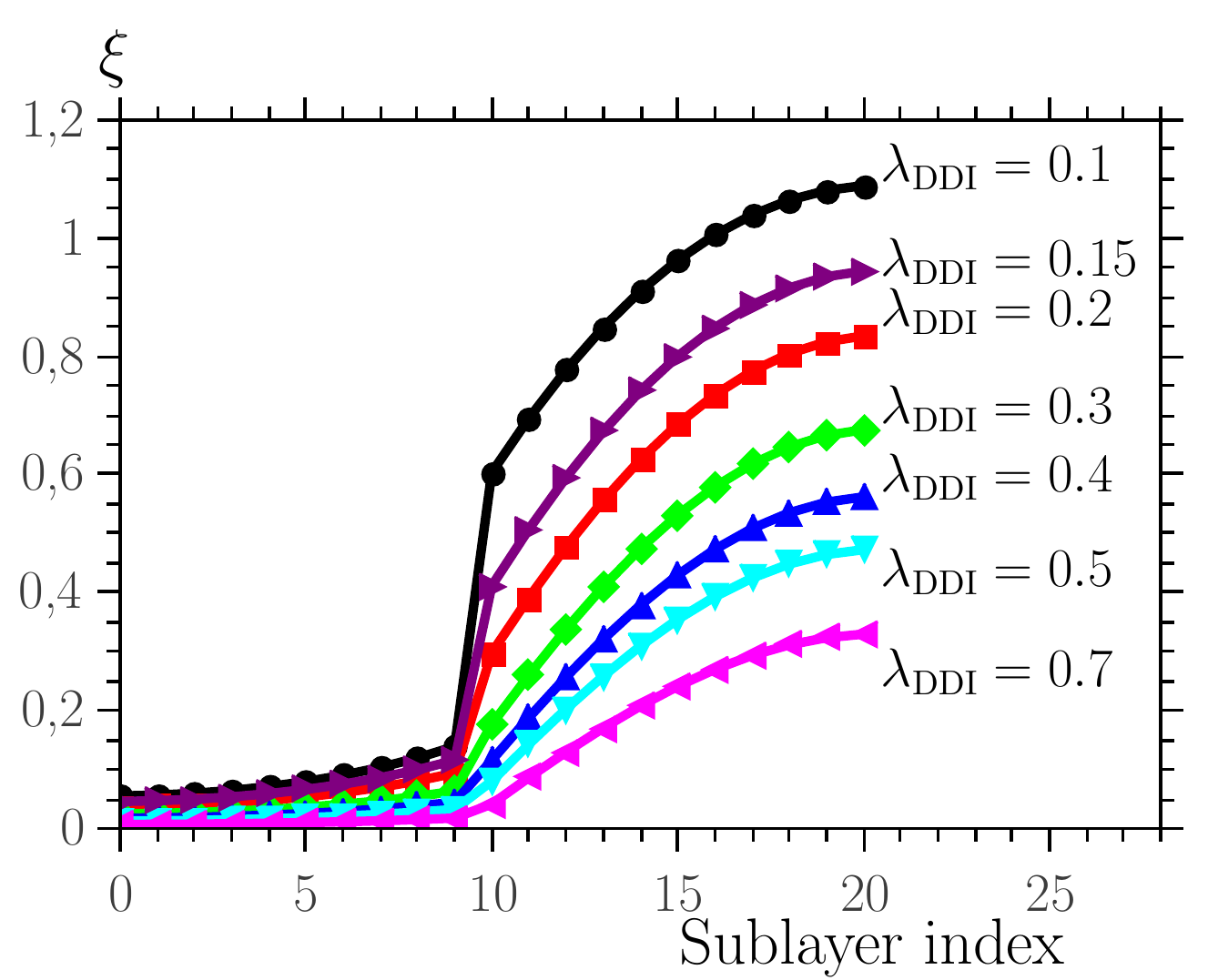}\label{fig:DDIMP}}
\subfigure[ ]{\includegraphics[width=8cm]{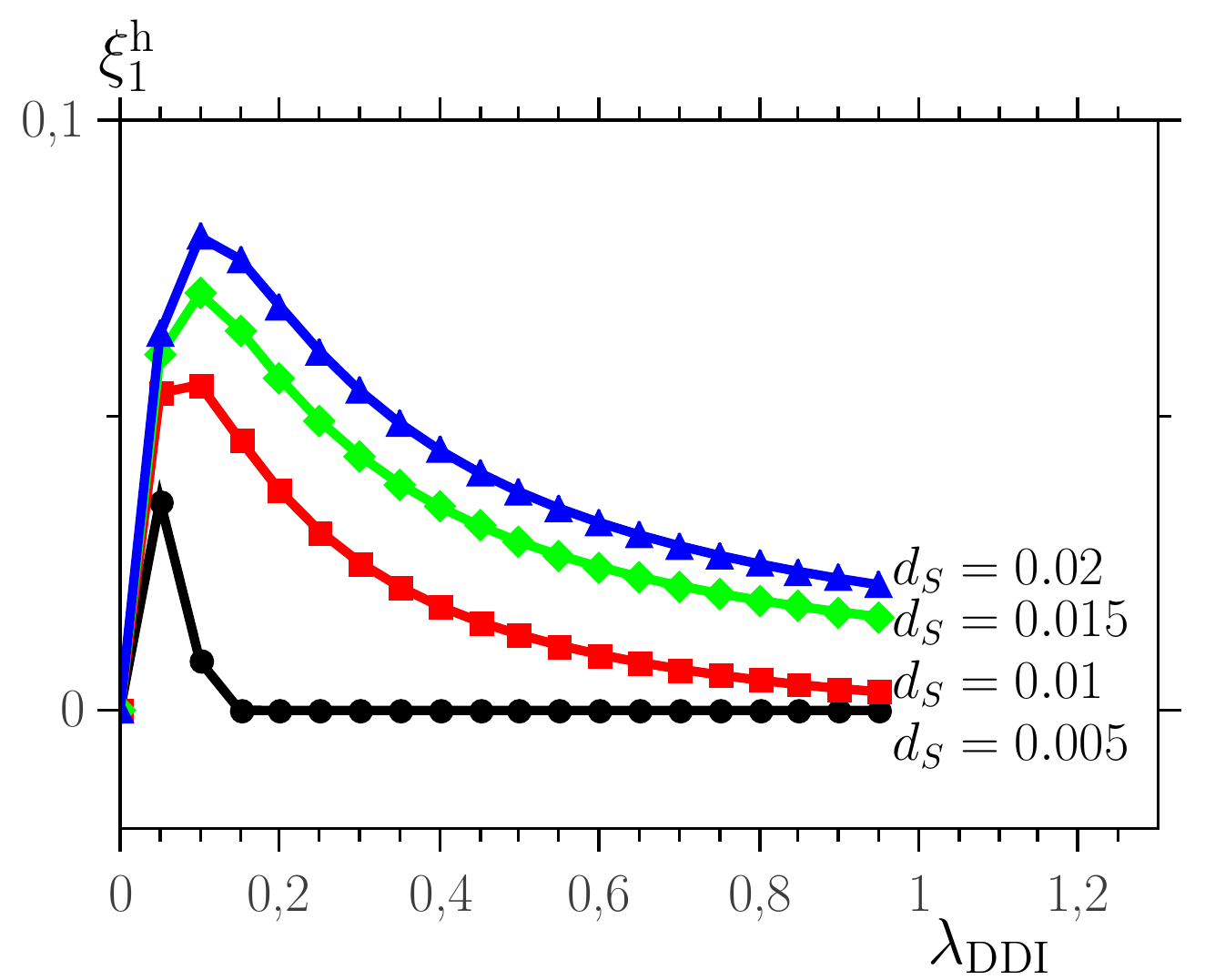}\label{fig:DDIxi0}}
\subfigure[ ]{\includegraphics[width=8cm]{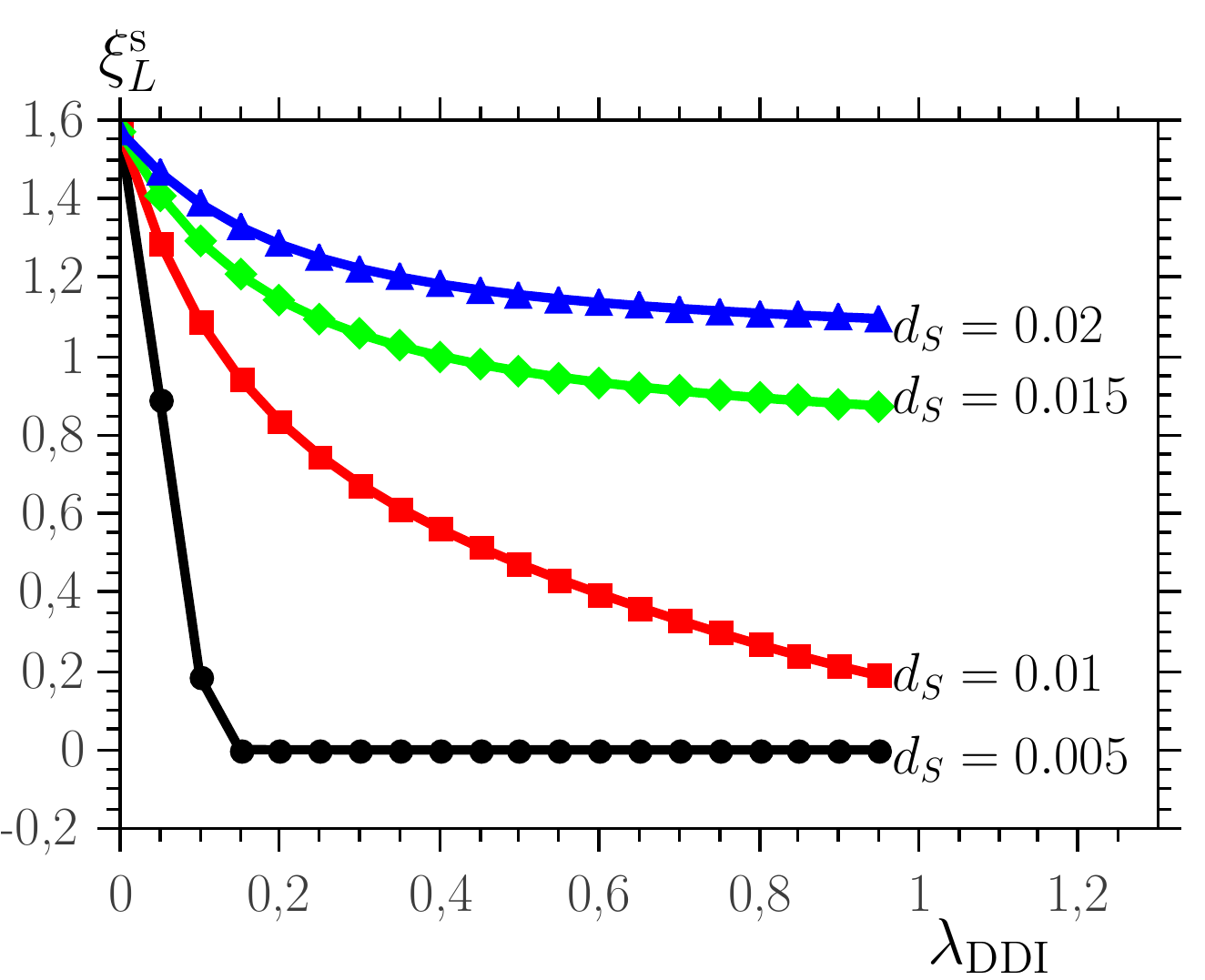}\label{fig:DDIxiL}}
\subfigure[ ]{\includegraphics[width=8cm]{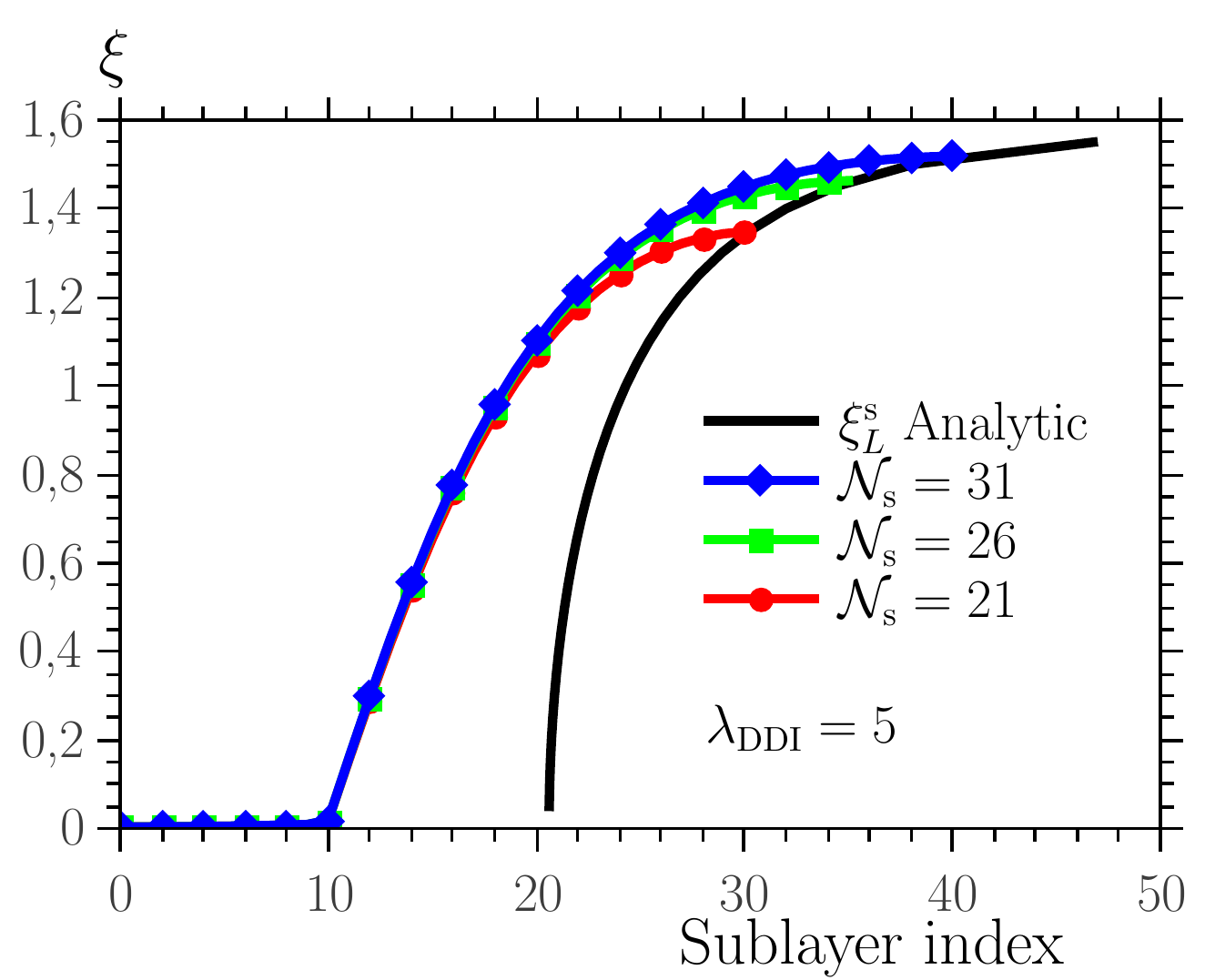}\label{fig:DDIxiLAnalytic}}
\caption{(a) System setup for a system with inter-slab dipolar interaction
with a dipolar vector parallel to the HMS anisotropy; (b) variation of
MP; (c) deviation of the magnetic moment of the bottom HMS
sub-layer $\xi_{1}^{\mathrm{h}}$ and (d) of the top SMS sub-layer
$\xi_{L}^{\mathrm{s}}$
with the inter-slab DDI $\lambda_{\mathrm{DDI}}$. (e) Comparison
between RI analytical expressions and strong DDI numerical calculations.
$\mathcal{N}_{h}=10$ and $\mathcal{N}_{s}=11$.}

\label{fig:DDIMPandxiL} %
\end{figure*}

In Fig. \ref{fig:DDIMP} are plotted the MP for different values of
the DDI inter-slab coupling $\lambda_{\mathrm{DDI}}$. Apart from
the obvious shift upwards as $\lambda_{\mathrm{DDI}}$ decreases,
there is an abrupt change at the interface especially for small $\lambda_{\mathrm{DDI}}$,
mainly due to the fact that the in-plane anisotropy $d_{s}$ has a
stronger effect than DDI.

Figs. \ref{fig:DDIxi0} and \ref{fig:DDIxiL} show that the system
tends towards an asymptote as $\lambda_{\mathrm{DDI}}$ increases.
In our case, since the DDI vector is parallel to the HMS anisotropy
(along the $z$ axis), a strong enough interaction aligns the magnetic
moments at the interface (and all the magnetic moments in the HMS)
in the direction $\theta=0$, thus driving the system into an effective
RI state. As such, the asymptote can be found by calculating $\xi_{L}^{{\rm s}}$
using the analytical expressions \citep{sousaetal10prb}, i.e. Eqs.
(\ref{eq:magnetProfileContinuum}) and (\ref{eq:SMSTopLayer}). Indeed,
Fig. \ref{fig:DDIxiLAnalytic}, where the MP is plotted for different
values of $\mathcal{N}_{s}$ with very strong DDI ($\lambda_{\mathrm{DDI}}\sim5$),
shows a perfect agreement. Furthermore, if we examine the analytic
curve (black line) near $n=20$ we observe a regime where a slight
variation in $\mathcal{N}_{\mathrm{s}}$ induces a large change in
$\xi_{L}^{\mathrm{s}}$, indicating that $\mathcal{N}_{s}=10$ is
approximately the critical length of the SMS chain.\textbf{ }A similar
regime starts to be seen for stronger values of DDI in Fig.~\ref{fig:DDIMP}\textbf{.
}Then, if we take into account that this behavior is not observed
for weak DDI, it suggests that the critical length of the chain increases
with increasing DDI.

\begin{figure*}[!htbp]
 \subfigure[ $\mathcal{N}_s=10$]{\includegraphics[width=8cm]{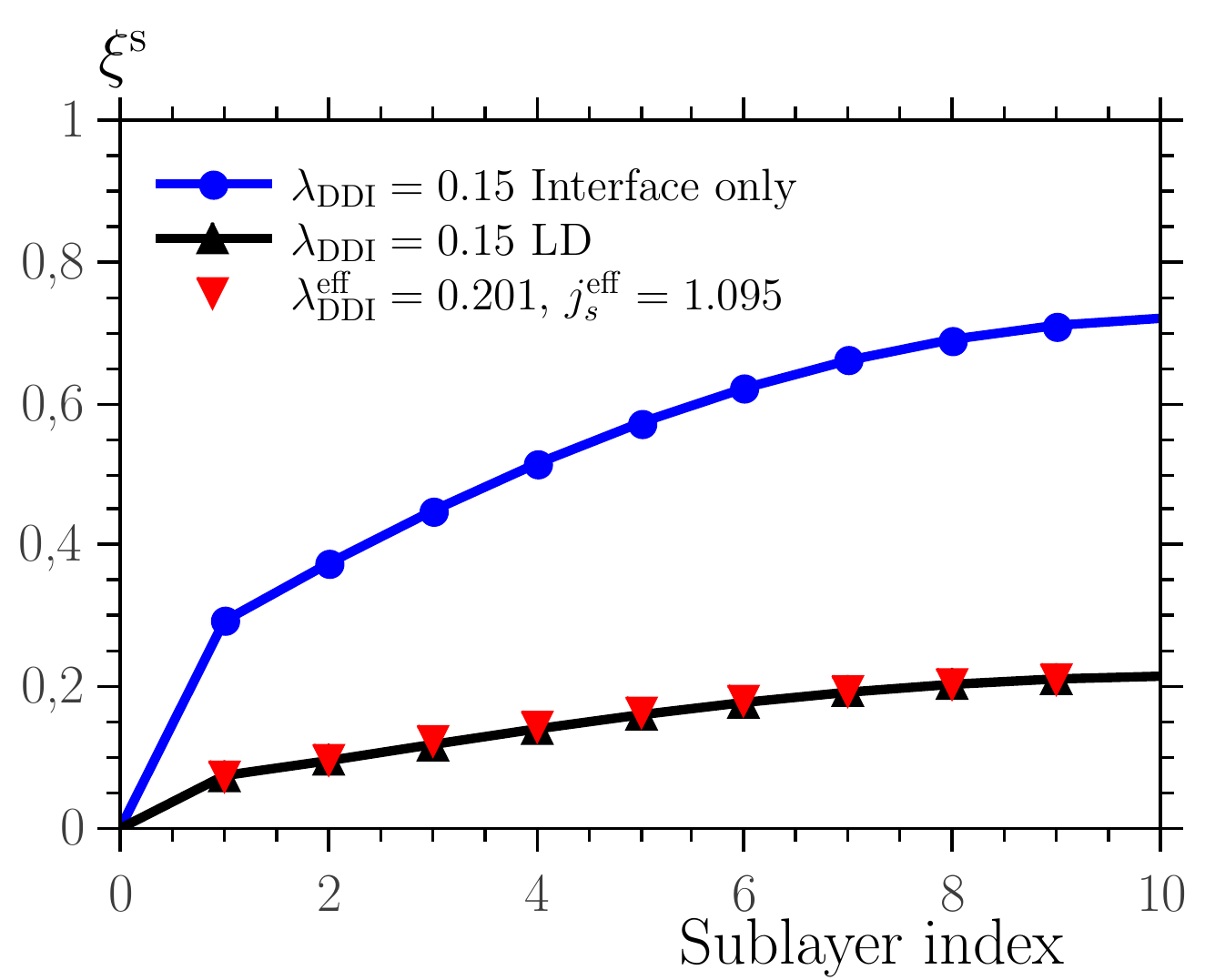}\label{fig:DDIALLxi015n10}}
\subfigure[ $\mathcal{N}_s=20$]{\includegraphics[width=8cm]{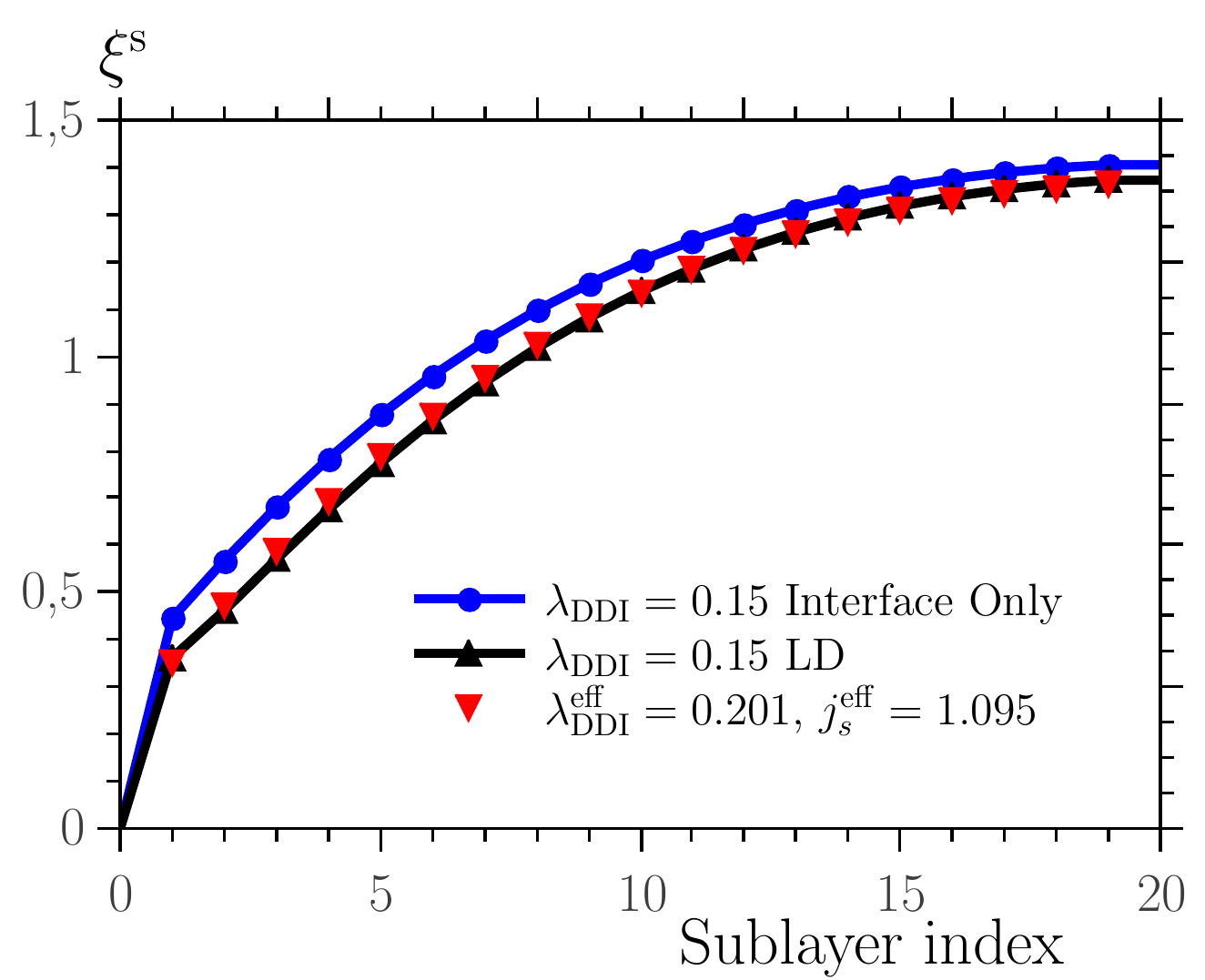}\label{fig:DDIALLxi015n20}}
\subfigure[ $\mathcal{N}_s=30$]{\includegraphics[width=8cm]{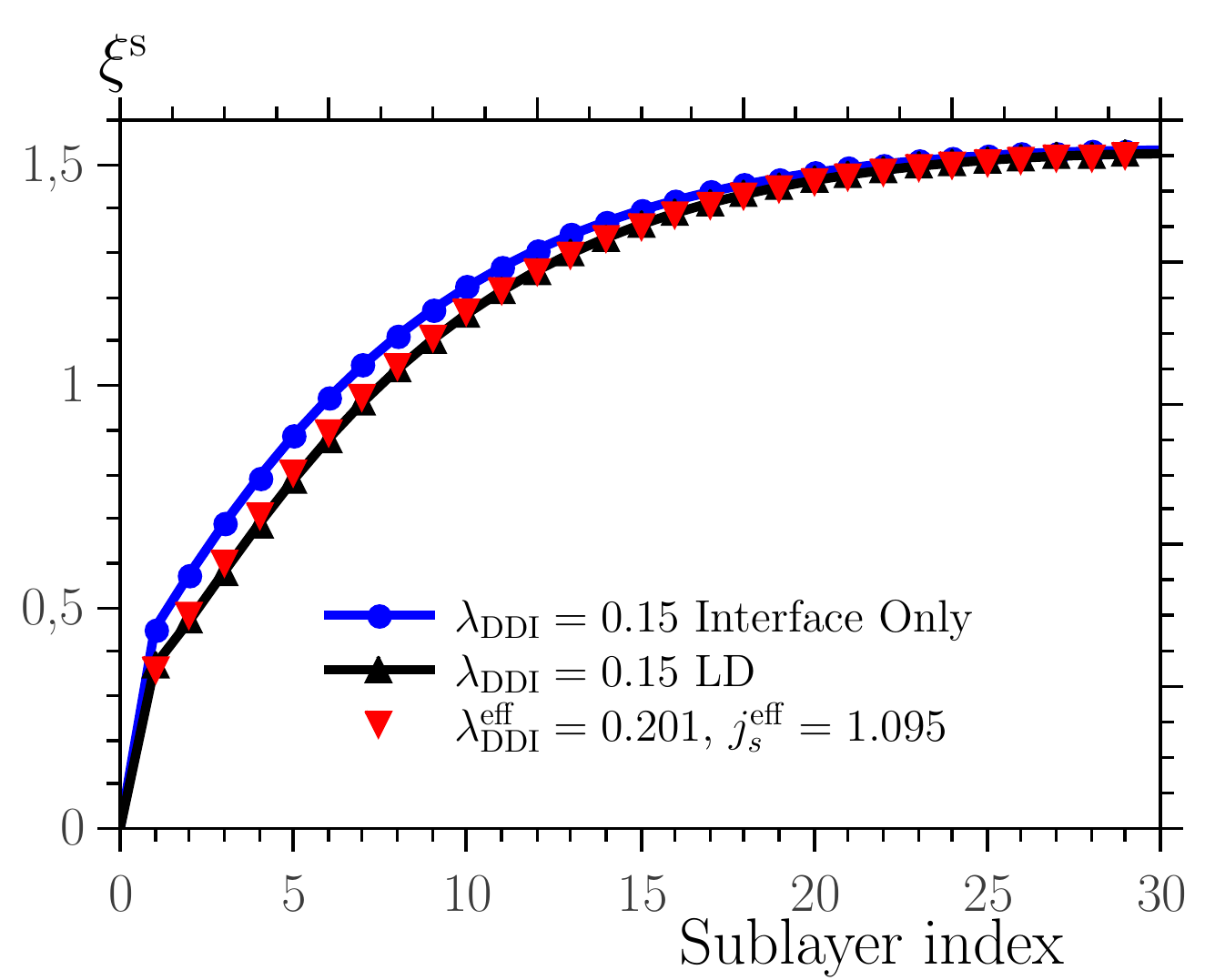}\label{fig:DDIALLxi015n30}}
\caption{Comparison between different MP for a RI system with DDI at the interface
only and with an effective value for both $\lambda_{\mathrm{DDI}}$
and $j_{s}$. $\lambda_{\mathrm{DDI}}=0.15$, $\lambda_{\mathrm{DDI}}^{\mathrm{eff}}=0.201$,
$j_{s}^{\mathrm{eff}}=1.095$.}

\label{fig:DDIALLxi015} %
\end{figure*}

\begin{figure*}[!htbp]
 \subfigure[$\lambda_{\mathrm{DDI}}=0.1$, $\lambda_{\mathrm{DDI}}^{\mathrm{eff}}=0.1285$, $j_S^{\mathrm{eff}}=1.0778$]{\includegraphics[width=8cm]{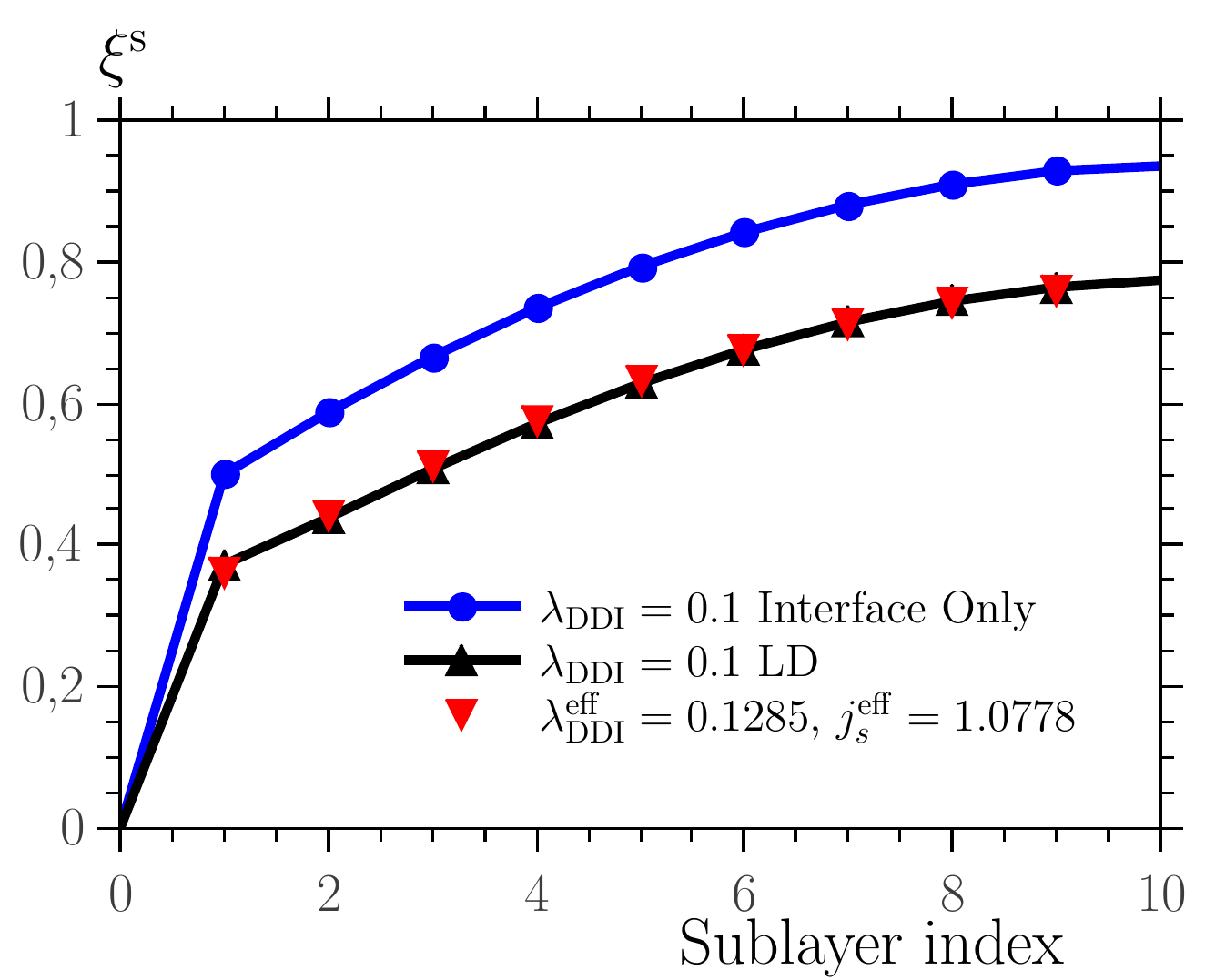}\label{fig:DDIALLxi01}}
\subfigure[$\lambda_{\mathrm{DDI}}=0.2$, $\lambda_{\mathrm{DDI}}^{\mathrm{eff}}=0.2815$, $j_S^{\mathrm{eff}}=1.104$]{\includegraphics[width=8cm]{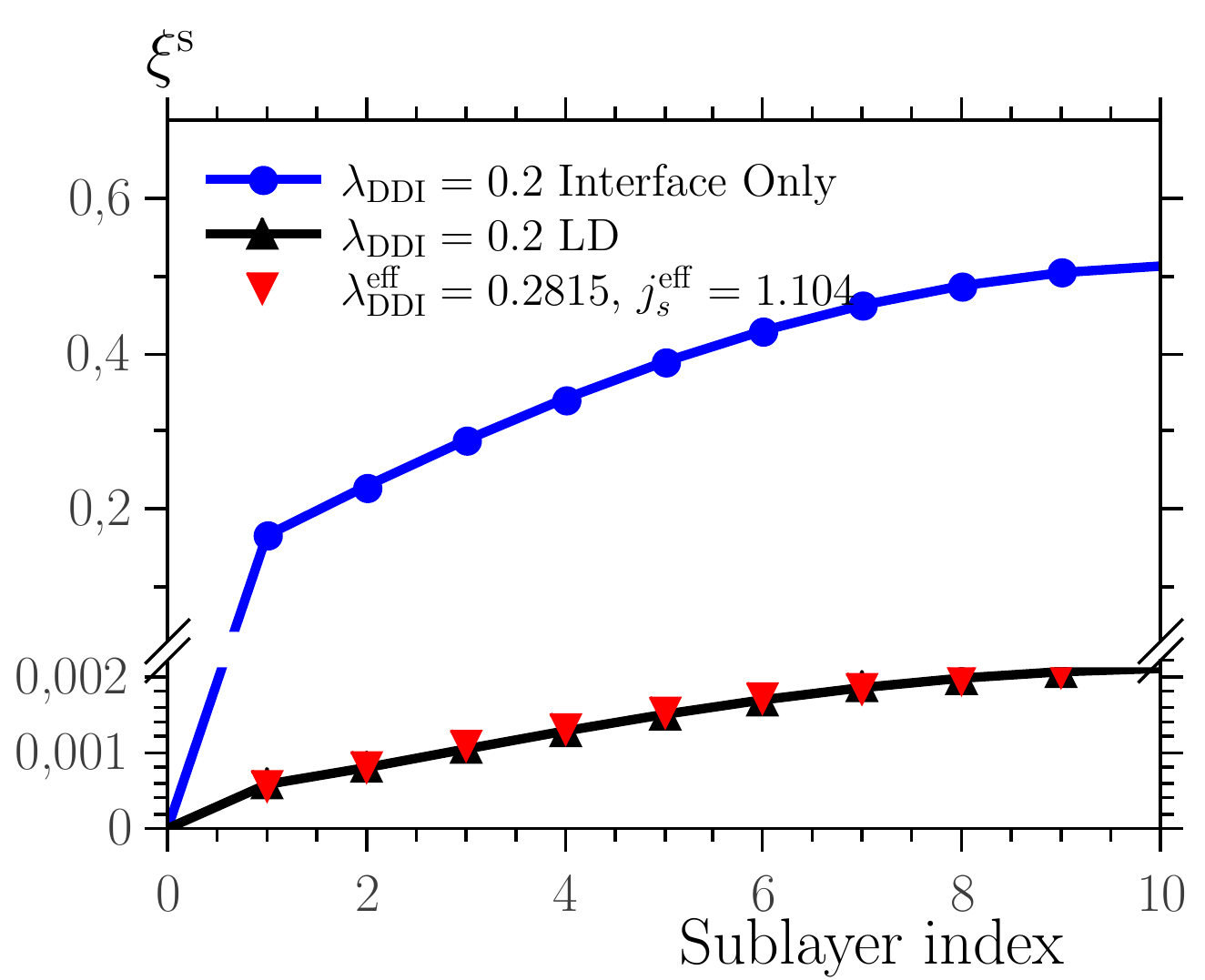}\label{fig:DDIALLxi02}}
\caption{Comparison between MP for a RI system with DDI at the interface
only, LD regime, and for the case with an effective values for
$\lambda_{\mathrm{DDI}}$
and $j_{s}$ yielding the same MP as for the LD regime. $\mathcal{N}_{s}=10$.}

\label{fig:DDIALLns10} %
\end{figure*}

DDI is a long-ranged interaction that can penetrate through the SMS
leading a \emph{priori} to a coupling\emph{ }of the HMS interface
with every SMS sub-layer, and vice-versa, with a strength $\xi\sim d^{-3}$
where $d$ is the distance between two magnetic moments. We also show
in Fig.~\ref{fig:DDIMPandxiL} that with $\lambda_{\mathrm{DDI}}$
at the interface with the bond $\mathbf{e}_{\mathrm{DDI}}$ along
the $z$ axis, we can approximate the configuration of our system
by a RI. Upon taking long distance interaction into account, the HMS interface
will be coupled with every SMS sub-layer, but the SMS interface will
only be coupled with the HMS at the interface, due to the latter being
in a RI configuration. We call this a long distance (LD) configuration.
Fig.~\ref{fig:DDIALLxi015} shows a comparison between the MP for
$\lambda_{\mathrm{DDI}}=0.15$ only at the interface and LD configuration.
The effect of the coupling penetration is a global decrease in the
deviation of the magnetic moments of the SMS sub-layers. A similar
behavior can be obtained with the effective couplings
$\lambda_{\mathrm{DDI}}^{\mathrm{eff}}=0.201$ 
and $j_{s}^{\mathrm{eff}}=1.095$ with interaction only at the interface.
This means that the effect of LD configuration is equivalent to that
of an interaction which is limited to the interface, but with the
re-normalized couplings $\lambda_{\mathrm{DDI}}^{\mathrm{eff}}$ and
$j_{s}^{\mathrm{eff}}$. Figs. \ref{fig:DDIALLxi015} (a)-(c) show
that $\lambda_{\mathrm{DDI}}^{\mathrm{eff}}$ and $j_{s}^{\mathrm{eff}}$
do not change with the number of SMS sub-layers. However, Fig.
\ref{fig:DDIALLns10} shows that they do depend on $\lambda_{\mathrm{DDI}}$.

\subsubsection{\label{sec:DMI}DM Interaction}

As discussed in the introduction, it is relevant in the present study
to investigate the effect of DMI on the dynamics of the MD, on the
same footing as the (symmetrical) effective exchange coupling and
anisotropic dipolar coupling. DM interaction may be induced by spin-orbit
coupling between two ferromagnetic layers separated by a paramagnetic
layer \cite{xiaetal97prb,crelac98jmmm}. It plays an important role
in the presence of roughness and disorder at the interface of multi-layer
systems. For a simple cubic lattice, on the (1 0 0) surface the DMI
vector $\bm{D}$ lies in the layer plane and thus induces a perpendicular
anisotropy. In the present work, we consider two situations where
the DMI vector $\bm{D}$ lies either along the $x$ direction (thus
in the $xy$ plane) or along the $z$ direction (\emph{i.e. }normal to
the $xy$ plane).

\paragraph{DM vector in the $x$ Direction\protect \protect \protect \protect
\protect \protect \\
 }

\begin{figure*}[!htbp]
 \subfigure[ ]{\includegraphics[width=5cm]{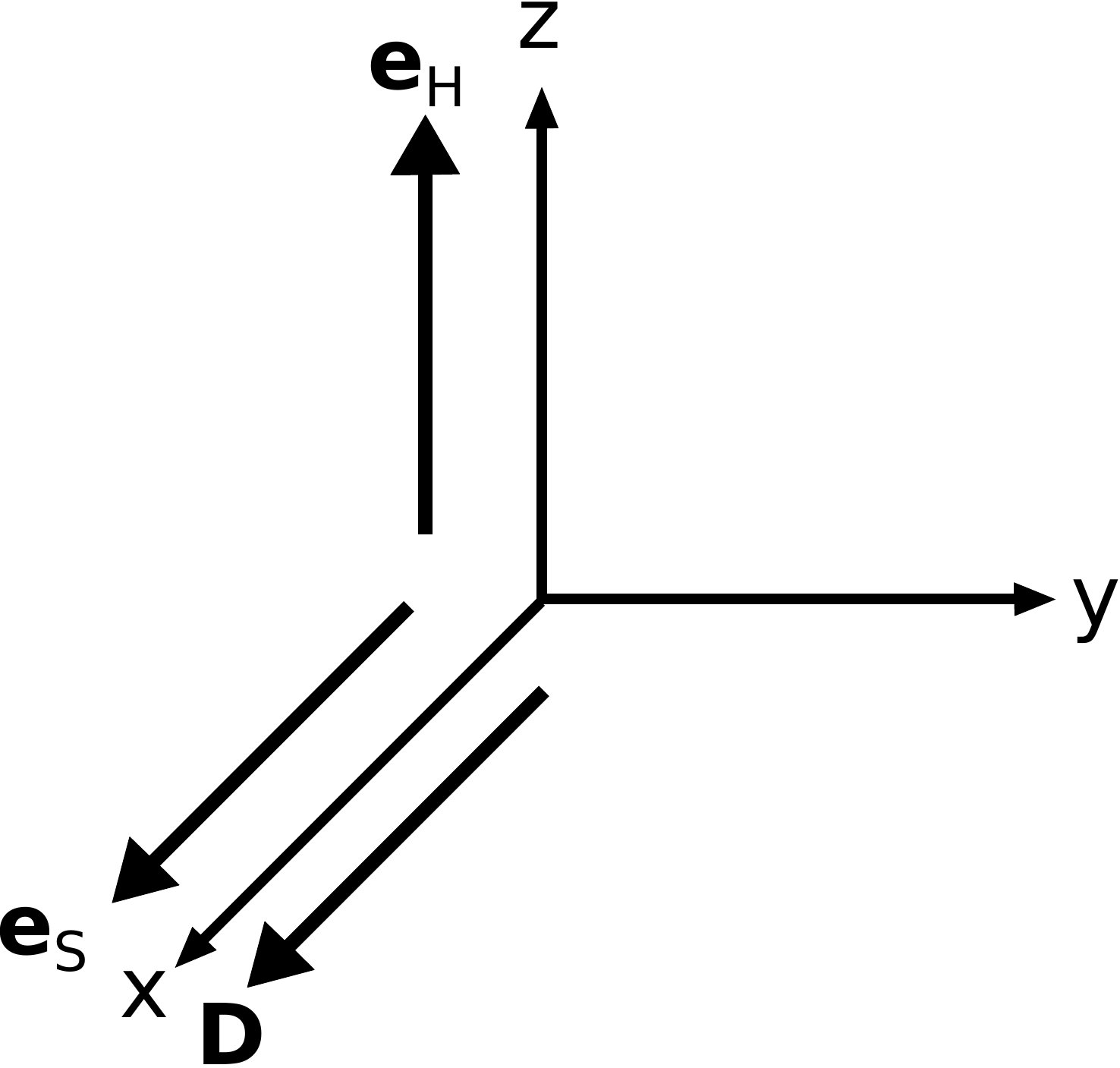}\label{fig:systemConfigESDMx}}
\subfigure[ ]{\includegraphics[width=7cm]{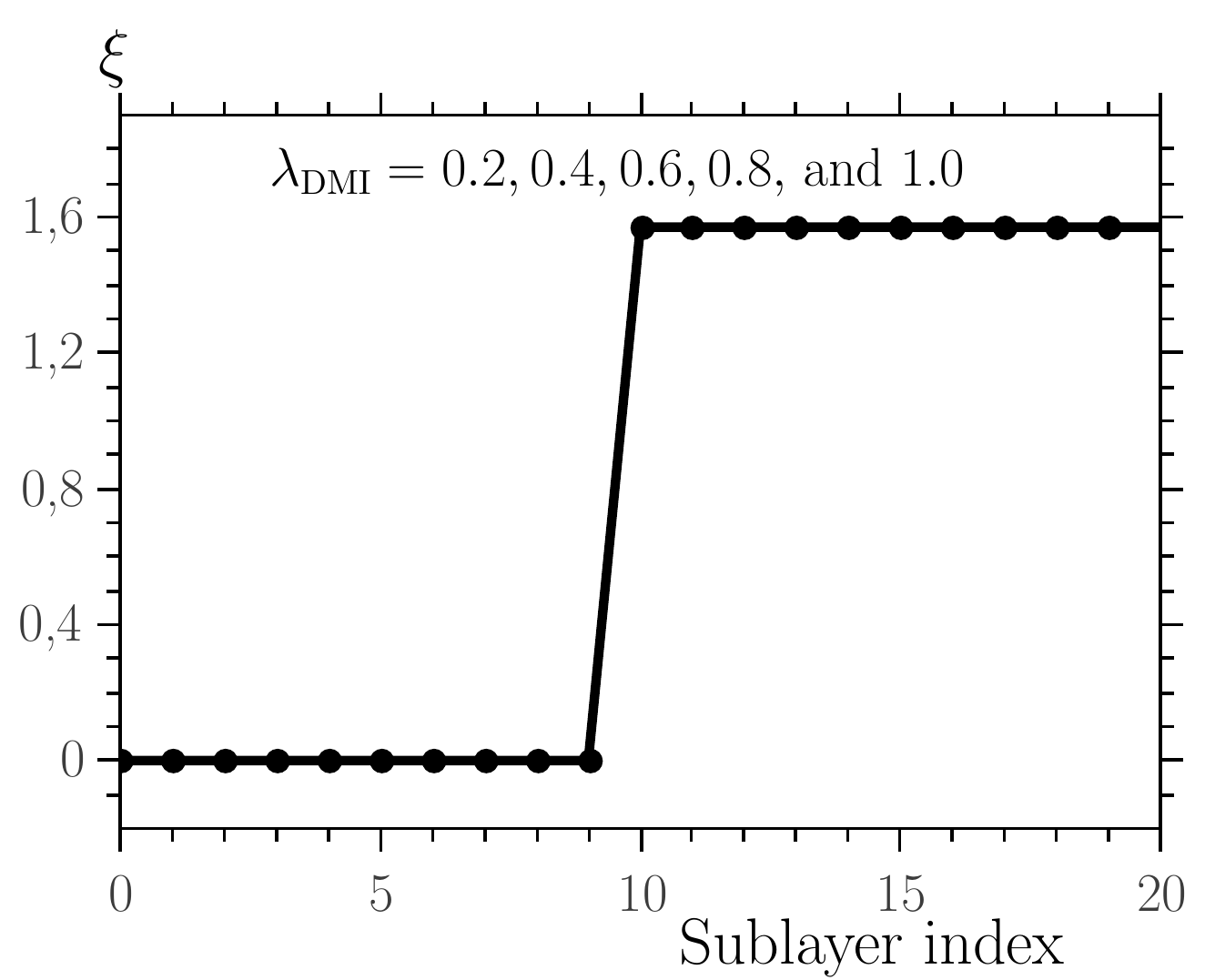}\label{fig:DMIxMPxi}}
\subfigure[ ]{\includegraphics[width=7cm]{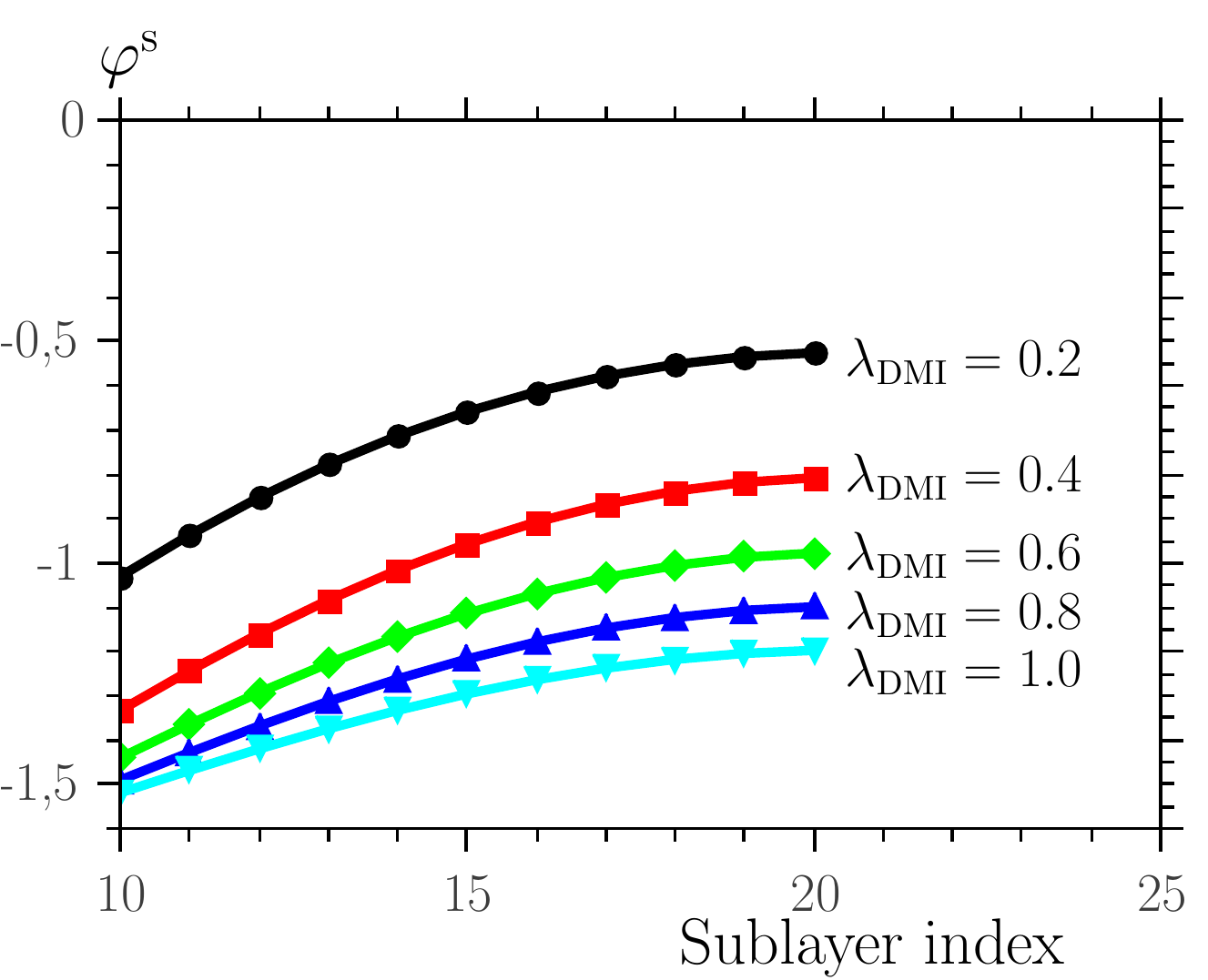}\label{fig:DMIxMPphi}}
\subfigure[ ]{\includegraphics[width=7cm]{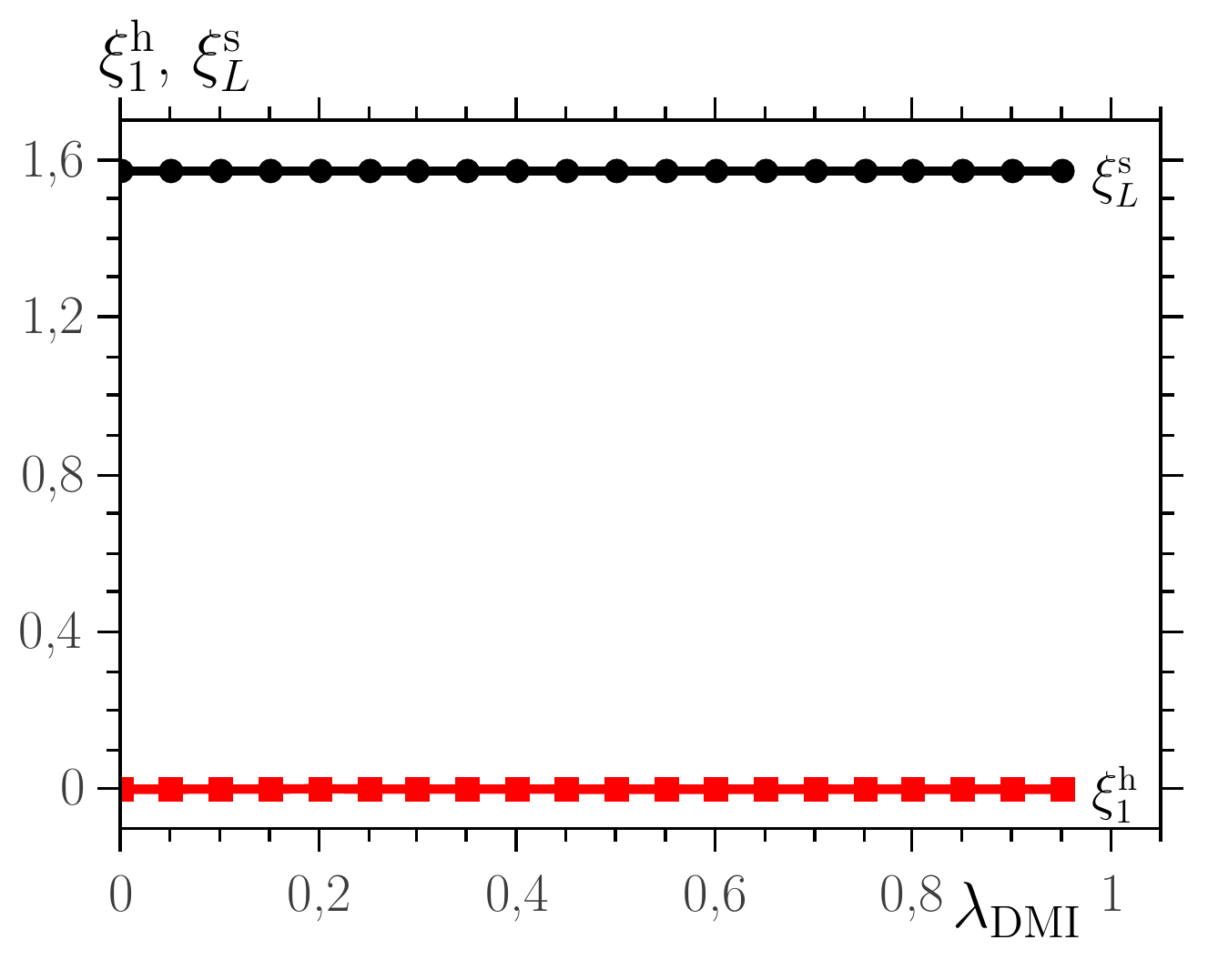}\label{fig:DMIxxiL}}
\subfigure[ ]{\includegraphics[width=7cm]{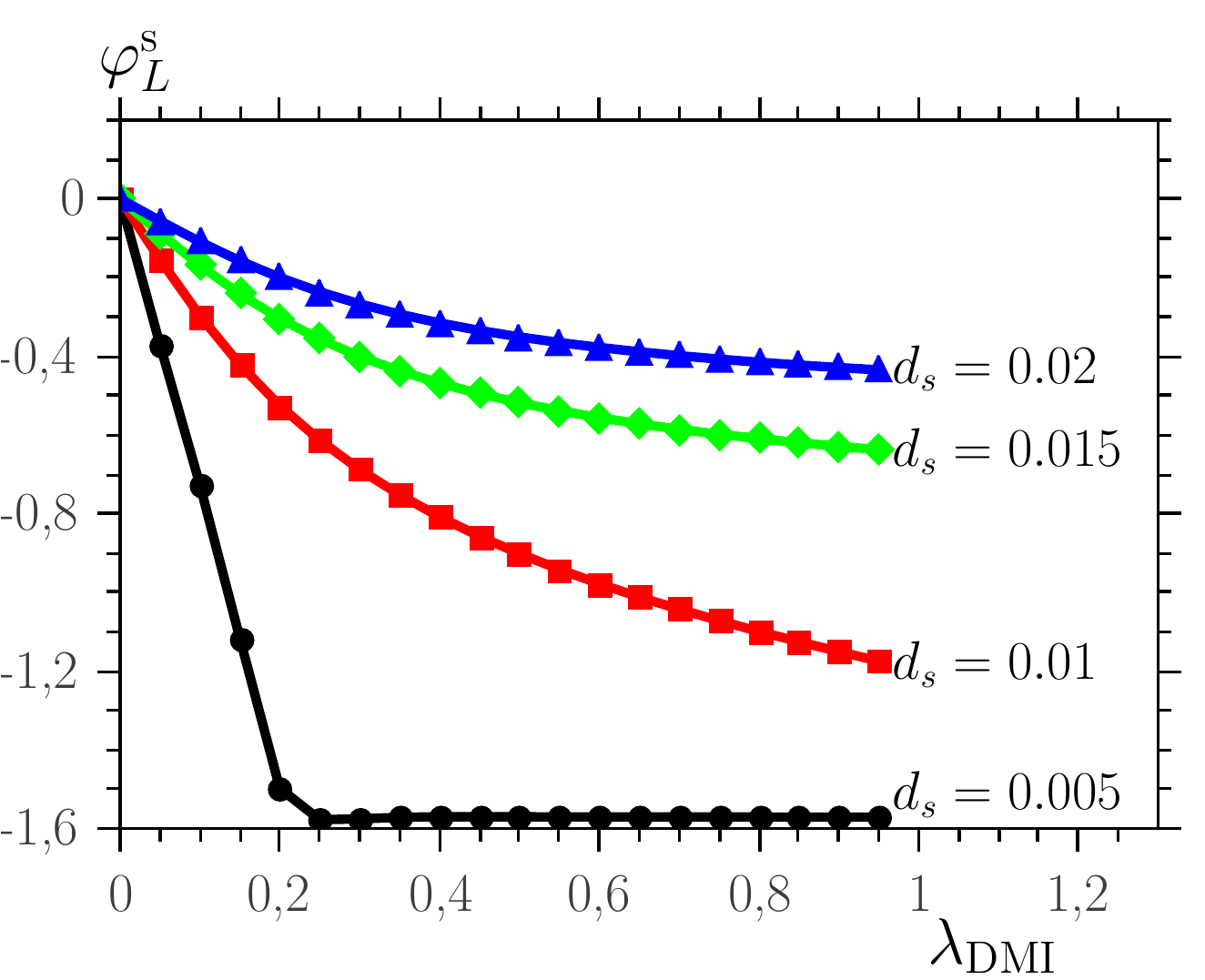}\label{fig:DMIxphiL}}
\caption{(a) System setup with inter-slab DMI interaction with a vector $\mathbf{D}$
perpendicular to the HMS easy axis; (b) variation of the polar MP ($\xi$);
(c) azimuthal MP ($\varphi$), (d) the polar and (e) azimuthal deviations
of the magnetic moments at the bottom HMS sub-layer $\xi_{1}^{\mathrm{h}}$
and the top SMS sub-layer $\xi_{L}^{\mathrm{s}}$ with the inter-slab
DMI $\lambda_{\mathrm{DDI}}$, for various values of $d_{s}$.}

\label{fig:DMIxMPandxiL} %
\end{figure*}

According to Eq. (\ref{eq:DM-DM}), the DMI tends to orientate the
magnetic moments in such a way that they are normal to each other
and perpendicular to the vector $\mathbf{D}$. Fig. \ref{fig:DMIxMPxi}
shows that with $\mathbf{D}$ along the $x$ axis, the HMS magnetic
moments at the interface align along their easy axis $\mathbf{e}_{z}$,
as it is perpendicular to the vector $\mathbf{D}$ and there are no
other fields that could lead to a deviation from it. Therefore, all
magnetic moments of the HMS slab will stick up along the direction
$\theta=0$. On the other hand, the SMS magnetic moment at the interface,
because it has to be perpendicular both to the HMS magnetic moment
($z$ axis) and the vector $\mathbf{D}$ ($x$ axis), lies in the $xy$
plane. However, in the SMS slab there is a competition between the
DMI and the anisotropy $d_{s}$ in the $xy$ plane leading to a variable
azimuthal angle $\varphi^{\mathrm{s}}$ as can be seen in Fig. \ref{fig:DMIxMPandxiL}
(c). With weaker DMI the anisotropy has a stronger effect and the SMS
magnetization at the interface lies near the $x$ axis. As DMI increases
the magnetization turns towards the $y$ axis as it has now to be
perpendicular to the vector $\mathbf{D}$.

In Fig. \ref{fig:DMIxxiL} we see that the HMS magnetization stays
along the $z$ axis ($\xi_{1}^{\mathrm{h}}=0$) while that of the
SMS slab remains in the $xy$ plane ($\xi_{L}^{\mathrm{s}}=\pi/2$)
for values of $d_{s}=0.005,0.01,0.015$ and $0.02$. This means that
the SMS magnetic moment at the interface is not affected by a varying
SMS anisotropy in this setup. However, as is seen in
Fig. \ref{fig:DMIxphiL} the $\varphi^{\mathrm{s}}$ dependence on
the SMS-HMS coupling is affected by the anisotropy, exhibiting higher
deviations for lower values of the anisotropy. Indeed, it is more
favorable for the interaction to win the competition and drive the
SMS interface magnetization onto the $y$ axis. For a strong enough
interaction the SMS magnetization at the interface is pinned along
the $y$ axis and the system can be modeled as a rigid magnet where
the magnetization of the SMS varies in-plane with no out-of-plane
component.\textbf{ }In this situation, the top SMS magnetization can
be calculated using Eq.~(\ref{eq:SMSTopLayer}) with now the angle
variation $\xi_{L}$ referring to $\varphi_{L}^{\mathrm{s}}$ instead,
because $\xi^{\mathrm{s}}$ is fixed at $\pi/2$.

\begin{figure*}[!htbp]
 \subfigure[ ]{\includegraphics[width=8cm]{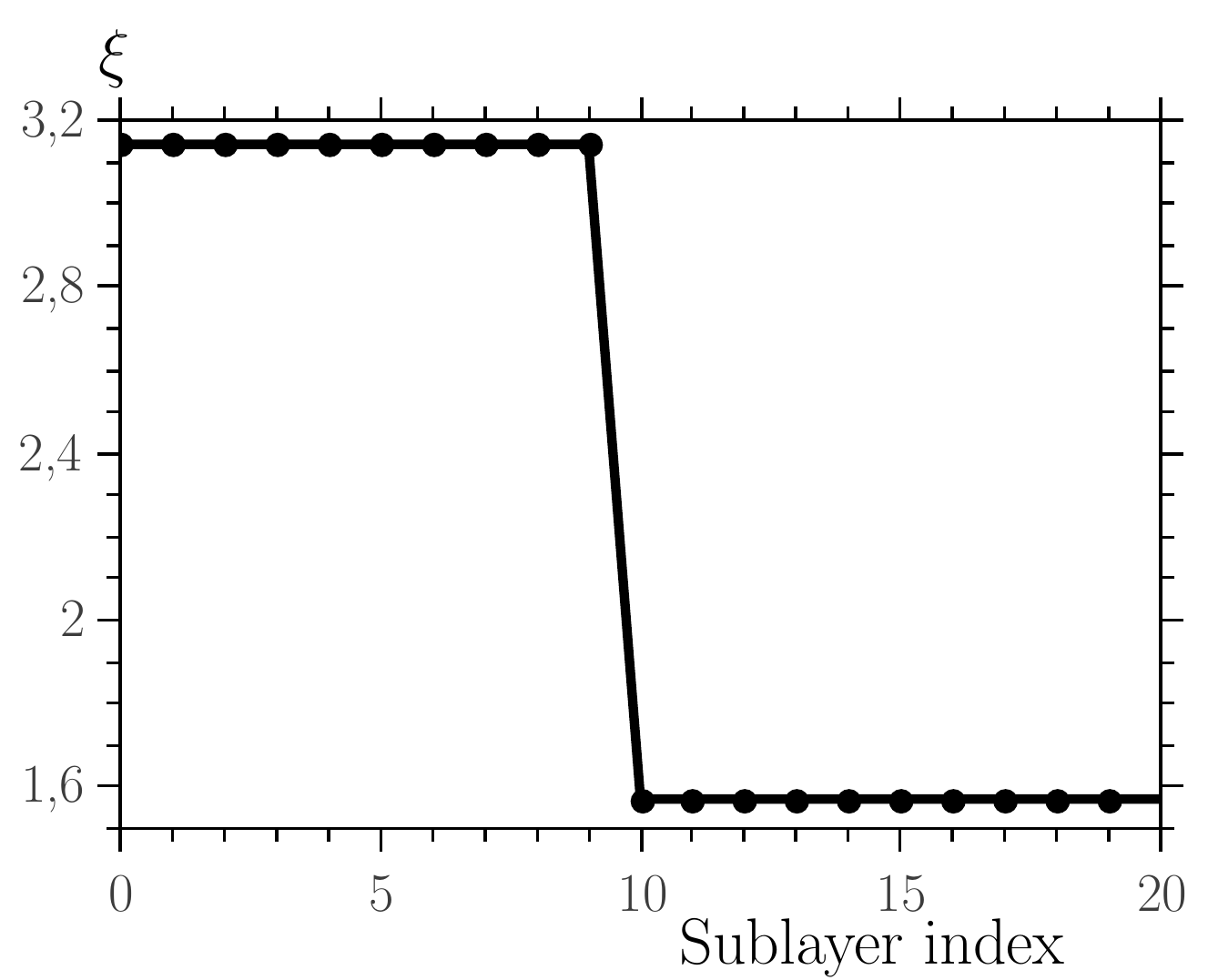}\label{fig:MagSwitchDMxTh}}
\subfigure[ ]{\includegraphics[width=8cm]{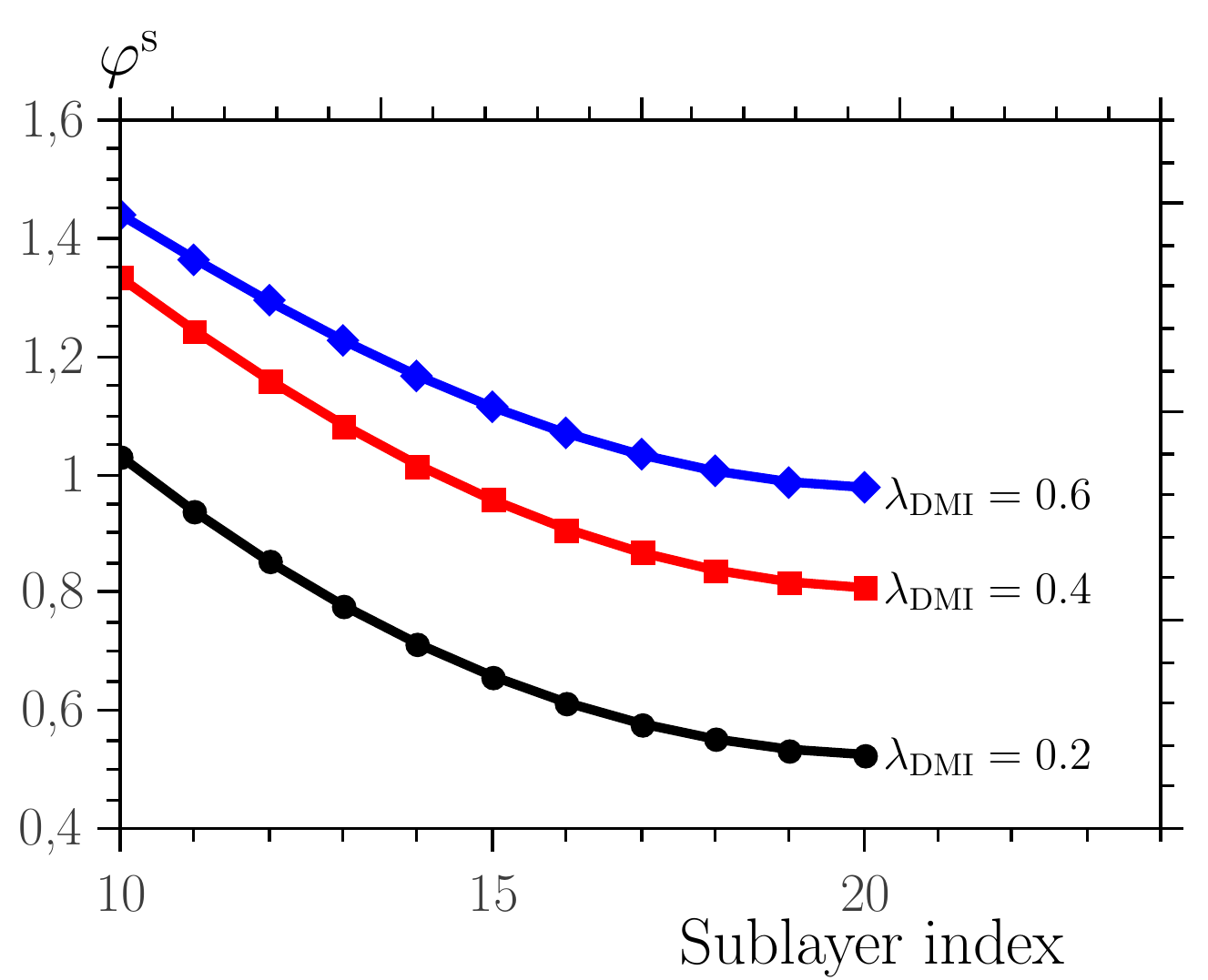}\label{fig:MagSwitchDMxPh}}
\caption{Magnetic {}``switch'' behavior in the (a) out-of-plane ($\xi$)
and (b) in-plane ($\varphi$) directions of the system with DMI at
the interface and vector $\mathbf{D}$ along the $x$ axis. $\mathcal{N}_{h}=10$
and $\mathcal{N}_{s}=11$.}

\label{fig:MagSwitchDMx} %
\end{figure*}

Fig. \ref{fig:MagSwitchDMx} shows the MP similar to the one shown
in Figs. \ref{fig:DMIxMPandxiL} (b) and (c), but in this case we
force the HMS magnetic moments to have a deviation $\xi^{\mathrm{h}}=\pi$,
opposite to that of Fig. \ref{fig:DMIxMPandxiL} (b), and we let the
system evolve. This causes the orientation of the SMS magnetic moments
to switch as well, thus demonstrating that for a system with this
specific setup, it can be used as a ``magnetic switch'' since
switching only the magnetic moment of the HMS slab induces a switching
of the SMS magnetic moment. This is also true the other way round as is the case in
exchange-spring systems where one attempts to achieve the reversal of the hard layer by
smaller DC fields upon acting on the soft layer.

\paragraph{DM vector in the z Direction\protect \protect \protect \protect
\protect \protect \\
 }

\begin{figure*}[!htbp]
 \subfigure[ ]{\includegraphics[width=6cm]{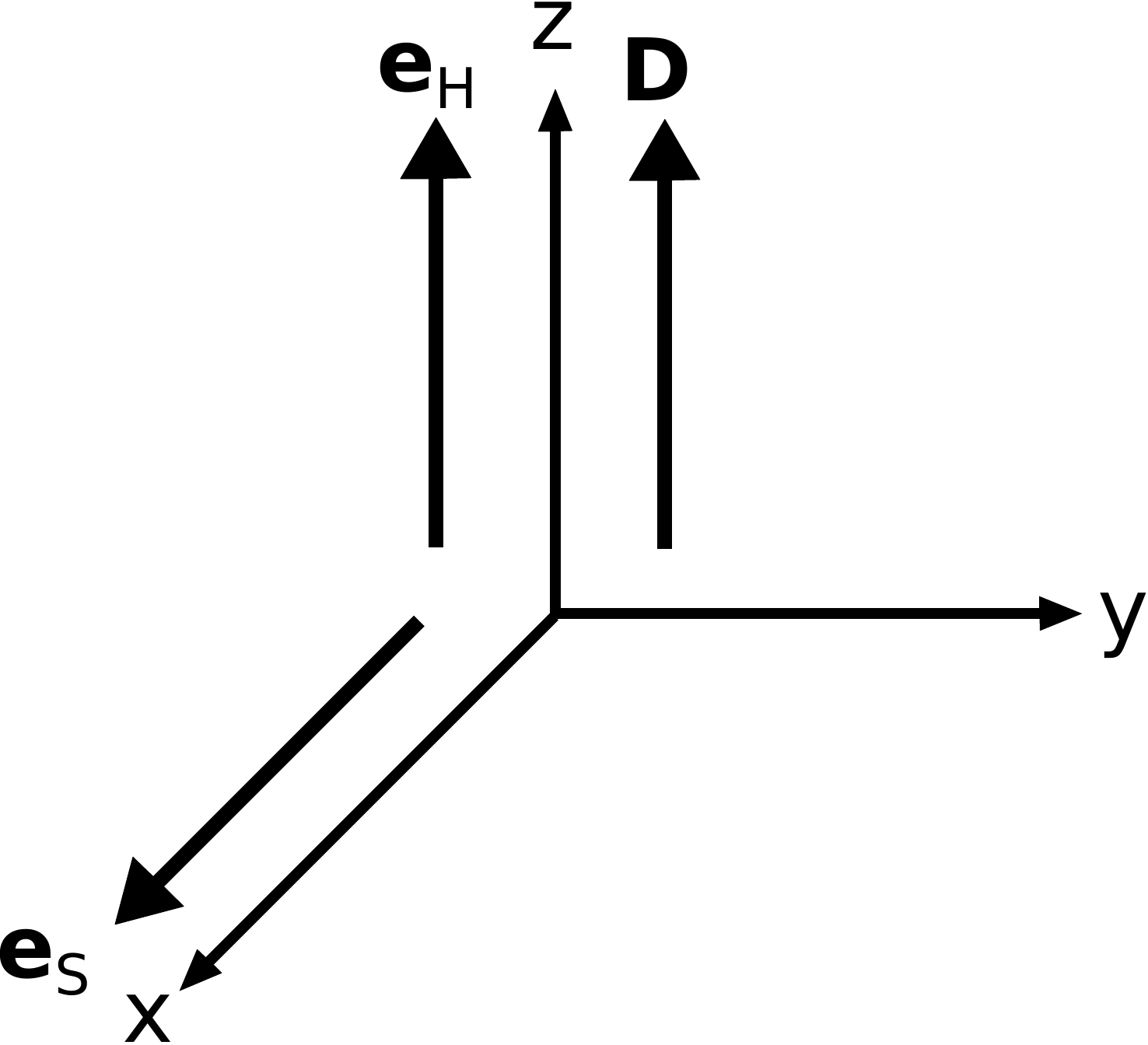}\label{fig:systemConfigESDMz}}
\subfigure[ ]{\includegraphics[width=8cm]{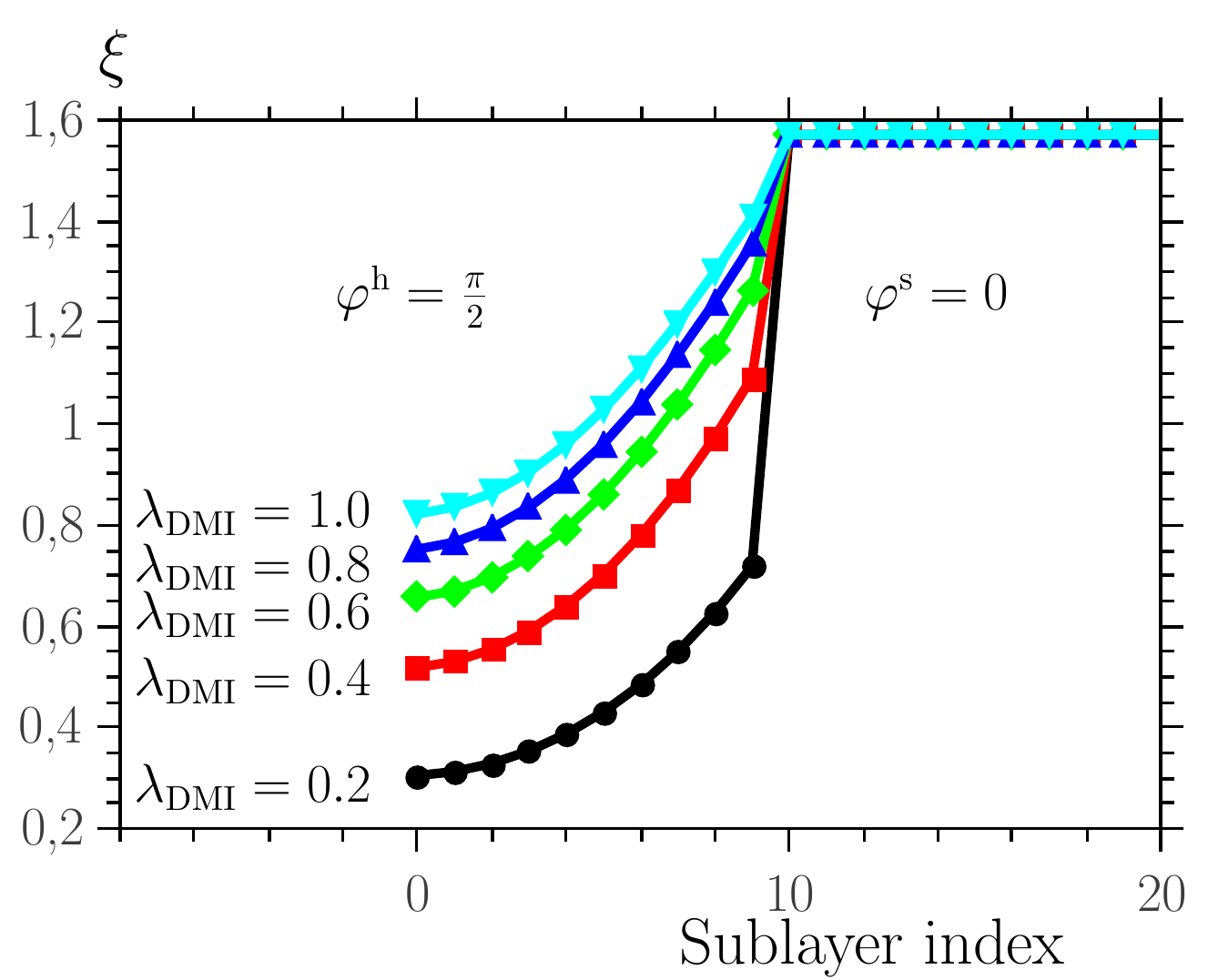}\label{fig:DMIzMPxi}}
\subfigure[ ]{\includegraphics[width=8cm]{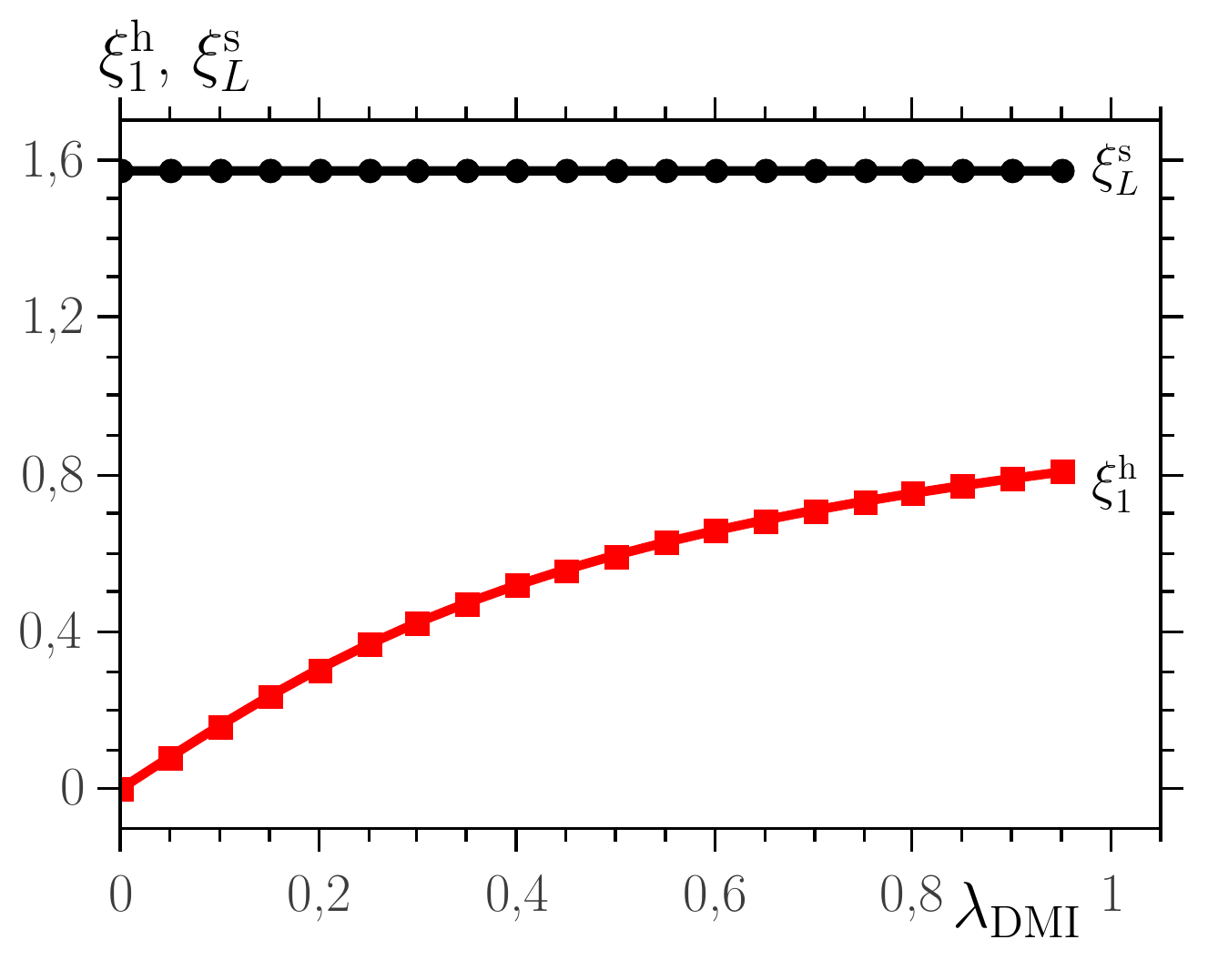}\label{fig:DMIzxiL}}
\caption{(a) System setup with inter-slab DMI interaction with a vector $\mathbf{D}$
parallel to the HMS anisotropy; (b) variation of the MP; (c) deviations of the
magnetic moments at the bottom HMS sub-layer $\xi_{1}^{\mathrm{h}}$
and the top SMS sub-layer $\xi_{L}^{\mathrm{s}}$ with the inter-slab
DMI $\lambda_{\mathrm{DDI}}$. $\mathcal{N}_{h}=10$ and $\mathcal{N}_{s}=11$.}

\label{fig:DMIzMPandxiL} %
\end{figure*}

Similarly to the previous case, the DMI will align the magnetic moment
of one of the layers at the interface along the anisotropy axis of
the corresponding slab, whereby its magnetization will not vary with
the interaction\textbf{.} In the present case of $\bm{D}\parallel\bm{e}_{z}$,
however, it is the SMS magnetic moment that remains constant because
its anisotropy axis is normal to the vector $\mathbf{D}$. On the
other hand, the HMS (net) magnetic moment becomes pinned in the $yz$
plane, implying that the angles $\varphi$ for both layers will remain
constant and equal to $\varphi^{\mathrm{h}}=\pi/2$ and $\varphi^{\mathrm{s}}=0$.
As such, the deviation at the interface for the HMS slab will vary
according to the competition between the HMS anisotropy and the DMI, see Fig. \ref{fig:DMIzMPxi}.
Hence, changes in the SMS anisotropy will not affect the MP of the
system. Indeed, calculations of the $\xi_1^h$ and $\xi^s_L$ for various values
of $d_{s}$ were
performed and the same results, shown in Fig. \ref{fig:DMIzxiL},
were obtained for all of them, thus confirming that the system is
not affected by varying the value of the SMS anisotropy.

For larger values of DMI the system can be viewed as an ``inverted''
rigid magnet, where the SMS and the HMS (net) magnetic moments at
the interface are pinned in-plane. The out-of-plane variation of the
HMS magnetic moment results from the competition between the HMS intra-slab
exchange interaction and anisotropy. In this situation, the variation
in the HMS magnetization can be calculated by using Eq.~(\ref{eq:SMSTopLayer}),
upon substituting $d_{h}$ for $d_{s}$.

\subsubsection{Comparison between the inter-slab couplings}

\begin{figure*}[!htbp]
 \subfigure[ ]{\includegraphics[width=8cm]{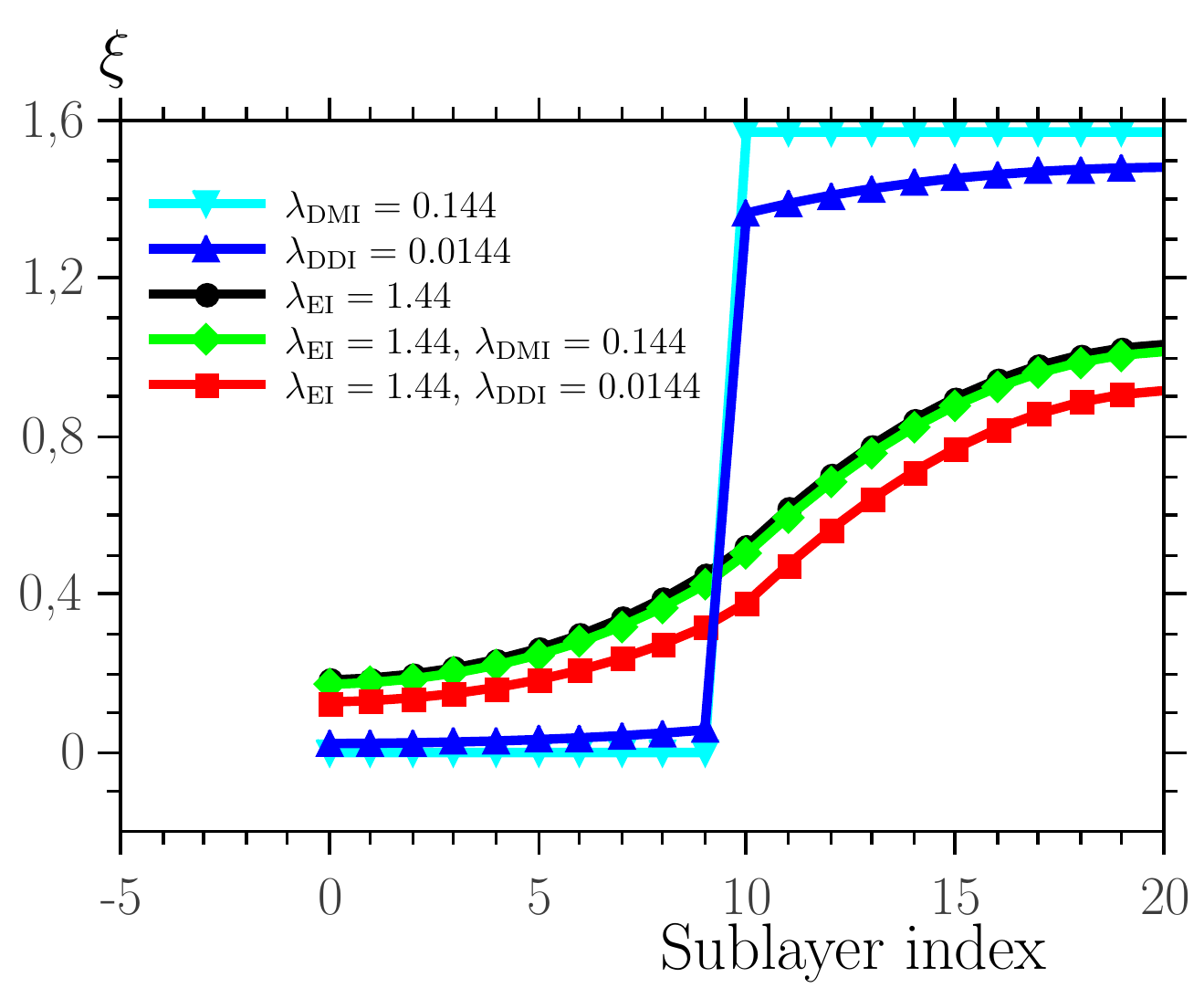}} \subfigure[
]{\includegraphics[width=8cm]{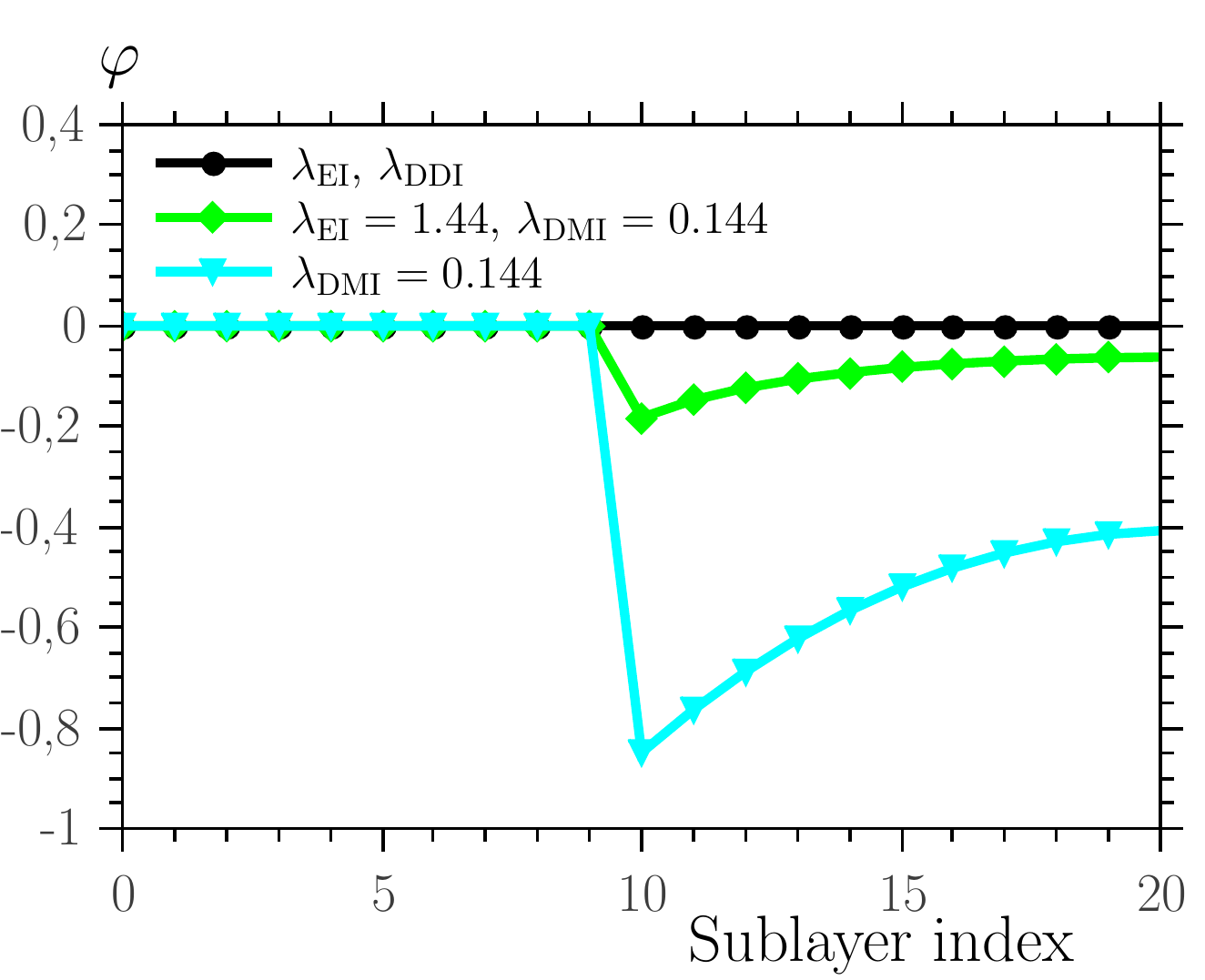}} 
\caption{\label{fig:jvsxivsDM_MP_RI}Effect of EI, DDI and DMI ($x$ direction)
on the magnetic profile of a exchange spring, with $j=1.44$, $\lambda_{DDI}=0.0144$
and $\delta=0.144$. $\mathcal{N}_{h}=10$ and $\mathcal{N}_{s}=11$.}
\end{figure*}

In Fig. \ref{fig:jvsxivsDM_MP_RI} we present different magnetic profiles
for typical values of the EI ($\lambda_{EI}=1.44$), DDI ($\lambda_{\mathrm{DDI}}/\lambda_{EI}\approx10^{-2}$)
and DMI ($\lambda_{\mathrm{DMI}}/\lambda_{EI}\approx0.1$), as can
be found in Refs. \citep{levfer81prb,crelac98jmmm,sousaetal10prb}.
Figs. \ref{fig:jvsxivsDM_MP_RI} (left, right) present the MP in the
polar ($\xi$) and azimuthal ($\varphi$) directions, respectively.
In the polar MP the black curve with circles is the MP with only EI
at the interface and serves as a reference. The red curve with squares
is the MP obtained as we add DDI. Compared to the EI strength, the
DDI is two orders of magnitude weaker and thereby its contribution
to the alignment of the magnetic moments at the interface is very
small. Nonetheless, it tends to align the interface along the $z$
axis and this effect propagates throughout all sub-layers by the in-plane
exchange interaction, causing a global decrease in the MP. The dark blue curve
(triangles up) represents the MP for DDI only. We see that the weak
DDI is not sufficient to overcome the anisotropy of each sub-layer,
leading only to a slight deviation near the interface but away from the
latter each slab remains parallel to its easy axis. 
We observe an
increased (induced) deviation in the SMS as compared to that of the
HMS. This is due to the stronger anisotropy of the HMS enhanced by
the DDI-induced anisotropy in the $z$ direction, making the out-of-plane
direction more favorable at the interface than the in-plane direction. 
On the other hand, for the reasons given earlier, the DMI with its vector
$\bm{D}$ along the $x$ axis, leads to a similar result with each slab aligned
along its own easy axis even near the interface.

The MP in the azimuthal direction shown in Fig. \ref{fig:jvsxivsDM_MP_RI}
(right) compares the three inter-slab couplings. The EI and DDI are
two interactions that do not induce an azimuthal rotation and the
corresponding MP always lies in the same plane ($\varphi=0$). The
green curve (with diamonds) is the MP obtained after adding DMI with
its vector $\mathbf{D}$ in the $x$ direction. In Fig. \ref{fig:jvsxivsDM_MP_RI}
we see that the DMI induces a slight decrease in the deviation $\varphi^{\mathrm{s}}$
because the DMI tends to align the magnetic moments of the SMS slab along the $y$ axis, 
as is seen previously in Fig. \ref{fig:DMIxphiL}.
Even though the magnitude of this interaction was not high enough
to overcome the anisotropy and internal exchange, it induced a deviation
at the interface in the azimuthal direction. This deviation is, however,
constrained by the EI at the interface and the anisotropies of the
slabs. The light blue curve (triangles down) represents the MP for
DMI with no exchange at the interface. As it could be expected, the
magnetic moments of the HMS are all aligned along the $z$ axis (recall
that $\mathbf{D}$ is along the $x$ axis), while those of the SMS
slab are all in the $xy$ plane with a higher $\varphi$ deviation
at the interface than in the previous case. This is obviously due
to the fact that now the DMI competes only with the anisotropy of
the SMS slab.

\subsubsection{Hysteresis loops}

Here we present a succinct study of the hysteresis loop for different
values of the interaction. It helps us to understand how the switching
mechanism of the multi-layer system changes with the nature and strength
of the inter-slab coupling. All the hysteresis curves presented are
plots of the normalized magnetization $M_{z}/M_{s}$ versus reduced
applied magnetic field $h$, along the HMS anisotropy axis ($z$
axis). $\mathcal{N}_{h}=12$ and $\mathcal{N}_{s}=16$. We start in
a state of zero magnetization (\textit{i.e.} the magnetic moments
of all sub-layers are on the plane with no applied field) and iterate
the LLE (\ref{eq:LLE}) until we reach the state of minimal energy. Next, we
increase
the applied field slightly and wait for the system to reach the new
equilibrium state. We keep increasing the applied field until we reach
saturation of the magnetization. Then, we ramp the field down and
up again thus closing the hysteresis loop. Obviously, for each value
of $h$, we wait until the state of minimal energy is reached.

\begin{figure*}[!htbp]
 \subfigure[ ]{\includegraphics[width=8cm]{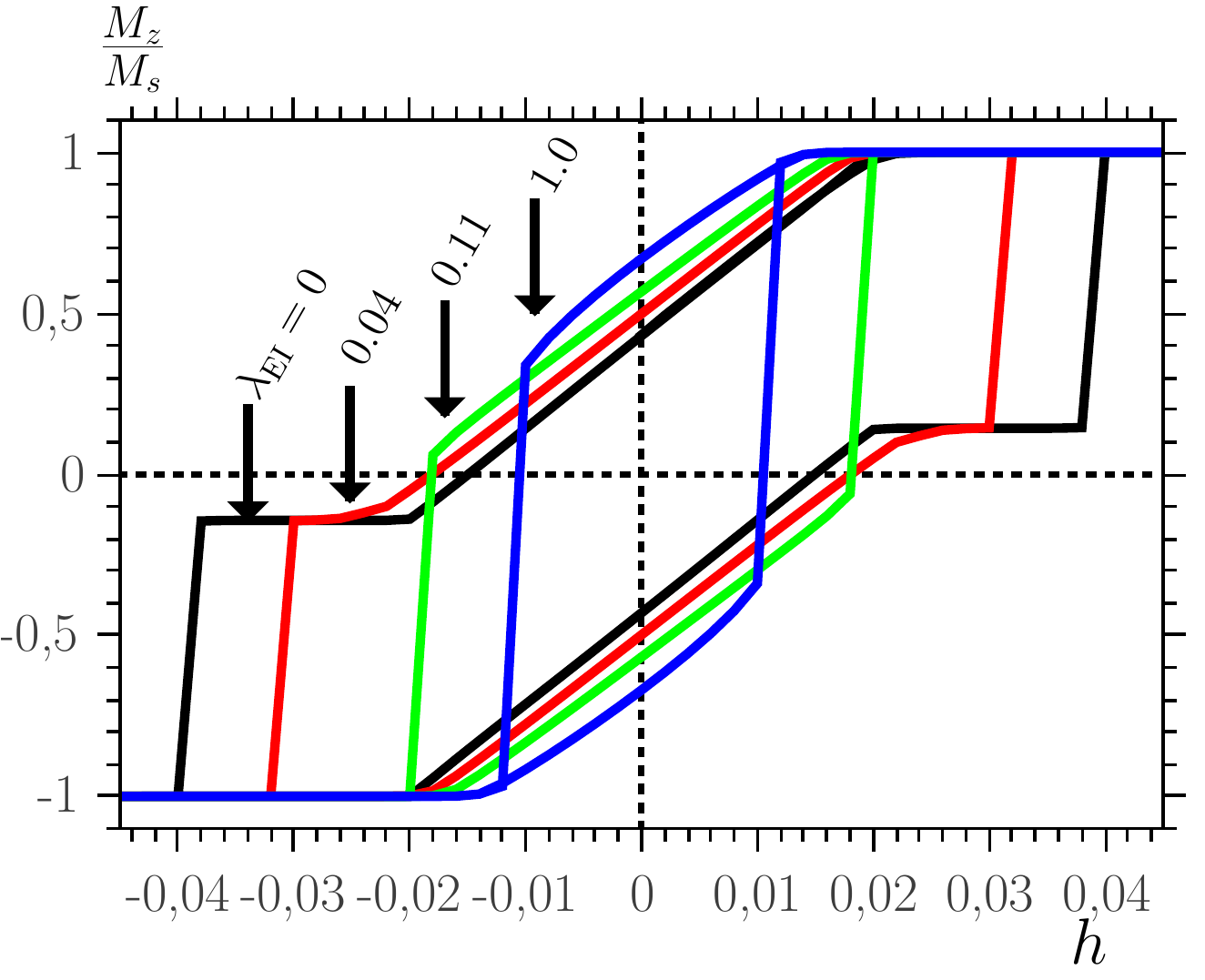}} \subfigure[ ]{\includegraphics[width=8cm]{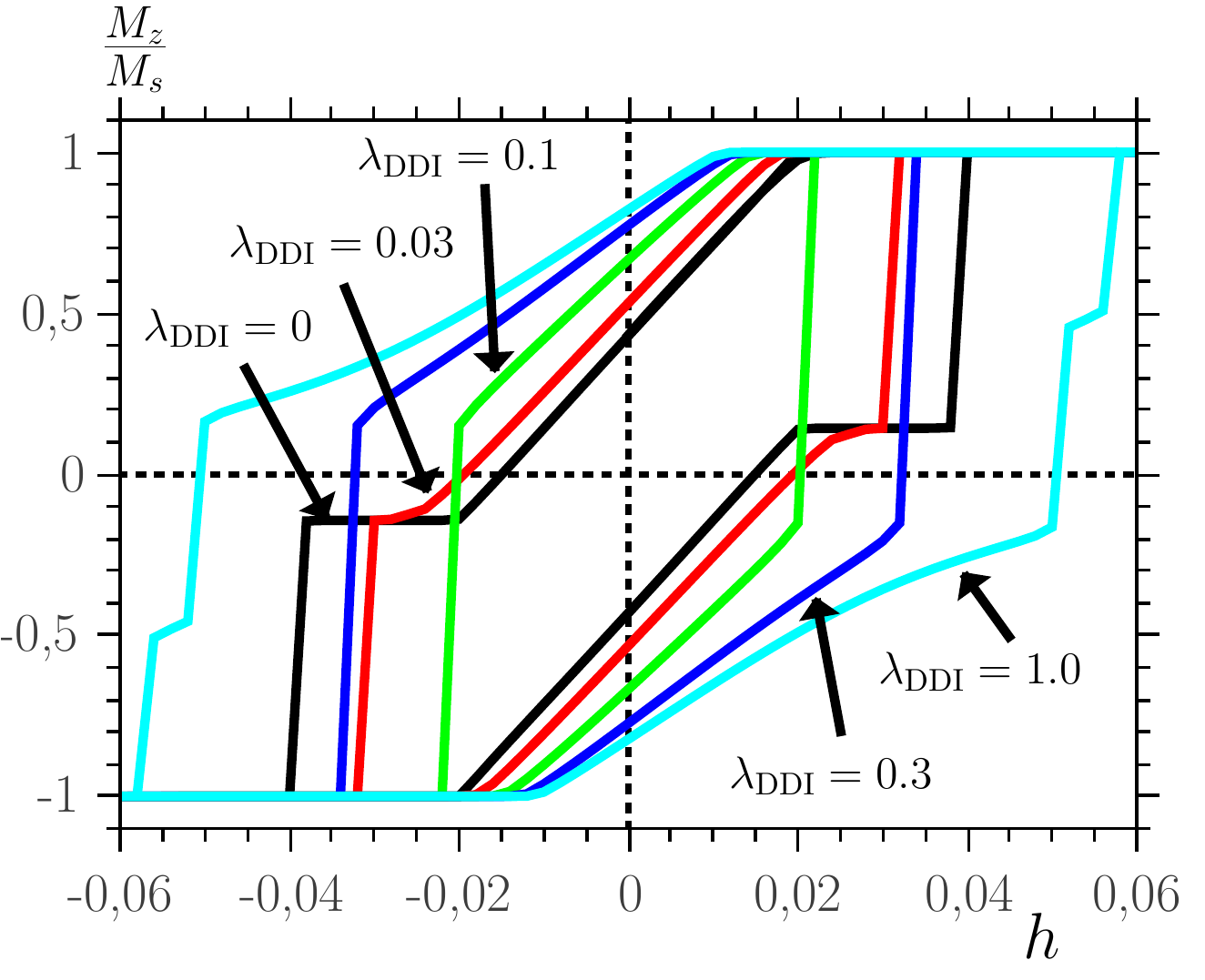}}

\subfigure[ ]{\includegraphics[width=8cm]{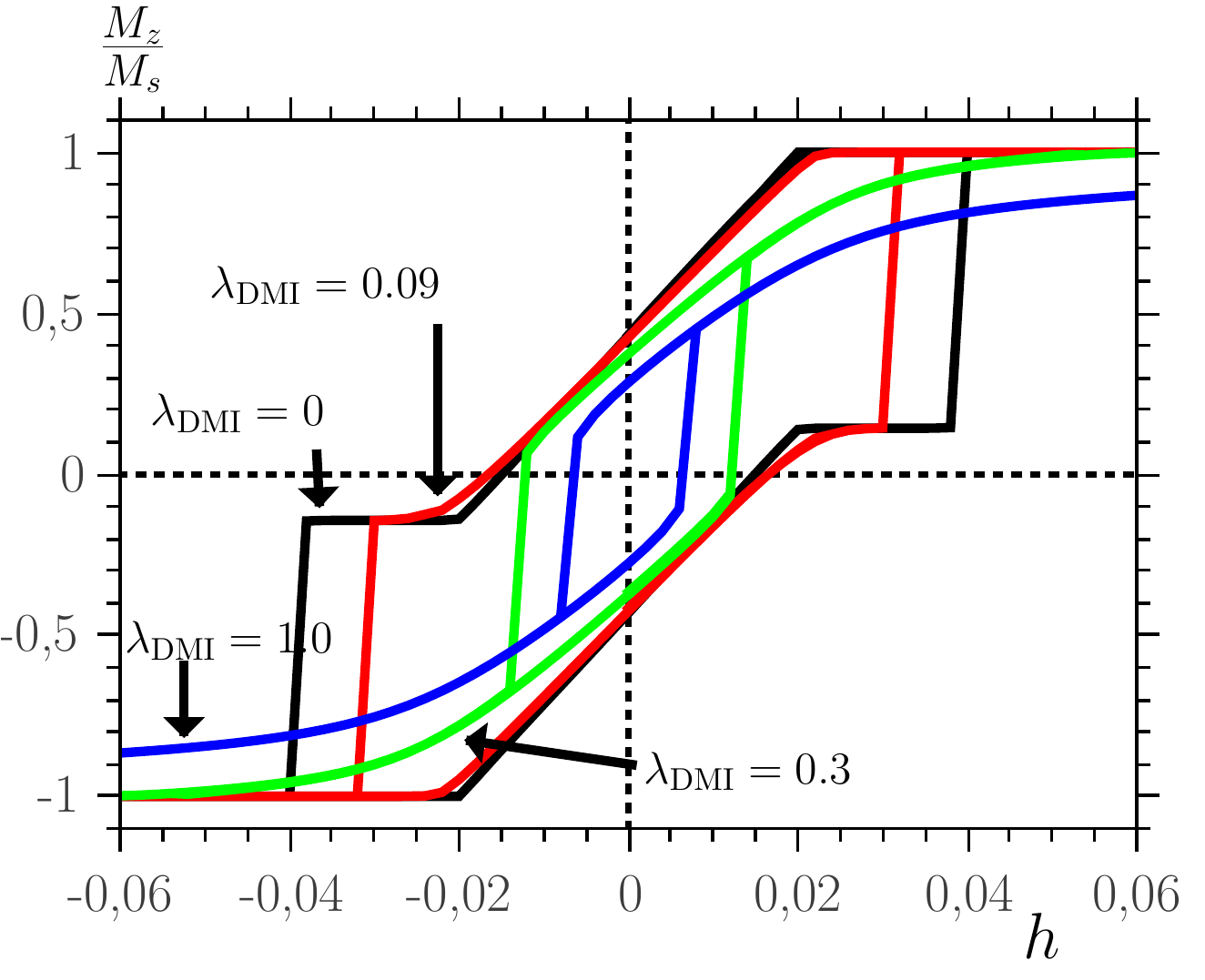}}

\caption{(a) Effect of EI, (b) of DDI, and (c) of DMI (along the $z$ axis)
at the HMS-SMS interface on the hysteresis cycle and switching mechanisms
of the HMS/SMS-SI multi-layer systems with $\mathcal{N}_{h}=12$ and
$\mathcal{N}_{s}=16$.}

\label{fig:HysteresisHz} %
\end{figure*}

Figure \ref{fig:HysteresisHz} (a) shows hysteresis cycles for
different values of EI. Let us denote by $h_{s}^{c}$ the SMS switching
field, by $h_{h}^{c}$ the HMS switching field, and by $h_{\mathrm{sat}}$
the saturation field of the whole multi-layer system. With no interaction
($\lambda_{\mathrm{EI}}=0$) and for $h=0$ the equilibrium state
is that where the magnetic moments (of HMS and SMS) are oriented along
their respective easy axes, \emph{i.e.} $\theta_{k}=0$ and $\theta_{i}=\pi/2$
for all $k\in\mathrm{HMS},i\in\mathrm{SMS}$. The net magnetization
of the system is that given by the HMS net magnetic moment projected
on the $z$ axis. When $h$ is increased the magnetic moments of the
SMS start to rotate towards the applied field in a reversible process until
the system reaches saturation ($\theta_{k}=\theta_{i}=0$ for all
$k,i$) at $h=h_{\mathrm{sat}}$. As we ramp down the field across
zero until it reaches the SMS anisotropy field ($h_{s}^{c}=2d_{s}=0.02$),
the SMS magnetic moments switch towards the field direction $\theta_{i}=\pi$
for all $i\in\mathrm{SMS}$. Further increase of the applied field
(in the opposite direction) induces a slight reversible deviation
of the HMS magnetization. In Fig.~\ref{fig:HysteresisHz} (a) this
corresponds to the plateau (on the $\lambda_{\mathrm{EI}}=0$ curve) from $h=-0.02$ to
$h\simeq-0.04$, until $h$ reaches the HMS anisotropy field
$h_{h}^{c}=h_{\mathrm{sat}}\simeq2d_{h}=0.04$.
At this value of the field, the HMS magnetic moments coherently switch
from $\theta_{k}\approx0$ to $\theta_{k}=\pi$, thus achieving negative
saturation. Along the lower branch the system follows the same switching
process from $\theta=\pi$ to $\theta=0$.

For nonzero but weak coupling, $\lambda_{\mathrm{EI}}=0.04$, the
system follows the same behavior, but neither the HMS nor the SMS
goes through a coherent rotation because of the inter-slab interaction.
The SMS saturation field $h_{s}^{c}$ is now higher because its magnetization
is stabilized by the interaction with the HMS. On the other hand,
$h_{h}^{c}=h_{\mathrm{sat}}$ decreases because the molecular field
of the already reversed SMS acts against the HMS anisotropy field.
For stronger EI ($\lambda_{\mathrm{EI}}=0.11,1$) the plateau disappears
completely ($h_{s}^{c}=h_{h}^{c}=h_{\mathrm{sat}}$), indicating that
once the SMS magnetic moments reach a certain deviation, the EI induces
a cascade effect in the HMS that causes the switching of the latter.
The same observations of nonuniform magnetization switching leading to a smaller coercive
field were made for magnetic recording exchange-spring media \cite{wangetal05apl,
bergeretal08apl}.
Finally, an increase of the EI increases the remanent magnetization.
The coercive field increases as long as $h_{\mathrm{sat}}>h_{s}^{c}$.
As soon as the EI becomes strong enough to observe the cascade effect
($h_{\mathrm{sat}}=h_{s}^{c}$), the coercive field becomes a decreasing
function of EI.
Now, since the inter-layer exchange coupling decreases for an increasing spacer
thickness, the coercive field passes by a minimum when the latter increases, again in
agreement with what has been observed in Refs. \onlinecite{wangetal05apl,
bergeretal08apl}.

Weak ($\lambda_{\mathrm{DDI}}=0,0.03$) or medium ($\lambda_{\mathrm{DDI}}=0.1,0.3$)
DDI has a similar behavior to that of the EI, as can be seen in Fig.~\ref{fig:HysteresisHz}
(b). There is some difference, however, between the EI and DDI regarding
the evolution of the hysteresis cycle as the inter-slab coupling is
varied. While increasing EI decreases $h_{\mathrm{sat}}$, the DDI
induces an increase of $h_{\mathrm{sat}}$ with increasing $\lambda_{\mathrm{DDI}}$.
Indeed, DDI whose bond is along the $z$ axis induces an additional anisotropy at
the interface along this axis and thereby tends to stabilize the
magnetization in this direction. 
For $\lambda_{\mathrm{DDI}}=1.0$ the induced anisotropy field is stronger
than the HMS anisotropy field. This implies that it is possible for
the magnetic moments of the HMS that are far from the interface to
switch before those at the interface. When the applied field becomes
in excess of the induced-anisotropy field a complete switching is
achieved. Finally, both the remanent magnetization and coercive
field increase with DDI.

The effect of the DMI along the $z$ axis on the hysteresis cycle
is shown in Fig.~\ref{fig:HysteresisHz} (c). A similar behavior
to those exhibited by EI and DDI for weak interaction ($\lambda_{\mathrm{DMI}}=0.09$)
is observed. Again, we observe a progressive rotation of the SMS magnetization
before the applied field becomes strong enough to induce a reversal
of the HMS magnetization. As we discussed earlier, with no applied
field the DMI along the $z$ axis induces a variation in the deviation
of the magnetic moments of the HMS, while pinning the SMS (net) magnetic
moment along the $x$ axis. The slight deviation induced in the magnetic
moments of the HMS leads to a reduced HMS switching field $h_{h}^{c}=h_{\mathrm{sat}}$.
Further increase of the interaction ($\lambda_{\mathrm{DMI}}=0.3,1$)
induces again a cascade effect in the HMS that causes both slabs to
switch in the same field ($h_{s}^{c}=h_{h}^{c}$). However, we have
to note that unlike EI and DDI, the HMS/SMS switching field is not
equal to $h_{\mathrm{sat}}$. This is again due to the deviation induced
in the HMS magnetic moments by DMI. Both the remanent magnetization
and the coercive field decrease with increasing DMI.

We can see that each kind of interaction induces a specific {}``single
magnetic moment''-like behavior for strong coupling $\lambda$. A
strong EI tends towards a system with no anisotropy, whereby the area
of the loop tends to vanish, reaching saturation with very low applied
fields in a switch-like behavior. DDI on the other hand tends to induce
a square loop, typical of a very high uniaxial anisotropy with easy
axis parallel to the applied field. Finally, the DMI tends to narrow
the cycle and the magnetization follows the applied field, typical
of a system with uniaxial anisotropy perpendicular to the field. Obviously,
the perfect {}``single magnetic moment'' behavior is never reached
because the inter-slab interaction is present only at the interface
and the deviation of the outer sub-layers is limited only by the intra-slab
exchange interaction and the applied field.

\section{Conclusion}

We have studied a magnetic multi-layer system composed of a hard magnetic
slab with out-of-plane uniaxial anisotropy and a soft magnetic slab
with in-plane uniaxial anisotropy, separated by a nonmagnetic spacer.
We have considered three cases of inter-slab coupling, namely exchange,
dipolar or Dzyaloshinski-Moriya interactions. The soft magnetic slab
has been modeled as a stack of atomic (e.g. Fe) layers, while the
hard magnetic slab has been modeled either as a macroscopic magnetic
moment or as another stack of atomic (e.g. FePt) layers. Each atomic
layer is modeled as a macroscopic magnetic moment representing its
net magnetic moment, and is coupled to adjacent layers by exchange
coupling.

We have investigated the effect of the external magnetic field, the
in-plane or out-of-plane anisotropy, and the three interactions on
the deviation angle (relative to the hard slab anisotropy easy axis).
We have computed the magnetization profile through the whole multi-layer
system. Our computing method consists in solving the set of (coupled)
Landau-Lifshitz equations for the net magnetic moments of the layers. We
first validate this method by comparing the corresponding results
to the previously obtained analytical expressions for the case of
exchange inter-slab coupling.

For the effect of the applied magnetic field, we found that with sufficient
number of layers in the soft slab, the system behaves according to
the Stoner–Wohlfarth regime and that there exists a critical field
at which the whole system aligns along the applied field.

For exchange and dipolar interactions, there is an asymptotic value
that depends on the anisotropy and the intra-layer exchange. For the
dipolar interaction with a bond along the hard slab anisotropy, this
asymptotic value is given by the analytical expression for rigid interface,
where the hard slab is modeled as a single pinned magnetic moment,
and only the variation in the soft slab is relevant. The dipolar interaction
was next extended through the whole soft slab with a rigid interface.
The ensuing effect on the magnetization profile of the soft slab has
been recovered by an effective dipolar coupling at the interface only,
upon re-scaling the intra-slab exchange coupling. The two corresponding
effective coupling parameters depend on the initial dipolar interaction,
but not on the number of SMS layers.

For the Dzyaloshinski-Moriya interaction, we found that the magnetic
moments of one of the slabs are pinned in a given direction, whereas
those of the remaining slab rotate in either the polar angle or the azimuthal
angle, depending on the direction of the vector $\mathbf{D}$. Large
values of the DM coupling lead to a system with rigid interface, with
the magnetic moments at the interface being perpendicular to each
other and to the vector $\mathbf{D}$. In addition, a switch-like
mechanism can be achieved with this interaction. Indeed, an indirect
reversal of the SMS magnetization can be achieved by directly forcing
a reversal of the HMS magnetization with the help of a magnetic field.
It is obviously also possible to obtain the desired effect for the exchange-spring system
by achieving the reversal of HMS via the switching of the SMS magnetization with the help
of a smaller magnetic field.

A comparison between the three interactions with typical
orders of magnitude has been given. The exchange coupling shows the
strongest effect, and when added, the dipolar and Dzyaloshinski-Moriya
interactions induce a slight (but non negligible) deviation in either
the polar or azimuthal magnetization profile.

Finally, hysteresis cycles for the three different interactions are
computed. A typical exchange spring behavior, where the soft slab
switches first, followed by the hard slab at a stronger field, is
observed for weak interaction in all cases. Strong coupling causes
both slabs to switch under the same field. Exchange and Dzyaloshinski-Moriya
interactions tend to narrow the cycle, while the dipolar interaction
leads to squared cycles. For application purposes such as magnetic recording using
vertical exchange-spring media, the Dzyaloshinski-Moriya inter-layer coupling
strongly reduces the coercive field while keeping high values of the anisotropy
which thus ensure good thermal stability.

A work in progress consists in treating each atomic layer as a two-dimensional
lattice with the aim to compute spin correlations as functions of
the various energy parameters and to determine the spin-wave spectrum.
In the near future, this zero-temperature study will be extended to
finite temperature with the aim to investigate thermal effects with
special emphasis on the calculation of activation rates of such multi-layer
systems and thereby assess their thermal stability.

%\bibliography{hkbib}

\end{document}